\pdfoutput=1 

\documentclass[usenatbib]{mn2e}
\usepackage{graphicx}

\usepackage[total={17.8cm,24.0cm},centering]{geometry}
\usepackage{times,deluxetable}

\newcommand{\apj}{ApJ}           
\newcommand{\apjl}{ApJ}           
\newcommand{\mnras}{MNRAS}       
\newcommand{\nat}{Nature}
\newcommand{\aap}{A\&A}

\newcommand{\araa}{ARA\&A}
\newcommand{\aj}{AJ}
\newcommand{\pasp}{PASP}
\newcommand{\apjs}{ApJS}           

\newcommand{\na}{New Astr.}

\newcommand{\sauron}{\texttt{SAURON}}
\newcommand{\atl}{ATLAS$^{\rm 3D}$}
\newcommand{\kms}{\hbox{km s$^{-1}$}}
\newcommand{\msun}{\hbox{$M_\odot$}}
\newcommand{\lsun}{\hbox{$L_\odot$}}
\newcommand{\re}{\hbox{$R_{\rm e}$}}
\newcommand{\rmaj}{\hbox{$R_{\rm e}^{\rm maj}$}}
\newcommand{\se}{\hbox{$\sigma_{\rm e}$}}
\newcommand{\mjam}{\hbox{$M_{\rm JAM}$}}
\newcommand{\mljam}{\hbox{$(M/L)_{\rm JAM}$}}
\newcommand{\plotone}[1]{\includegraphics[width=\columnwidth]{#1}}
\newcommand{\refsec}[1]{Section~\ref{#1}}
\newcommand{\reffig}[1]{Fig.~\ref{#1}}
\newcommand{\refeq}[1]{equation~(\ref{#1})}

\title[The \atl\ project -- XX. Mass-size]
{The \atl\ project -- XX. Mass-size and mass-$\sigma$ distributions of early-type galaxies: bulge fraction drives kinematics, mass-to-light ratio, molecular gas fraction and stellar initial mass function}

\author[M.~Cappellari et al.]
{Michele Cappellari,$^1$\thanks{E-mail: cappellari@astro.ox.ac.uk}
Richard M. McDermid,$^{2}$
Katherine Alatalo,$^3$
Leo Blitz,$^3$\and
Maxime Bois,$^{4}$
Fr\'ed\'eric Bournaud,$^{5}$
M.~Bureau,$^1$
Alison F. Crocker,$^6$
Roger L. Davies,$^1$\and
Timothy A. Davis,$^{1,7}$
P. T. de Zeeuw,$^{7,8}$
Pierre-Alain Duc,$^{5}$
Eric Emsellem,$^{7,9}$\and
Sadegh Khochfar,$^{10}$
Davor Krajnovi\'c,$^7$
Harald Kuntschner,$^{7}$
Raffaella Morganti,$^{11,12}$\and
Thorsten Naab,$^{13}$
Tom Oosterloo,$^{11,12}$
Marc Sarzi,$^{14}$
Nicholas Scott,$^{1,15}$
Paolo Serra,$^{11}$\and
Anne-Marie Weijmans$^{16}$
and Lisa M. Young$^{17}$\\
$^1$ Sub-department of Astrophysics, Department of Physics, University of Oxford, Denys Wilkinson Building, Keble Road, Oxford OX1 3RH\\
$^2$ Gemini Observatory, Northern Operations Centre, 670 N. A`ohoku Place, Hilo, HI 96720, USA\\
$^3$ Department of Astronomy, Campbell Hall, University of California, Berkeley, CA 94720, USA\\
$^4$ Observatoire de Paris, LERMA and CNRS, 61 Av. de l'Observatoire, F-75014 Paris, France\\
$^5$ Laboratoire AIM Paris-Saclay, CEA/IRFU/SAp, CNRS, Universit\'e Paris Diderot, 91191 Gif-sur-Yvette Cedex, France\\
$^6$ Department of Astrophysics, University of Massachusetts, 710 North Pleasant Street, Amherst, MA 01003, USA\\
$^7$ European Southern Observatory, Karl-Schwarzschild-Str. 2, 85748 Garching, Germany\\
$^8$ Sterrewacht Leiden, Leiden University, Postbus 9513, 2300 RA Leiden, the Netherlands\\
$^9$ Universit\'e Lyon 1, Observatoire de Lyon, Centre de Recherche Astrophysique de Lyon and\\ Ecole Normale Sup\'erieure de Lyon, 9 avenue Charles Andr\'e, F-69230Saint-Genis Laval, France\\
$^{10}$ Max-Planck Institut f\"ur extraterrestrische Physik, PO Box 1312, D-85478 Garching, Germany\\
$^{11}$ Netherlands Institute for Radio Astronomy (ASTRON), Postbus 2, 7990 AA Dwingeloo, The Netherlands\\
$^{12}$ Kapteyn Astronomical Institute, University of Groningen, Postbus 800, 9700 AV Groningen, The Netherlands\\
$^{13}$ Max-Planck Institut f\"ur Astrophysik, Karl-Schwarzschild-Str. 1, 85741 Garching, Germany\\
$^{14}$ Centre for Astrophysics Research, University of Hertfordshire, Hatfield, Herts AL1 9AB, UK\\
$^{15}$ Centre for Astrophysics \& Supercomputing, Swinburne University of Technology, PO Box 218, Hawthorn, VIC 3122, Australia\\
$^{16}$ Dunlap Institute for Astronomy \& Astrophysics, University of Toronto, 50 St. George Street, Toronto, ON M5S 3H4, Canada\\
$^{17}$ Physics Department, New Mexico Institute of Mining and Technology, Socorro, NM 87801, USA}

\date{Accepted 2013 April 15. Received 2013 March 24; in original form 2012 August 17}

\pagerange{\pageref{firstpage}--\pageref{lastpage}} \pubyear{2013}

\begin{document}
\label{firstpage}
\maketitle

\clearpage
\begin{abstract}
In the companion Paper~XV of this series we derive accurate total mass-to-light ratios $(M/L)_{\rm JAM}\approx(M/L)(r=\re)$ within a sphere of radius $r=\re$ centred on the galaxy, as well as stellar $(M/L)_{\rm stars}$ (with the dark matter removed) for the volume-limited and nearly mass selected (stellar mass $M_\star\ga6\times10^9 \msun$) \atl\ sample of 260 early-type galaxies (ETGs, ellipticals Es and lenticulars S0s). Here we use those parameters to study the two orthogonal projections $(\mjam,\se)$ and $(\mjam,\rmaj)$ of the thin Mass Plane (MP) $(\mjam,\se,\rmaj)$ which describes the distribution of the galaxy population, where $\mjam\equiv L\times (M/L)_{\rm JAM}\approx  M_\star$. The distribution of galaxy properties on both projections of the MP is characterized by: (i) the same zone of exclusion (ZOE), which can be transformed from one projection to the other using the scalar virial equation. The ZOE is roughly described by two power-laws, joined by a break at a characteristic mass $\mjam\approx3\times10^{10} \msun$, which corresponds to the minimum \re\ and maximum stellar density. This results in a break in the mean $\mjam-\se$ relation with trends $\mjam\propto\sigma_{\rm e}^{2.3}$ and $\mjam\propto\sigma_{\rm e}^{4.7}$ at small and large \se\ respectively; (ii) a characteristic mass $\mjam\approx2\times10^{11} \msun$ which separates a population dominated by flat fast rotator with disks and spiral galaxies at lower masses, from one dominated by quite round slow rotators at larger masses; (iii) below that mass the distribution of ETGs properties on the two projections of the MP tends to be  constant along lines of roughly constant $\sigma_{\rm e}$, or equivalently along lines with $\rmaj\propto\mjam$ respectively (or even better paralell to the ZOE: $\rmaj\propto M_{\rm JAM}^{0.75}$); (iv) it forms a continuous and parallel sequence with the distribution of spiral galaxies; (v) at even lower masses, the distribution of fast rotator ETGs and late spirals naturally extends to that of dwarf ETGs (Sph) and dwarf irregulars (Im) respectively. 

We use dynamical models to analyse our kinematic maps. We show that \se\ traces the bulge fraction, which appears the main driver for the observed trends in the dynamical $(M/L)_{\rm JAM}$ and in indicators of the $(M/L)_{\rm pop}$ of the stellar population like H$\beta$ and colour, as well as in the molecular gas fraction. A similar variation along contours of \se\ is also observed for the mass normalization of the stellar Initial Mass Function (IMF), which was recently shown to vary systematically within the ETGs galaxy population. Our preferred relation has the form  $\log_{10} [(M/L)_{\rm stars}/(M/L)_{\rm Salp}]=a+b\times\log_{10}(\se/130\, \kms)$ with $a=-0.12\pm0.01$ and $b=0.35\pm0.06$. Unless there are major flaws in all stellar population models, this trend implies a transition of the mean IMF from Kroupa to Salpeter in the interval $\log_{10}(\se/\kms)\approx1.9-2.5$ (or $\se\approx90-290$ \kms), with a smooth variation in between, consistently with what was shown in \citet{Cappellari2012}.
The observed distribution of galaxy properties on the MP provides a clean and novel view for a number of previously reported trends, which constitute special two-dimensional projections of the more general four-dimensional parameters trends on the MP.
We interpret it as due to a combination of two main effects: (i) an increase of the bulge fraction, which increases \se\ and decrease in \re, and greatly enhances the likelihood for a galaxy to have its star formation quenched, and (ii) dry merging, increasing galaxy mass and \re\ by moving galaxies along lines of roughly constant \se\ (or steeper), while leaving the population nearly unchanged.
\end{abstract}

\begin{keywords}
galaxies: elliptical and lenticular, cD --
galaxies: evolution --
galaxies: formation --
galaxies: structure --
galaxies: kinematics and dynamics
\end{keywords}

\section{Introduction}

Much of our understanding of galaxy formation and evolution comes from the study of dynamical scaling relations relating galaxy luminosity or mass, size and kinematic \citep[e.g.][]{Faber1976,Kormendy1977,Dressler1987,Faber1987,Djorgovski1987} or regular trends in the distribution of galaxy properties as a function of their scaling parameters \citep[e.g.][]{Bender1992,Burstein1997,Kauffmann2003sfh,Gallazzi2006}, and from the study of their evolution with redshift \citep[e.g.][]{vanDokkum1996,Kelson1997,vanDokkum1998,Treu2005,Franx2008}. 

The volume-limited \atl\ sample of nearby early-type galaxies (\citealt{Cappellari2011a}, hereafter Paper~I) constitute an ideal benchmark for studying scaling relations and the distribution of galaxy properties, given the  availability of a high-quality multi-wavelength dataset. In particular in \citet[hereafter Paper~XV]{Cappellari2012p15} we used the state-of-the-art integral-field kinematics, in combination with detailed axisymmetric dynamical models, to derive accurate masses and global dynamical parameter. We found that galaxies lie, with very good accuracy, on a thin Mass Plane (MP) describing galaxies in the parameter space defined by mass, velocity dispersion and projected half-light radius $(\mjam,\se,\rmaj)$. Here \mjam\ is a very good estimate of the galaxy total stellar mass, \rmaj\ is the major axis of the `effective' isophote containing half of the observed galaxy light and \se\ is the velocity dispersion measured measured within that isophote. The existence of this MP is mainly due to the virial equilibrium condition
\begin{equation}\label{eq:virial}
    M_{\rm JAM}\propto\sigma_{\rm e}^2 R_{\rm e}^{\rm maj}
\end{equation}
combined with a smooth variation of galaxy properties with \se. For this reason by itself it contains no useful information on galaxy formation. All the useful constraints on galaxy formation models come from the inhomogeneous distribution of galaxies in non-edge-on views of the MP and from the distribution of galaxy properties along the MP.

This paper is devoted to a study of the non edge-on projections of the MP to see what we learn from it on galaxy formation. This is done in the spirit of the classic papers by \citet{Bender1992} and \citet{Burstein1997}. However the fact that we have accurate dynamical masses implies that our MP is extremely thin and it follows the scalar virial relation~(\ref{eq:virial}) quite accurately. For this reason we can ignore edge-on views of the plane and focus on non edge-on projections only. The thinness of the MP implies that any inclined projection shows essentially the same information, after a change of coordinates. We can use standard and easy-to-understand observables as our main coordinates, instead of trying to observe the plane at a precisely face-on view.

In this paper, in Section~2 we summarize the sample and data, in Section~3 we present our projections of the MP. We illustrate the distribution of a number of quantities on the $(\mjam,\se)$ and $(\mjam,\rmaj)$ projection. We show the variation of the $(M/L)_{\rm JAM}$, as well as of population indicators of $M/L$. We show the variation of galaxy concentration, intrinsic shape, morphology and stellar rotation. The variations of the IMF are separately presented in Section~4, together with a review of previous results on the IMF variation. in Section~5 we discuss the implications of our findings for galaxy formation and briefly summarize our paper in Section~6.

\section{Sample and data}

\subsection{Selection and completeness confirmation}

The galaxies studied in this work are the 260 early-type galaxies which constitute the volume-limited and nearly mass-selected \atl\ sample (Paper I). The object were morphologically selected as early-type according to the standard criterion \citep{Hubble1936,deVaucouleurs1959,Sandage1961} of not showing spiral arms or a disk-scale dust lane (when seen edge-on). The early-types are extracted from a parent sample of 871 galaxies of all morphological types brighter than $M_K=-21.5$ mag, using 2MASS photometry \citep{Skrutskie2006}, inside a local ($D<42$ Mpc)  volume of $1.16\times10^5$ Mpc$^3$ (see full details in Paper~I).

In Paper~I (section~2.3) we discussed possible incompleteness in the volume-limited parent sample of 871 galaxies, from which the \atl\ sample of ETGs was extracted, due to lack of redshift determinations at the faint end limit of our selection. We concluded that the parent sample was 99\% complete. We pointed out that the 2MASS redshift survey (2MRS) was going to change the situation by providing redshifts for the all galaxies brighter than the apparent total magnitude $K_T<11.6$ mag, which makes a galaxy a potential candidate for satisfying our absolute magnitude cut of $M_K<-21.5$ mag, within the adopted $D=42$ Mpc volume. The 2MRS was recently released with a completeness of 98\% for $K_T\le 11.75$ mag \citep{Huchra2012}. This allowed us to explicitly verify the completeness level estimated in Paper~I. For this we used the redshifts from the 2MRS, in combination with the available redshift-independent distance determinations, to repeat the selection of the \atl\ parent sample as described in Paper~I. We confirmed that our parent sample has a completeness of 98\%, quite close to our original estimation.

In Paper~XV we compared our sample to previous samples for which accurate dynamical masses have been determined either via gravitational lensing \citep[e.g.][]{Bolton2008,Auger2010} or dynamics \citep[e.g.][]{Magorrian1998,Cappellari2006,Thomas2009dm} and conclude that it provides a major step forward in sample size and accuracy.

\subsection{Stellar kinematics and imaging}

Various multi-wavelengths datasets are available for the sample galaxies (see a summary in Paper~I). In this work we make use of the \sauron\ \citep{Bacon2001} integral-field stellar kinematics within about one half-light radius \re. The kinematics of all galaxies in the \atl\ sample was homogeneously extracted as described in Paper~I, using the pPXF software\footnote{Available from http://purl.org/cappellari/idl} \citep{Cappellari2004} and the full MILES stellar library \citep{Sanchez-Blazquez2006,FalconBarroso2011miles} as templates. For the subset of 48 early-types in the \sauron\ survey \citep{deZeeuw2002}, the kinematics had already been presented in \citet{Emsellem2004}.

The photometry used in this work comes from the Sloan Digital Sky Survey (SDSS, \citealt{York2000}) data release eight \citep[DR8][]{Aihara2011} and was supplemented by our own photometry taken at the 2.5-m Isaac Newton Telescope in the same set of filters and with comparable signal to noise for the rest of the sample galaxies  \citep[hereafter Paper~XXI]{Scott2011}.

\subsection{Measuring galaxy global parameters: $M/L$, \re\ and $\sigma_{\rm e}$}
\label{sec:global_parameters}

The dynamical masses used in this paper were obtained with the dynamical models of the \atl\ sample presented in \citet{Cappellari2012} and described in more detail in Paper~XV. In brief the modelling approach starts by approximating the observed SDSS and INT $r$-band surface brightness distribution of the \atl\ galaxies using the Multi-Gaussian Expansion (\textsc{MGE}) parametrization \citep{Emsellem1994}, with the fitting method and software$^1$ of \citet{Cappellari2002mge}. Full details of the approach and examples of the resulting MGE fits are given in Paper~XXI. The \textsc{MGE} models are used as input for the Jeans Anisotropic MGE (\textsc{JAM}) modelling  method$^1$  \citep{Cappellari2008} which calculates a prediction of the line-of-sight second velocity moments $\langle v^2_{\rm los}\rangle$ for a given set of model parameters and fits this to the observed $V_{\rm rms}\equiv\sqrt{V^2+\sigma^2}$ using a Bayesian approach \citep{gelman2004bayesian}. The models assume axisymmetry and include a spherical dark halo, which is parametrised according to six different sets of assumptions (See Paper~XV for full details). In \citet{Cappellari2012} we showed that the adopted assumptions on the halo have insignificant influence on the measured trend of $M/L$. For this reason in this paper we only use the two simplest sets of models from that paper (using the same notation): 
\begin{description}
    \item[\bf (A)] Self-consistent axisymmetric JAM model, in which the dark matter is assumed to be proportional to the stellar one. We call the total (luminous plus dark) $M/L$ of this model $(M/L)_{\rm JAM}$;
    \item[\bf (B)] Axisymmetric JAM model with spherical \citet{navarro96} (NFW) dark matter halo. Here the $M/L$ of the stars alone $(M/L)_{\rm stars}$ is measured directly.
\end{description}

\citet[hereafter Paper~XII]{Lablanche2012} studied the accuracy in the $M/L$ obtained with the JAM method, using $N$-body simulations that resemble real galaxies, and concluded that the $M/L$ of unbarred galaxies can be measured with negligible bias $<1.5\%$, while the $M/L$ of barred galaxies that resemble typical bars found in the \atl\ sample can bias the determination by up to 15\% in our tests, depending on the position angle of the bars.

The $M/L$ of the stellar population  $(M/L)_{\rm Salp}$ in the SDSS $r$-band was presented in \citet{Cappellari2012}. It  was extracted using the penalized pixel-fitting (pPXF) method and software of \citet{Cappellari2004} for full spectrum fitting. In this case the model templates consisted of a regular rectangular grid of 26 ages and 6 metallicities $[M/H]$, assuming a \citet{Salpeter1955} IMF for reference, from the MILES models\footnote{Available from http://miles.iac.es/} of \citet{Vazdekis2012}. We used the pPXF keyword REGUL to enforce linear regularization \cite[eq.~18.5.10]{Press1992} in the recovered set of templates weights $w_j$ from the fit. The regularization parameter was chosen for every galaxy to obtain an increase $\Delta\chi^2=\sqrt{2\times N_{\rm pix}}$ in the $\chi^2$, with respect to a non-regularized fit. In this way the recovered $w_j$  solution constitutes the smoothest one consistent with the observed spectrum (see fig.~20 of \citealt{Onodera2012} for an example). This regularized approach helps reducing the noise in the $(M/L)_{\rm Salp}$ determination, however the results are quite insensitive to the precise value of this parameter and remain essentially unchanged for a range of plausible values. Once the weights for the $N$ model templates have been obtained with pPXF, the properly mass-weighted $M/L$ of the combined population is given by 
\begin{equation}
(M/L)_{\rm Salp,r}=\frac{\sum_{j=1}^N w_j M_j^{\rm nogas}}{\sum_{j=1}^N w_j L_{j,r}},
\label{eq:mlpop}
\end{equation}
where $M_j^{\rm nogas}$ is the mass of the $j$th model in living stars and stellar remnants (black holes and neutron stars), but excluding the gas lost during stellar evolution, and $L_{j,r}$ is the corresponding $r$-band luminosity of the population. The values are given in Table~1. The details on our spectral fitting of the \sauron\ data and the analysis of the resulting weight distributions is presented in McDermid et al. (in preparation).

Effective radii are defined as the projected half-light radii of the isophote  containing half of the total analytic light of the MGE models from Paper~XXI. Both circularized radii \re\ and the more robust major axes $R_{\rm e}^{\rm maj}$ were extracted as described in Paper~XV and are taken from table~1 there.

Effective velocity dispersions $\sigma_{\rm e}$ were measured by co-adding all the spectra in the \sauron\ datacube contained within the half-light isophote. The resulting `effective' spectrum was fitted with pPXF using the full MILES stellar library as templates, and assuming a Gaussian line-of-sight velocity distribution. Gas emission around the possible H$\beta$ and O\,{\sc iii} emission lines were systematically excluded from the fits and the CLEAN keyword of pPXF was used to reject possible remaining outliers from the spectra. The values for the sample are given in table~1 of Paper~XV.

\section{Projections of the Mass Plane}
\label{sec:vp_projections}

\subsection{Total $M/L$ variations}

We have shown in Paper~XV that the existence of the FP is due, with good accuracy, to a virial equilibrium condition combined with a smooth variation of galaxy properties, mainly the total mass-to-light ratio $M/L$, with velocity dispersion.
Once this is clarified, the edge-on projection of the MP becomes uninteresting from the point of view of the study of galaxy formation, as it merely states an equilibrium condition satisfied by galaxies and it does not encode any memory of the formation process itself. This is in agreement with previous findings with simulations \citep{Nipoti2003,Boylan-Kolchin2006}. All information provided by scaling relations on galaxy formation is now encoded in the non edge-on projections of the MP, and first of all in the distribution of $M/L$ on that plane. In Paper~XV we also confirmed that $M/L$ correlates remarkably tightly with $\sigma_{\rm e}$ \citep{Cappellari2006}. This is especially true (i) for slow rotators, (ii) for galaxies in clusters and (iii) at the high-end of the $\sigma_{\rm e}$ range. Here we look at the entire MP and try to clarify the reason for these and other galaxy correlations.

In a classic paper \citet{Bender1992} studied the distribution of hot stellar systems in a three-dimensional space, they called $\kappa$ space, defined in such a way that one of the axes was empirically defined to lie nearly orthogonal to the plane. This made it easy to look at both the edge-on and face-on versions of the plane. In this paper, thanks to the availability of state-of-the-art integral-field kinematics and the construction of detailed dynamical models, we can use mass as one of the three variables $(M_{\rm JAM},\sigma_{\rm e},\re)$. We have shown in Paper~XV that in these variables the plane is extremely thin and follows the scalar virial equation $\mjam\propto\sigma_{\rm e}^2 \rmaj$ within our tight errors. This implies that any projection of the plane contains the same amount of information, except for a change of coordinates. Instead of looking at the plane precisely face-on, we constructed special projections which correspond to physically-meaningful and easy-to-interpret quantities.

\begin{figure*}
\includegraphics[width=0.7\textwidth]{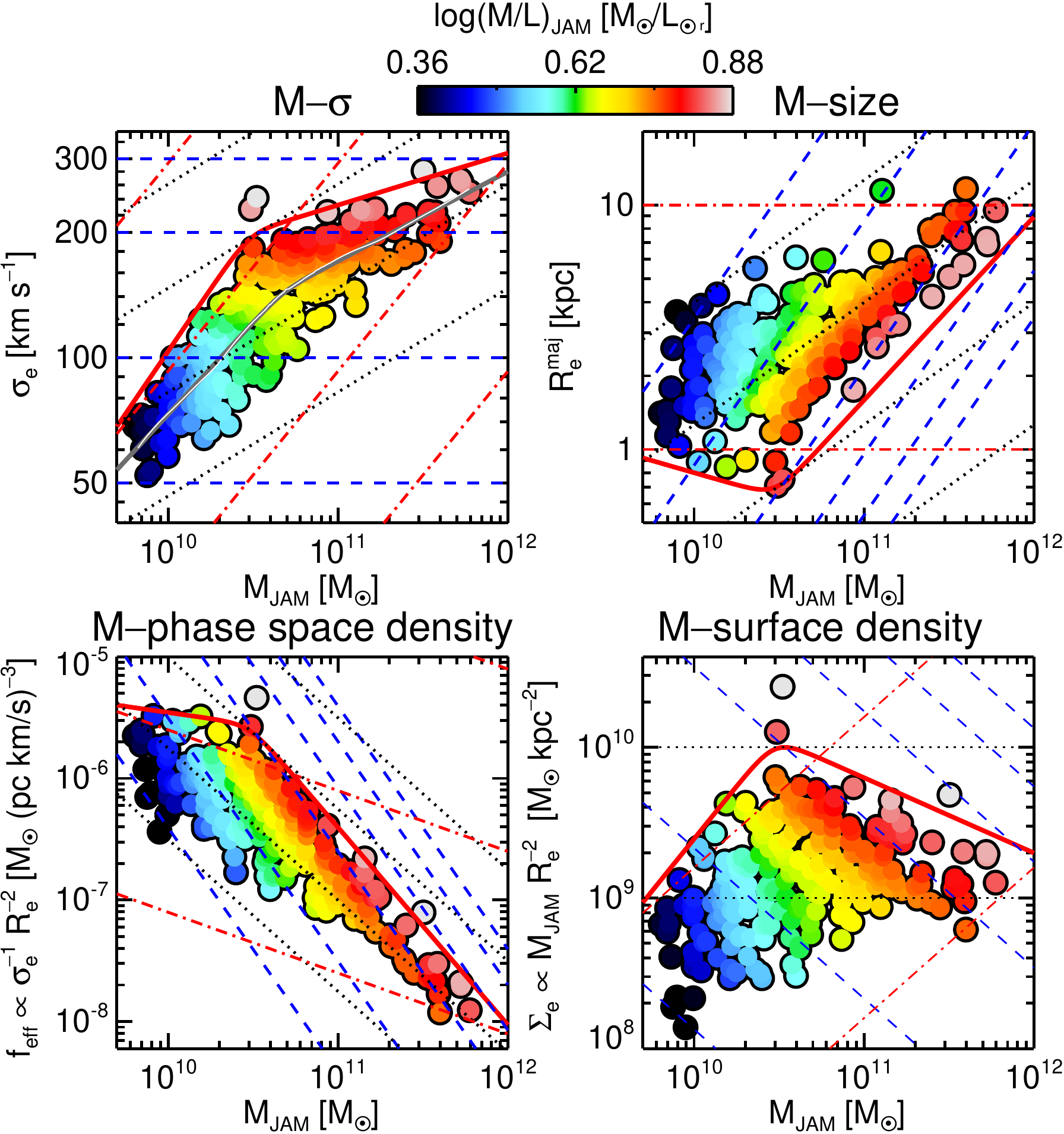}
\caption{The MP and it projections. The top two panels show the two main projections of the MP in the $(M_{\rm JAM},\sigma_{\rm e})$ and $(M_{\rm JAM},R_{\rm e}^{\rm maj})$ coordinates. Overlaid are lines of constant $\sigma_{\rm e}=50,100,200,300,400,500$ \kms\ (dashed blue), constant $R_{\rm e}^{\rm maj}=0.1,1,10,100$ kpc (dot-dashed red) and constant $\Sigma_{\rm e}=10^8,10^9,10^{10},10^{11}$ \msun\ kpc$^{-2}$ (dotted black) transformed in different panels using the scalar virial relation $M_{\rm JAM}=5.0\times\sigma_{\rm e}^2 \re/G$ . The observed $(M_{\rm JAM},\sigma_{\rm e},R_{\rm e}^{\rm maj})$ points follow the scalar virial relation  so closely that the virial coordinates provide a unique mapping on these diagram and one can reliably infer all characteristics of the galaxies from any individual projection. In each panel the galaxies are coloured according to the (LOESS smoothed) $\log (M/L)_{\rm JAM}$, as shown in the colour bar at the top. The LOESS method tries to estimate the mean values of the underlying galaxy population, namely the values one should expect to obtain by averaging much larger galaxy samples.  In all panels the thick red line shows the same ZOE relation given by \refeq{eq:zoe}, again transformed in different panels according to the scalar virial relation. The thin white line in the top-left panel is a LOESS smoothed version of the $\mjam-\se$ relation, while the thick dark grey line is a double power-law fit (equation~\ref{eq:m-sigma}) with trends $\mjam\propto\sigma_{\rm e}^{2.3}$ and $\mjam\propto\sigma_{\rm e}^{4.7}$ below or above $\se\approx140$ \kms\ respectively. While the top two panels contain different observable quantities, the bottom two panels merely apply a coordinate transformation to the quantities in the top two panels, to show the effective phase-space density $f_{\rm eff}\equiv 1/(\sigma \re^2)$ and effective mass surface density $\Sigma_{\rm e}\equiv M_{\rm JAM}/(2\pi\re^2)$. Two galaxies stand out for being significantly above the ZOE in the $(\mjam,\se)$ and $(\mjam,\Sigma_{\rm e})$ projections. The top one is NGC~5845 and the bottom one is NGC~4342.}
\label{fig:virial_plane_projections_ml}
\end{figure*}

Our selection of meaningful projections of the MP is shown in \reffig{fig:virial_plane_projections_ml}. We use as horizontal axis in all plots our main mass variable
\begin{equation}
    M_{\rm JAM}\equiv L\times (M/L)_{\rm JAM} \approx 2\times M_{1/2} \approx M_{\rm stars},
\end{equation}
where $(M/L)_{\rm JAM}$ is the total (luminous plus dark) dynamical $M/L$ obtained using self-consistent JAM models, $L$ is the total galaxy luminosity (both quantities were taken from table~1 of Paper~XV) and $M_{1/2}$ is the total mass within a sphere of radius $r_{1/2}$ enclosing half of the total galaxy light, where $r_{1/2}\approx1.33\re$ (\citealt{Hernquist1990,Ciotti1991,Wolf2010}; Paper~XV). Although the self-consistency assumption, where the total mass is proportional to the stellar mass, is not justified at large radii, it is accurate within the region where we have stellar kinematics (about 1\re). It was shown in \citet{Williams2010} and in Paper~XV that $(M/L)_{\rm JAM}$ closely reproduces the total $(M/L)(r=\re)$ inside a sphere (actually an iso-surface) with mean radius the projected half-light radius  \re, derived using models which explicitly include dark matter. Different assumptions on the dark halo produce minor differences in $(M/L)(r=\re)$. Given that $(M/L)(r=\re)$ is nearly insensitive to the choice of the halo assumptions, we choose the self-consistent one, being the simplest.

As illustrated in Paper~XV (see also \citealt{Williams2009}), most of the galaxies in our sample are consistent with having small fractions of dark matter within a sphere of radius $r=\re$, with a median value of just 13\% of DM within that radius for our standard models (B). This implies that $M_{1/2}$ is  dominated by the stellar mass. For this reason $M_{\rm JAM}$ is a quantity that very closely represents and is directly comparable to the total stellar galaxy mass used in numerous previous studies. Stellar mass seems to relate well with galaxy properties and is often used to study galaxy formation \citep[e.g.][]{Kauffmann2003mass,Kauffmann2003sfh,Hyde2009fp}. The difference of our mass parameter is that it does not suffer from the uncertainties related to the stellar population models \citep[e.g.][]{Maraston2006,Gallazzi2009,Conroy2009,Longhetti2009,Wuyts2009} moreover it automatically includes the effects of a non-universal IMF \citep{vanDokkum2010,Cappellari2012}. Being a measure of the total enclosed mass within a spherical region, and thus being directly related the dynamics, $M_{\rm JAM}$ is the ideal parameter to use in scaling relations. Note that $M_{\rm JAM}$, unlike mass determinations obtained via strong lensing, does not include the possible contribution of dark matter along the cylinder parallel to the line-of-sight \citep{Dutton2011swells}, which provides an useful additional constraint on dark matter at large radii, but complicates the interpretation of scaling relations in the galaxy centres.

\begin{figure}
\plotone{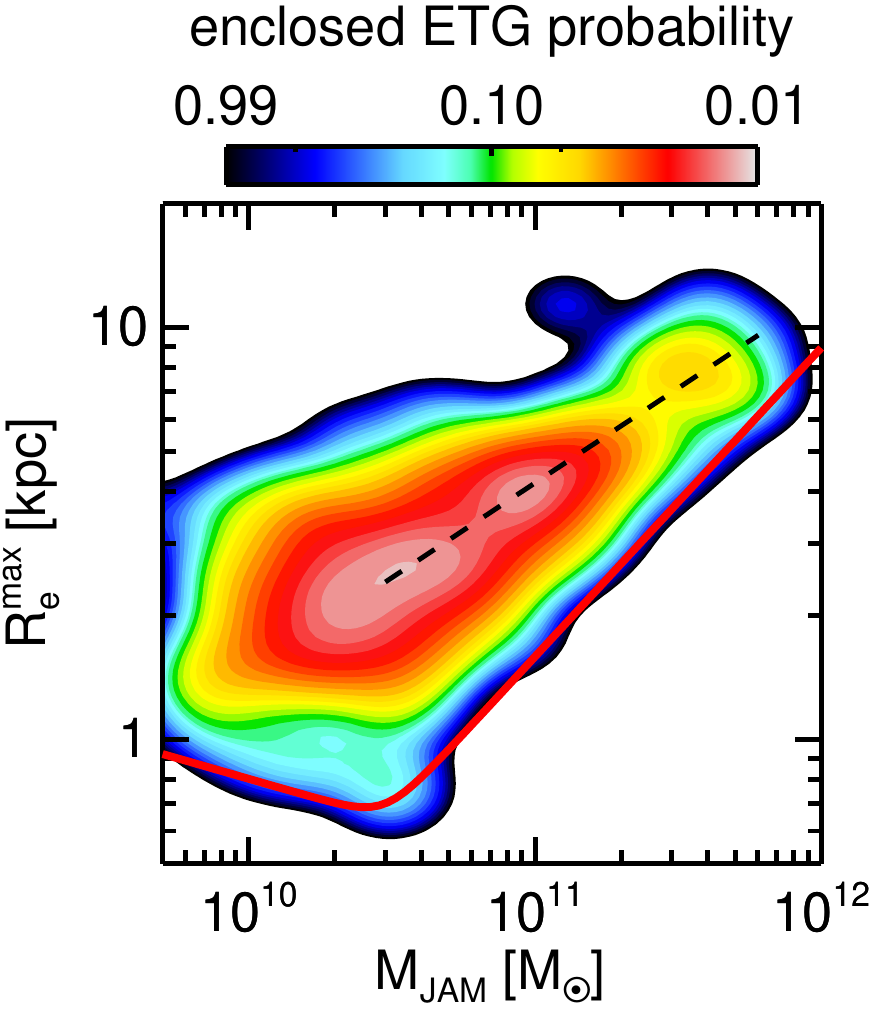}
\caption{Probability density contours for ETGs on the M-size plane. Colours indicate the contours enclosing a given fraction of the total probability, with the outermost (lowest) contour enclosing 99\% of the total probability, above our survey mass limit ($M_{\rm JAM}\ga6\times10^9$ \msun). The thick red curve is the same as in \reffig{fig:virial_plane_projections_ml} and provides a convenient approximation for the lowest 99\% contour. The dashed line indicates the relation $(R_{\rm e}^{\rm maj}/{\rm kpc})=4.2\times[M_{\rm JAM}/(10^{11} M_\odot)]^{0.46}$, which approximates the peak ridge-line of the ETGs M-size distribution.}
\label{fig:probability_density}
\end{figure}

In the top-left and top-right panels of \reffig{fig:virial_plane_projections_ml} we show the projections of the MP along the $(M_{\rm JAM},\sigma_{\rm e})$ and $(M_{\rm JAM},R_{\rm e}^{\rm maj})$ axes respectively. The colour in these diagrams represent the dynamical (total) $M/L$ inside a sphere of radius \re. Three important results are clear from these plots:
\begin{enumerate}

\item Both projections are equivalent and provide basically the same picture, apart from a coordinate transformation. This is expected from the tightness of the MP. Moreover, given that the MP nearly follows the scalar virial relation, the expression $M_{\rm JAM}=5.0\times\sigma_{\rm e}^2\re/G$ \citep{Cappellari2006} can be used to provide an accurate estimate of lines of constant \re\ on the $(\mjam,\se)$ projection or the lines of constant \se\ on the $(\mjam,\rmaj)$ projection;

\item Galaxies define a clear zone-of-exclusion (ZOE) at small sizes or large densities, as already pointed out by \citet{Bender1992} and \citet{Burstein1997}, however we find a clear break in the ZOE at a mass $\mjam\approx3\times 10^{10} \msun$ and two nearly power-law regimes above or below this value. The ZOE is approximated by the equation
\begin{equation}\label{eq:zoe}
\rmaj=R_{\rm e,b} \left(\frac{M_{\rm JAM}}{M_{\rm JAM,b}}\right)^\gamma
\left[
\frac{1}{2} + \frac{1}{2} \left(\frac{M_{\rm JAM}}{M_{\rm JAM,b}}\right)^\alpha
\right]^{(\beta-\gamma)/\alpha}
\end{equation}
with $R_{\rm e,b}=0.7$ kpc, $\alpha=8$, $\beta=0.75$, $\gamma=-0.20$. The relation has an asymptotic trend $\re\propto M_{\rm JAM}^{0.75}$ above $M_{\rm JAM,b}=3.0\times10^{10} \msun$, and a sharp transition into $\re\propto M_{\rm JAM}^{-0.20}$ below this break. The values were determined by simultaneously matching the observed boundary in the galaxy distribution in all MP projections (but see \reffig{fig:probability_density} for a quantitative definition).
These values are nearly the same to those we already reported in Paper~I (eq.~4 there), using $K$-band luminosity in place of mass and 2MASS \re\ instead of MGE ones. The maximum in the galaxy density at $M_{\rm JAM,b}$ that we infer from our sample is also clearly visible in much larger SDSS samples \citep[e.g.\ fig.~2 of][]{vanDokkum2008}. This shows that the relation is valid for the general galaxy population.  The ZOE relation of \refeq{eq:zoe} can be converted into a $(M_{\rm JAM},\sigma_{\rm e})$ one, or into other projections using the scalar virial relation $M_{\rm JAM}=5.0\times\sigma_{\rm e}^2 \re/G$. The corresponding trend becomes $\se\propto M_{\rm JAM}^{0.6}$ ($\mjam\propto\sigma_{\rm e}^{1.67}$) below the break, which flattens to $\se\propto M_{\rm JAM}^{0.125}$ ($\mjam\propto\sigma_{\rm e}^{8.0}$) above the break. The location of the break we find agrees with the value at which scaling relations of large sample of SDSS galaxies show a subtle deviation from a straight line \citep{Hyde2009curv}. Interestingly the characteristic mass at the cusp also coincides with the value reported by \citet{Kauffmann2003sfh}, as the fundamental dividing line between the two distinct families of passive and star-forming galaxies.

\item The contours of constant $(M/L)_{\rm JAM}$  closely follow lines of constant $\sigma_{\rm e}$ in the $(\mjam,\se)$ projection (above $\sigma_{\rm e}\ga120$ \kms) and equivalently $\rmaj\propto\mjam$ in the $(\mjam,\rmaj)$ projection. A comparable agreement is reached using lines parallel to the ZOE $R_{\rm e}^{\rm maj}\propto M_{\rm JAM}^{0.75}$. Neither mass $M_{\rm JAM}$, nor size \re\ nor surface mass density $\Sigma_{\rm e}$ provide a comparably good approximation to the $(M/L)_{\rm JAM}$ contours, although $\Sigma_{\rm e}$ provides a better approximation than the other two. This is consistent and explains the finding by \citet{Cappellari2006} that $\sigma_{\rm e}$ and not mass or luminosity, is the best tracer of $(M/L)_{\rm JAM}$. At lower $\sigma_{\rm e}\la120$ \kms\ the $(M/L)_{\rm JAM}$ contours start deviating from the lines of constant \se\ and tend to lie closer to lines of constant $M_{\rm JAM}$. Given that the $M/L$ describes the deviations between the FP and the MP, this twist in the contours demonstrates that the FP is warped with respect to the MP, and explains the sensitivity of the FP parameters to the region included in the fit \citep{dOnofrio2008,Gargiulo2009,Hyde2009fp}.

\end{enumerate}

To quantitatively define the meaning of the ZOE, which is shown in various plots of this paper, we estimated the probability density underlying the set of discrete measurements for our galaxies in the M-size plane. For this we used the kernel density estimator method. We adopted an Epanechnikov kernel and determined the optimal smoothing parameter in each dimension following \citet{silverman1986density}. The resulting estimate of the ETGs density is shown in \reffig{fig:probability_density}. The plot shows that the adopted functional form of the ZOE in \refeq{eq:zoe} provides a convenient approximation of the lower contour enclosing 99\% of the ETG population, above our survey mass limit $M_{\rm JAM}\ga 6\times 10^9$ \msun. We argue that, for volume-limited samples, this line constitutes a more meaningful reference to define the lower limit of the sizes of the normal galaxy population than the mean M-size relation and its scatter \citep[e.g.][]{Shen2003}. The reason is that the ZOE takes the galaxies mass function into account. Low-mass galaxies are more numerous than high-mass ones. This implies that, for a given rms scatter $\sigma_{\re}$, low-mass galaxies are more likely than high-mass galaxies to deviate more than $\sigma_{\re}$ from the mean M-size relation.

Two galaxies stand out for being significantly above the ZOE in the $(M_{\rm JAM},\sigma_{\rm e})$ and $(M_{\rm JAM},\Sigma_{\rm e})$ projections. The top one is NGC~5845 and the bottom one is NGC~4342. The two objects have very high surface brightness and consequently excellent kinematic and photometric data. They are genuine examples of dense objects in the nearby Universe. These two outliers were already presented in \cite{Cappellari2011dur} and their compactness was later discussed by \citet{Jiang2012}. However they have smaller masses ($M_{\rm JAM}\approx3\times10^{10}$ \msun) than their high-redshift counterparts \citep{Cimatti2008,vanDokkum2008}.

In the top-left panel of \reffig{fig:virial_plane_projections_ml} we overplot a version of the $\mjam-\se$ relation which was adaptively smoothed using the one-dimensional Locally Weighted Regression robust technique (dubbed LOESS\footnote{Our implementations \textsc{cap\_loess\_1d} and \textsc{cap\_loess\_2d} of the one-dimensional and two-dimensional LOESS method as used in this paper are available from http://purl.org/cappellari/idl}) of \citet{cleveland1979robust}. The relation is equivalent to the \citet{Faber1976} but uses dynamical mass instead of luminosity. Like the ZOE, the mean $\mjam-\se$ relation appears well approximated  by a double power-law of the form:
\begin{equation}\label{eq:m-sigma}
\se=\sigma_{\rm e,b} \left(\frac{M_{\rm JAM}}{M_{\rm JAM,b}}\right)^\gamma
\left[
\frac{1}{2} + \frac{1}{2} \left(\frac{M_{\rm JAM}}{M_{\rm JAM,b}}\right)^\alpha
\right]^{(\beta-\gamma)/\alpha},
\end{equation}
with best-fitting parameters (excluding NGC~5845 and NGC~4342) $\log (M_{\rm JAM,b}/\msun)=10.675\pm0.090$, $\log (\sigma_{\rm e,b}/\kms)=2.153\pm0.031$, $\alpha=8$, logarithmic slope $\beta=0.215\pm0.027$ at large masses and $\gamma=0.442\pm0.028$ at small masses. When the two power-laws are written as a function of \se, the trends become $\mjam\propto\sigma_{\rm e}^{1/\beta}=\sigma_{\rm e}^{4.65\pm0.58}$ for $\se\gg140$ \kms, and $\mjam\propto\sigma_{\rm e}^{1/\gamma}=\sigma_{\rm e}^{2.26\pm0.14}$ for $\se\ll140$ \kms. 

In all panels of \reffig{fig:virial_plane_projections_ml}, and in all subsequent similar figures for other parameters, the measured $(M/L)_{\rm JAM}$ have been adaptively smoothed using the LOESS technique, straightforwardly generalized to two dimension as in \citet{cleveland1988locally}. In all plots we adopt a regularization factor $f=0.6$, and a linear local approximation. To deal with the different scales of the axes and the elongated shape of the galaxy distributions, before applying LOESS we rotated and re-normalized the coordinates so that the ellipse of inertia of the galaxy distribution reduces to a circle in every projection$^3$. This change of coordinates improves the definition of neighbouring points during the LOESS smoothing. By design the LOESS smoothed maps try to remove observational errors and intrinsics scatter to estimate the mean values of the underlying galaxy population. The maps try to predict the mean values one would obtain via simple histograms of much larger samples of galaxies (see also discussion in \refsec{sec:population}).
We do not show here the original data, as the scatter in $(M/L)_{\rm JAM}$, projected along the \se\ coordinate, was already analysed in detail in Paper~XV. Given that most of the variation in the smooth surface happens orthogonally to the constant $\sigma_{\rm e}$ lines, the scatter from the smooth surface of 27\% is nearly unchanged from the value of 29\% in the global $(M/L)-\sigma_{\rm e}$ relation, further confirming that most of the variation happens orthogonally to $\sigma_{\rm e}$.

In the bottom left and right panels of \reffig{fig:virial_plane_projections_ml} we show the effective phase-space density, defined following \citet{Hernquist1993} as $f_{\rm eff}\equiv 1/(G\sigma_{\rm e}\re^2)$ and the mass surface density, defined as $\Sigma_{\rm e}\equiv M_{\rm JAM}/(2\pi\re^2)$. Note that, while we use $R_{\rm e}^{\rm maj}$ for the mass-size plane, one has to use \re\ for $f_{\rm eff}$ and $\Sigma_{\rm e}$. The two bottom panels do not plot new data, but are obtained by rearranging the three variables shown in the top panels, however they illustrate how the above trends relate to other physical quantities. In particular the phase-space density is interesting because it can only decrease during collisionless galaxy mergers, due to Liouville's theorem \citep{Carlberg1986,Hernquist1993}. Interestingly the ZOE we find is nearly flat in $f_{\rm eff}$ below $M_{\rm JAM,b}=3\times10^{10} \msun$ (a value $\gamma=1/3$ gives $f_{\rm eff}={\rm const}$ in the virial case) and starts decreasing at larger masses, as one would expect when dry mergers start becoming more important.

\subsection{Stellar population indicators of $(M/L)_{\rm pop}$}
\label{sec:population}

We have shown in Paper~XV that the stellar matter dominates in the regions we study. If this is indeed the case we would expect stellar population indicators of the $(M/L)_{\rm pop}$ of the stellar population to closely follow the behaviour of the dynamical $(M/L)_{\rm JAM}$. Before addressing this question with our own data, the reader is strongly encouraged to compare \reffig{fig:virial_plane_projections_ml} to the right panel of figure~15 of \citet{Gallazzi2006}, which shows the luminosity-weighted age versus the stellar mass in the $(\sigma,M_\star)$ projection. Even though our \reffig{fig:virial_plane_projections_ml} uses dynamical quantities, which are measured via dynamical models, while \citet{Gallazzi2006} derives population from line strengths, the resemblance of the two plots is striking. This includes rather subtle details like the change of orientation of the contours, which in both cases tend to be horizontal above $\sigma\ga120$ \kms\ and become nearly vertical at lower $\sigma$. This comparison already indicates a close link between $(M/L)_{\rm JAM}$ and stellar population (mainly age), in broad agreement with \citet{Cappellari2006}. Importantly, given that our sample is 100 times smaller than the one of \citet{Gallazzi2006}, this comparison also provides a validation of the adopted LOESS technique, which appears to recover as expected the underlying distribution of galaxy properties from our much smaller galaxy sample. This comparison confirms that the various LOESS smoothed plots in this paper should be interpreted as estimates of the average trends in the galaxy distribution we would observe if we had much larger samples of galaxies than we have in \atl.

We now address the relation between dynamical and population $M/L$ using our own dataset and a different approach. Two main simple tracers of $(M/L)_{\rm pop}$ have been proposed in the past: (i) the $B-R$ colour, which was shown by \citep{Bell2001} to trace $(M/L)_{\rm pop}$ alone, for a wide range of metallicities, and (ii) the line-strength index H$\beta$, which was shown to satisfy a similar property \citep[fig.~16 of][]{Cappellari2006} for a wide range of metallicities \citep[see also][]{Worthey1994}. Here we adopt the H$\beta$ determination for our \atl\ galaxies, as derived from the same very high $S/N$ `effective' \sauron\ spectra from which we derived \se. The homogeneous extraction of the line-strength parameters for the \atl\ sample is given in McDermid et al. (in preparation). The parameters for the \sauron\ survey \citep{deZeeuw2002} subset of 48 galaxies was already given in \citet{Kuntschner2006}.  As choice of colour we use the SDSS $g-i$ one, which is available for 223 \atl\ galaxies in SDSS DR8 and was shown by \citet{Gallazzi2009} to provide on average the smallest uncertainties and the most stable results, among the SDSS bands.

\begin{figure*}
\includegraphics[width=0.7\textwidth]{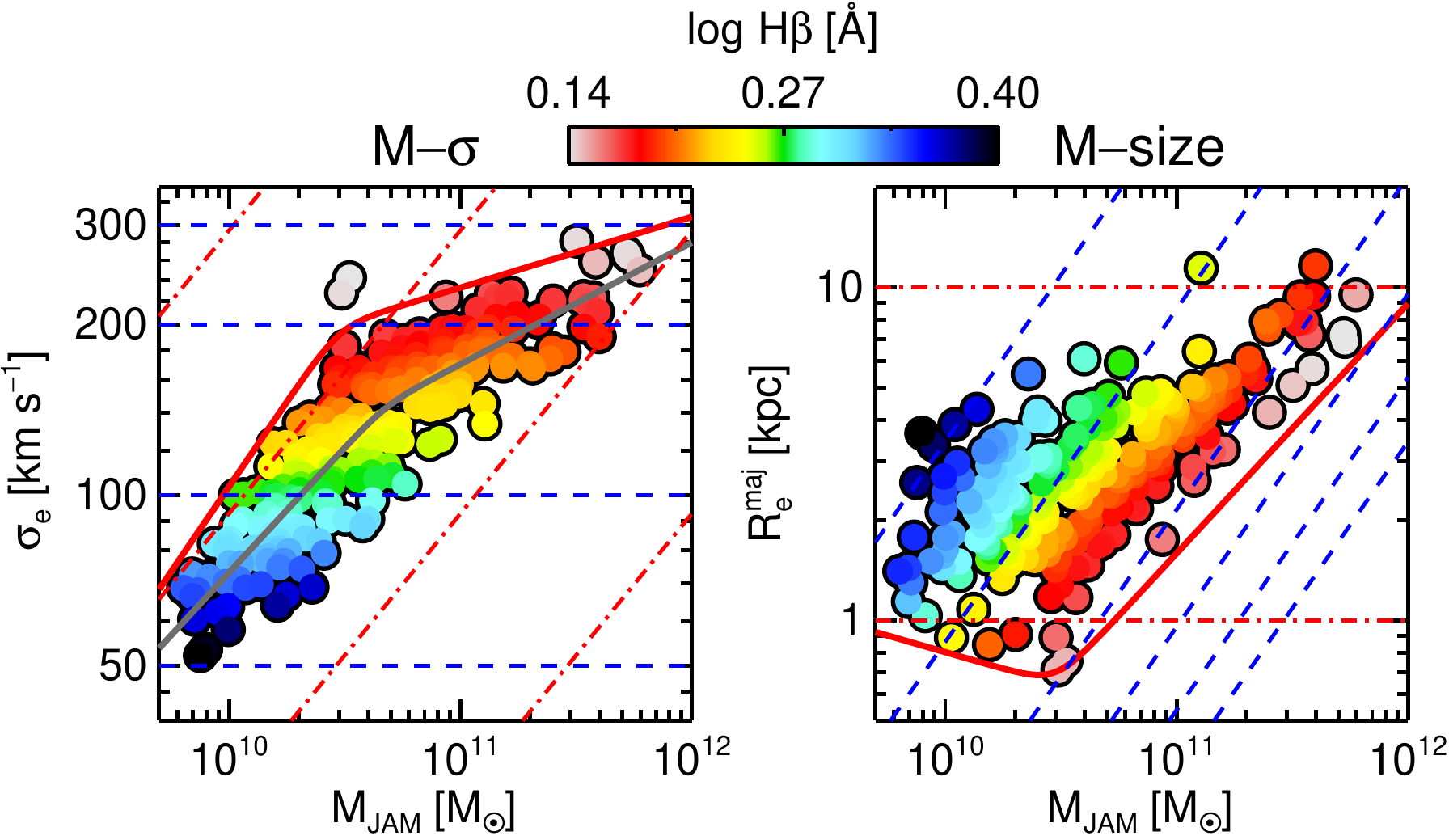}
\includegraphics[width=0.7\textwidth]{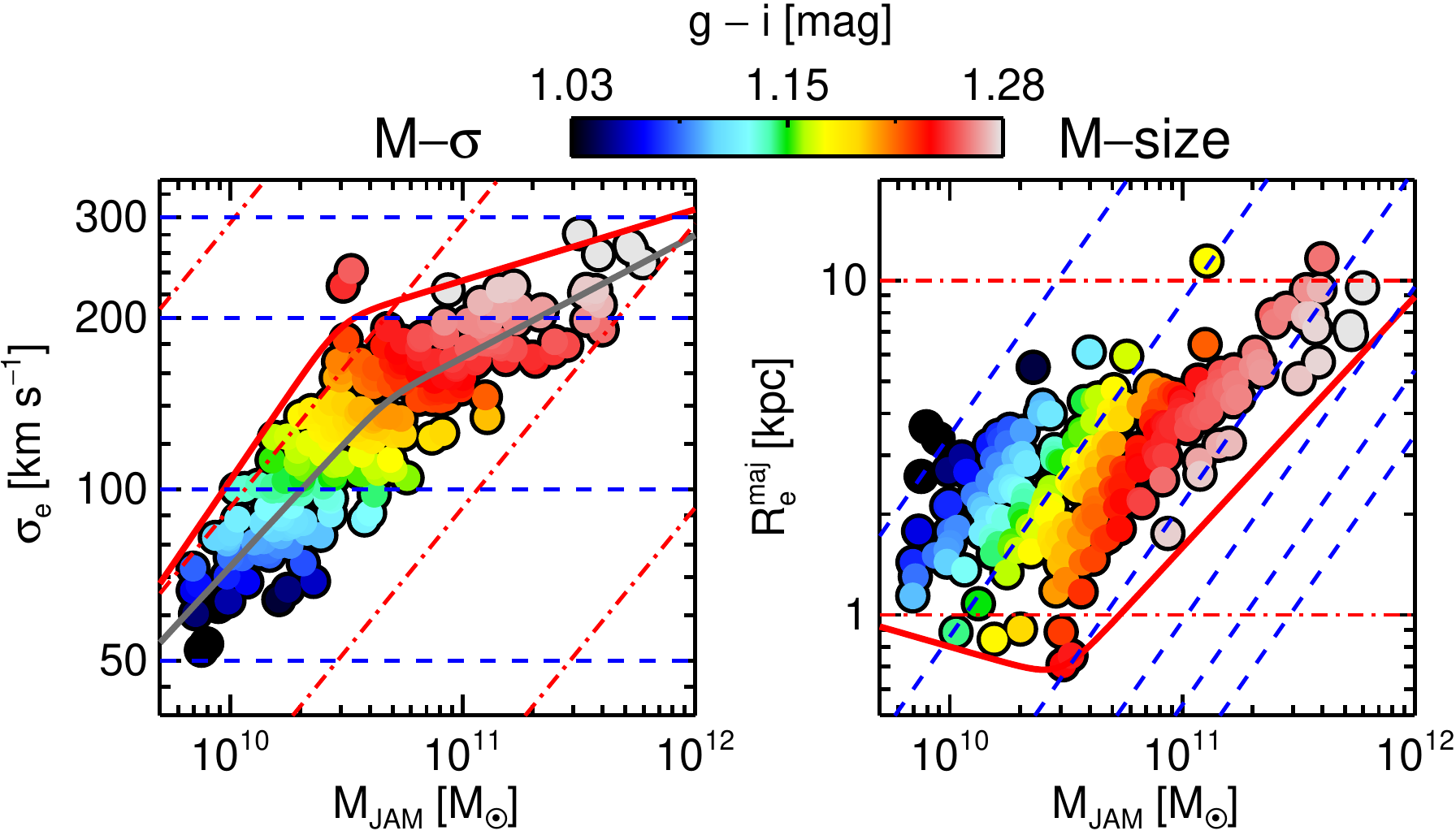}
\includegraphics[width=0.7\textwidth]{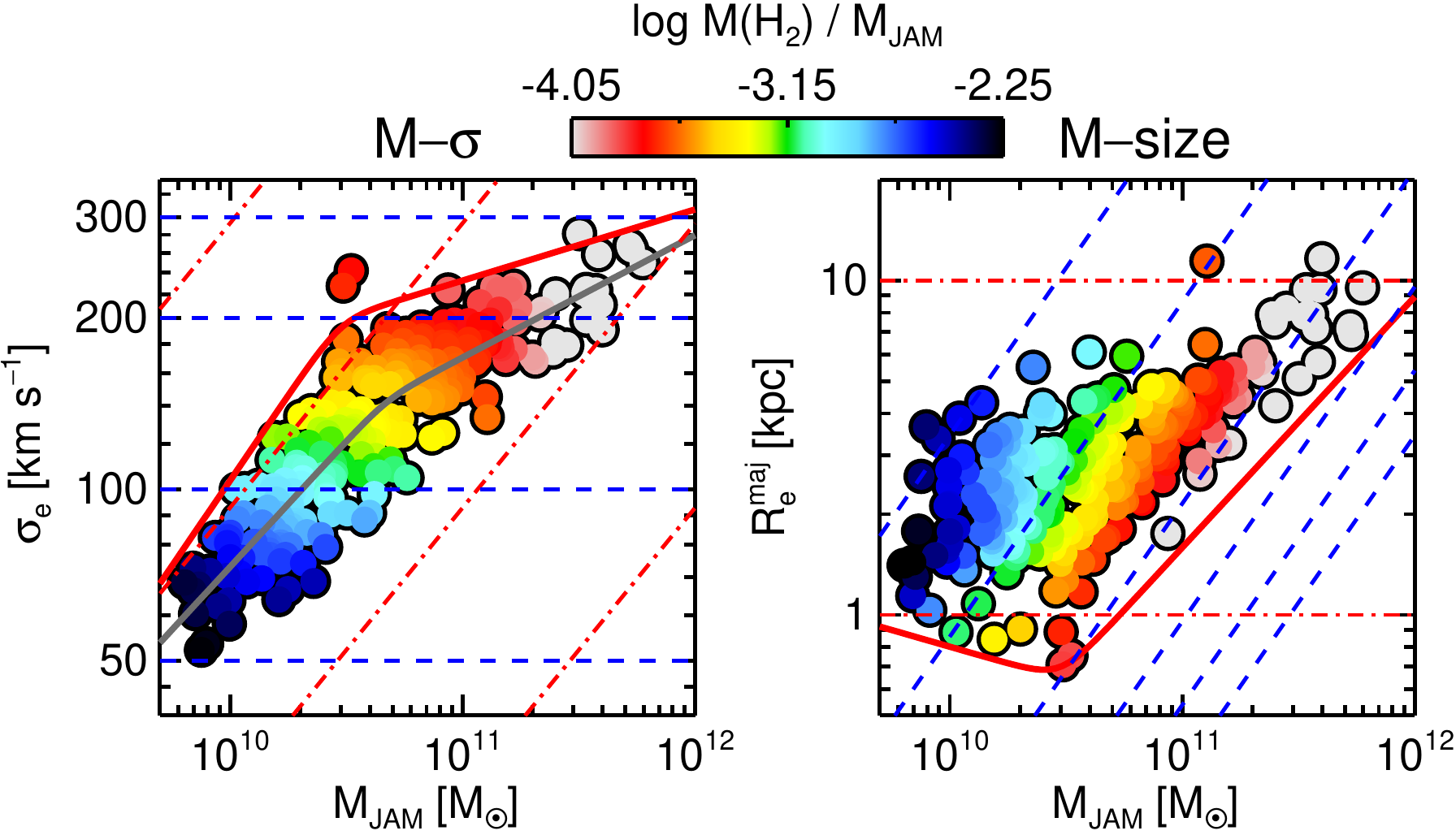}
\caption{Indicators of the stellar population $(M/L)_{\rm pop}$ on the MP. {\em Top Panels:} Same as in \reffig{fig:virial_plane_projections_ml}, with colours showing the $\log {\rm H}\beta$ stellar absorption line-strength, from McDermid et al. (in preparation). 
{\em Middle Panels:} Same as in \reffig{fig:virial_plane_projections_ml}, with colours indicating the SDSS galaxy colour $g-i$. 
{\em Bottom Panels:} Same as in \reffig{fig:virial_plane_projections_ml} with colour indicating the fraction of molecular hydrogen H$_2$ (from table~1 of Paper~IV) with respect to \mjam.
All three indicators H$\beta$, $g-i$ and the H$_2$ fraction show a similar trends as the $(M/L)_{\rm JAM}$ in the top panels of \reffig{fig:virial_plane_projections_ml}. This suggests that the stars dominate the region we observe and that the dynamical \mljam\ is driven by the stellar population, in agreement with the quantitative results via dynamical models in Paper~XV.}
\label{fig:virial_plane_projections_hbeta}
\end{figure*}

The results are presented in \reffig{fig:virial_plane_projections_hbeta} and show a quite good agreement between simple population estimators of $M/L$ and the dynamical one. Both H$\beta$ (top panel) and colour (middle panel) are nearly constant along lines of constant \se\ as it was the case for \reffig{fig:virial_plane_projections_ml}. This confirms the fact that the key driver of the total dynamical \mljam\ is the stellar population. We do notice some systematic differences at the low-$\sigma$ end, with the contours of constant \mljam\ not quite following the population trends. As the effect is visible both in colour and H$\beta$ we believe it is significant. It may be due to the presence of extended star formation episodes, which are expected to be more prevalent at low mass \citep{Heavens2004,Thomas2005daniel} and cause deviations from the simple one-to-one relations between population observables and $M/L$. The study of the relation between our parameters and the actual physical parameters of the population, like age, metallicity and abundance ratios, is presented in McDermid et al. (in preparation). Analysis of the stellar population parameters for the \sauron\ survey \citep{deZeeuw2002} subset of 48 galaxies was already presented in \citet{Kuntschner2010}.

In fig.~15 of \citet[hereafter Paper~IV]{Young2011} we showed that galaxies with the largest H$\beta$, and consequently younger mean ages, contain higher fractions of molecular hydrogen H$_2$ as traced by CO emission lines. The trend makes sense given that H$_2$ reservoir is expected to fuel star formation (see a more detailed discussion in Paper~IV). Given the trend we observed for H$\beta$ on the M-size plane, one should expect a similar trend for the H$_2$ fraction. This is indeed the case as presented in the bottom panel of \reffig{fig:virial_plane_projections_hbeta}, which shows the mass $M(H_2)$ as fraction of \mjam. Given that not all galaxies have a CO detection in Paper~IV, we assigned to undetected galaxies a uniformly distributed random H$_2$ mass between zero and the quoted upper limit. A similar plot is obtained by assigning to all undetected galaxies a fixed mass $M(H_2)=10^6$ \msun\ below our CO survey sensitivity, or by assigning the upper limit, showing that the trend does not depend critically on our treatment of the undetected galaxies. The plot shows that the molecular gas content, like other population indicators, also tends to follow contours of constant \se. However some notable differences exists between this plot and the \mljam\ or the H$\beta$ ones. In particular above the mass $\mjam\approx2\times10^{11}$ \msun, the likelihood of finding significant fractions of molecular gas seems to depend more on mass than \se. This mass is a characteristic one, where a number of other observables appear to change as we will show in \refsec{sec:rotation}.

\subsection{Inferring bulge fractions from stellar kinematics}

In the previous sections we showed that galaxy properties follow smooth trends in the projections of the MP. We have seen that the dynamical $M/L$ near the galaxy centre (1\re) is driven by a variation of the stellar population (including IMF), with the dynamical $M/L$ constituting an accurate tracer of the galaxy population! These trends distribute galaxies in the MP along lines of nearly constant $\sigma_{\rm e}$ on the $(\mjam,\se)$ plane, or equivalently $\rmaj\propto\mjam$ on the $(\mjam,\rmaj)$ plane. In Paper~I we showed that the luminosity-size relation changes gradually with morphological type. Galaxy size decreases as a function of the bulge fraction as indicated by the morphological classification. We explained the size decrease as an increase of the bulge fraction between different morphological types. In Paper~II and III, we showed that the vast majority of local ETGs is constituted by disk-like systems, the fast rotators, and in \citet[hereafter Paper~VII]{Cappellari2011b} we discussed how fast-rotators morphologically resemble spirals with a variety of bulge fractions. We concluded that fast rotators simply constitute the end point of a smooth sequence of disks with increasing bulges, and this lead us to propose a fundamental change in our view of galaxy morphology (Paper VII). If this picture is correct, we should be able to find direct evidence for a systematic change in the bulge fraction, in the fast rotators class, while moving from the region of the MP populated by spiral galaxies towards the ZOE, dominated by the oldest and reddest ETGs, with the largest $M/L$. This is what we show in this section.

A detailed photometric bulge-disk decomposition of the unbarred galaxies subset of the \atl\ sample is presented \citet[hereafter Paper XVII]{Krajnovic2012}. Although it provides a well established definition of the bulge luminosity, it is a complex technique which depends on the extraction details and suffers from degeneracies. In particular it is well established that, even assuming galaxies to be axisymmetric, the intrinsic stellar density cannot be uniquely recovered from the images unless galaxies are edge-on \citep{Rybicki1987,Gerhard1996,vandenBosch1997,Magorrian1998}. For this reason, unless we knew galaxies to be accurately described by \citet{Sersic1968} bulges and exponential disks \citep{Freeman1970}, we should not expect to be able to uniquely decompose bulges and disks of inclined galaxies, as this would imply we can uniquely infer their intrinsic densities. This is likely one of the reasons why we do not find trends in the bulge fraction (parametrized by the disk-over-total luminosity ratio $D/T$) of fast rotators on the M-size plane. An exception is the important and clear lack of disks above the characteristic $\mjam\ga2\times10^{11}$ \msun.

\begin{figure*}
\includegraphics[width=0.7\textwidth]{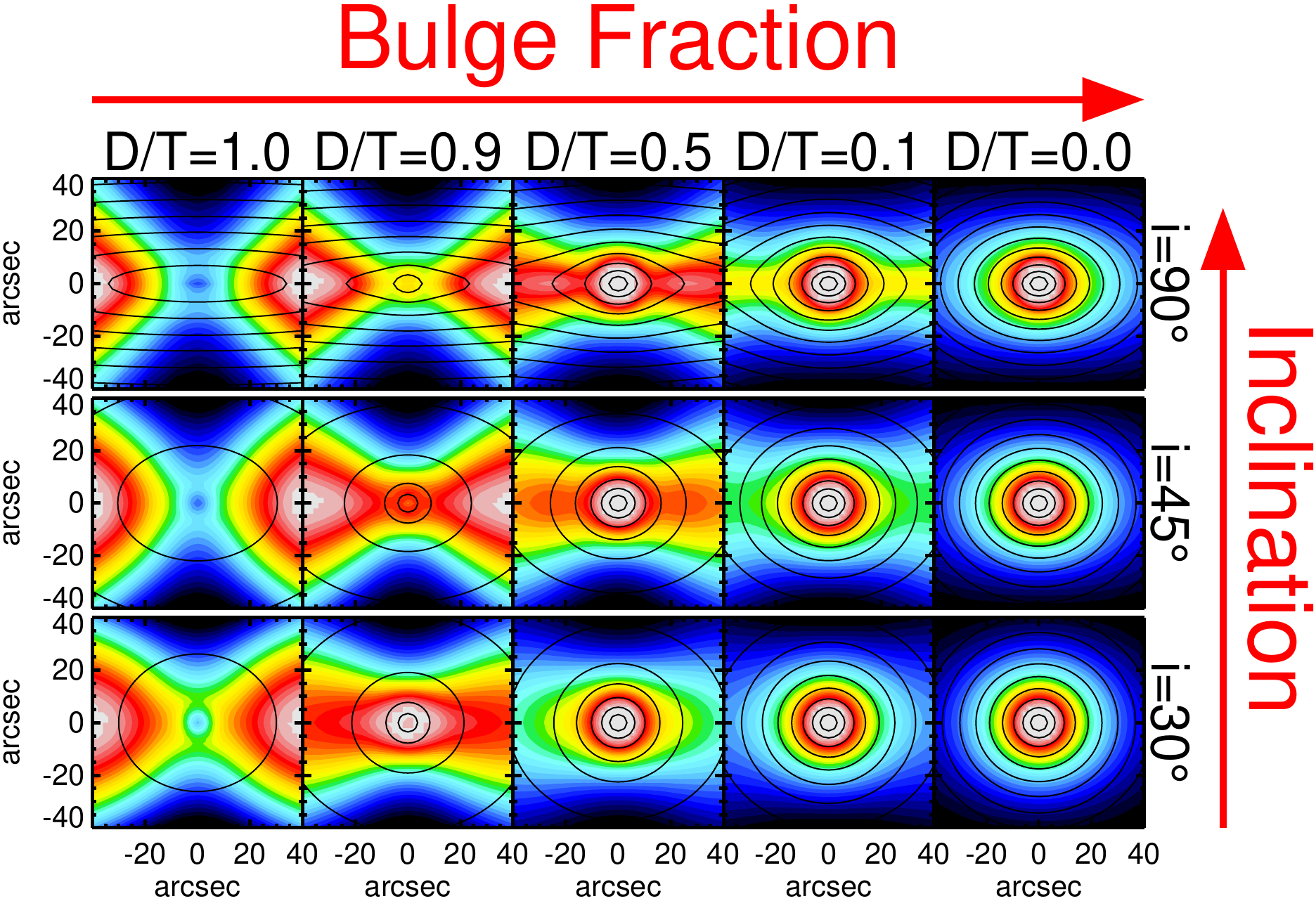}
\caption{Inferring bulge fraction from stellar kinematics. The top panels show the prediction for the second velocity moment $\langle v_{\rm los}^2\rangle^{1/2}$ using JAM dynamical models. This quantity approximates the observed $V_{\rm rms}\equiv\sqrt{V^2+\sigma^2}$, where $V$ is the mean stellar velocity and $\sigma$ the velocity dispersion. The top panels show the predicted kinematics for edge-on realistic galaxy models with different bulge fractions, quantified by the disk-over-total luminosity parameter $D/T$. The middle and bottom panels show the same models, projected at different inclinations. The plots show that for any inclination: (i) an increase of the bulge fraction produces a peak in the central $V_{\rm rms}$; (ii) the presence of a disk produces an horizontal elongation of the $V_{\rm rms}$ contours.}
\label{fig:bulge_fraction_from_jam_models}
\end{figure*}

In our survey we have \sauron\ integral-field kinematics (Paper~I) for all the galaxies in the sample. Here we exploit that extra information to robustly infer the bulge fractions even at low inclinations. The reason why this is possible is because the kinematics retains information on the {\em flattening} of the galaxy components even at low inclinations, when no such information can be inferred from the images. In this situation the photometric decomposition has to rely entirely on the radial profiles and the {\em assumption} of a specific functional form for the two components. The ability of the kinematics to uncover disks and bulges even at low inclination is illustrated in \reffig{fig:bulge_fraction_from_jam_models}. For the plot we constructed a model with a Sersic bulge and an exponential disk. To have a representative model, we adopted the median ratio $R_{\rm e}^{\rm disk}/R_{\rm e}^{\rm bulge}\approx5.2$ and the corresponding median Sersic index for the bulges $n\approx1.7$  (from table~C1 of Paper~XVII). We used the \textsc{mge\_fit\_1d} procedure$^1$ of \citet{Cappellari2002mge} to obtain an accurate MGE fit to the one-dimensional profiles of both the bulge and the disk. We then assigned a typical intrinsic axial ratio $q_{\rm d}=0.2$ to the disk and $q_{\rm b}=0.7$ to the bulge. The Gaussian components of bulge and disk were then combined into a single MGE model that was used as input for a JAM dynamical model \citep{Cappellari2008}, which provided a prediction for the second velocity moments $\langle v_{\rm los}^2\rangle^{1/2}$. It was shown in \citet{Cappellari2008}, \citet{Scott2009} and for the full \atl\ sample in Paper~XV, that the second moments calculated by the JAM method provide good predictions of the observed $V_{\rm rms}\approx\sqrt{V^2+\sigma^2}$ of real galaxies, where $V$ is the mean stellar velocity and $\sigma$ is the stellar velocity dispersion. The only free parameter in this model is the anisotropy in the $(v_R,v_z)$ plane, which is parametrized by $\beta_z\equiv1-\sigma_z^2/\sigma_R^2$, for which we adopted a typical value $\beta_z=0.25$ \citep{Cappellari2008}. While keeping the bulge and disk parameters fixed, we varied their relative contribution $D/T$ and we projected the models at different inclinations. \reffig{fig:bulge_fraction_from_jam_models} shows that the bulge fraction provides dramatic changes in the relative values of the $V_{\rm rms}$ in the galaxy centre with respect to its value at large radii. The presence of disks can still be inferred from the kinematics as a horizontal elongation in the $V_{\rm rms}$ maps, even for disk fractions as small as $D/T=0.1$, and even for inclination as low as $i=30^\circ$. Pure spheroids have $V_{\rm rms}$ maps roughly following the galaxy isophotes. Although the details of the $V_{\rm rms}$ maps are different for different galaxies, the qualitative trends shown in the figure are similar for quite different bulge and disk parameters.

\begin{figure*}
\includegraphics[width=0.7\textwidth]{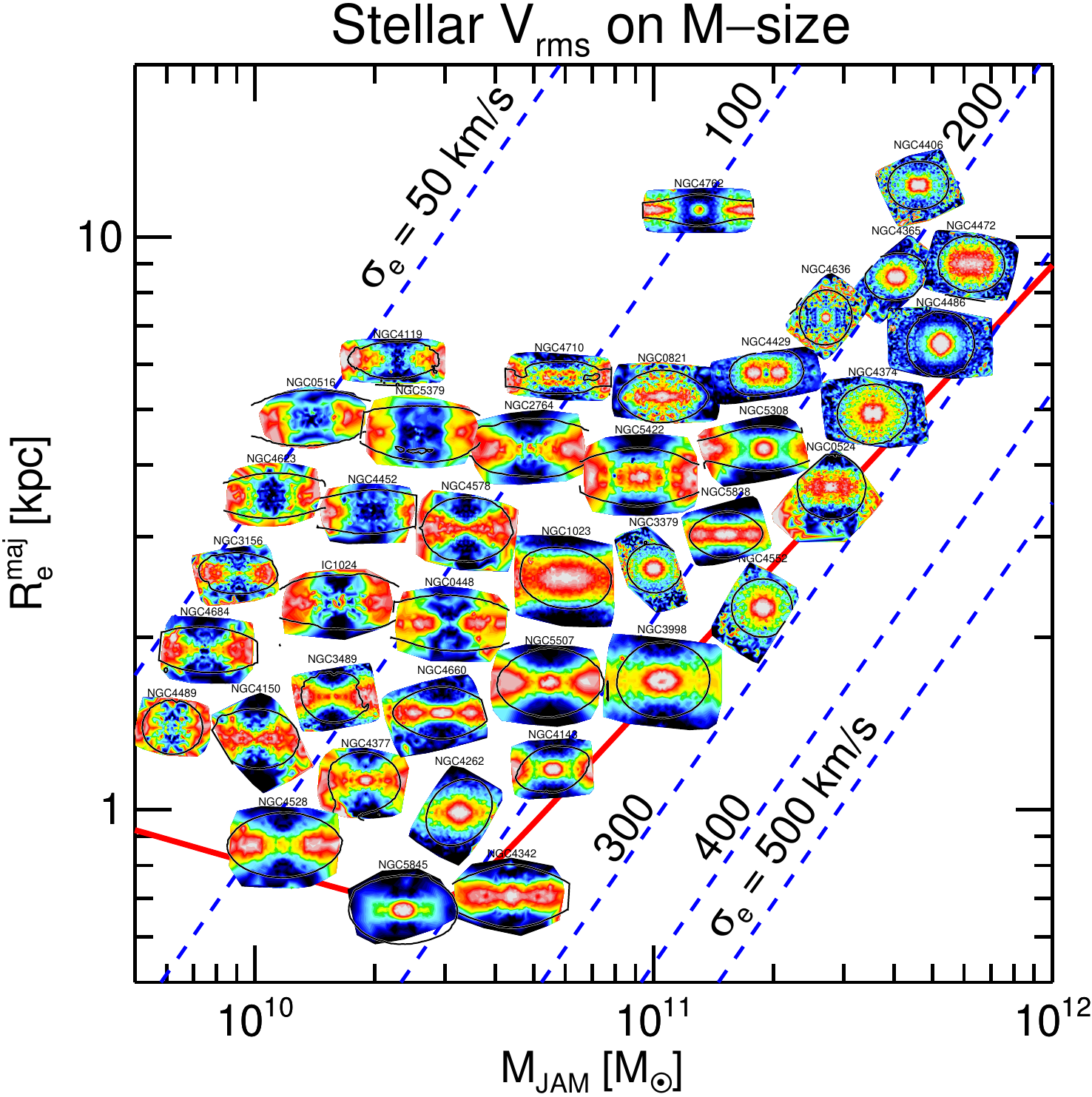}
\caption{Observed $V_{\rm rms}\equiv\sqrt{V^2+\sigma^2}$ maps on the M-size plane. All galaxies are oriented with their kinematical major axis PA$_{\rm kin}$ (from Paper~II) horizontal. A characteristic isophote is indicated by the thick black contour. The plot shows that: (i) near the $\se\approx50$ \kms\ line, galaxies show a drop in the central $V_{\rm rms}$ characteristic of pure disks (see \reffig{fig:bulge_fraction_from_jam_models}); (ii) near the $\se\approx100$ \kms\ line the maps show a butterfly-like shape characteristic of galaxies with small bulges; (iii) near the $\se\approx200$ \kms\ line the maps show strong peaks of $V_{\rm rms}$, indicative of bulges as massive or more massive than the disks, but horizontal elongations due to disks are still present; (iv) above the characteristic mass $\mjam\ga2\times10^{11}$ \msun\ no horizontal elongations of $V_{\rm rms}$ are seen any more, indicating galaxies are pure spheroids. The latter finding agrees with the classification of the objects as slow rotators (Paper~III) and the lack of disks in their bulge-disk decomposition (Paper~XVII).}
\label{fig:vrm_fields_on_mass-size}
\end{figure*}

In \reffig{fig:vrm_fields_on_mass-size} we show the maps of $V_{\rm rms}$ actually observed with \sauron\ for a representative subsample of 40 galaxies in the \atl\ sample, uniformly sampling the M-size plane (the full set of maps is shown in Paper~XV). The plot shows that near the line $\se=50$ \kms\ all maps present the minimum of $V_{\rm rms}$, which is characteristic of pure disk galaxies. The minimum disappears around the line $\se=100$ \kms, where the maps show the butterfly-like shape expected for galaxies with small bulges and dominant disks. For larger \se\ the central peaks in $V_{\rm rms}$ become the norm and finally, near the line $\se=200$ \kms, all galaxies show prominent $V_{\rm rms}$ peaks, indicating bulges as massive or more massive than the disks.
The situation is, once more, qualitatively different above the characteristic mass $\mjam\ga2\times10^{11}$ \msun. Above that mass the $V_{\rm rms}$ maps lack the horizontal $V_{\rm rms}$ elongation. This indicates they don't have disks and consist of pure spheroid, in agreement with the bugle-disk decompositions of Paper~XVII.

\begin{figure*}
\includegraphics[width=0.7\textwidth]{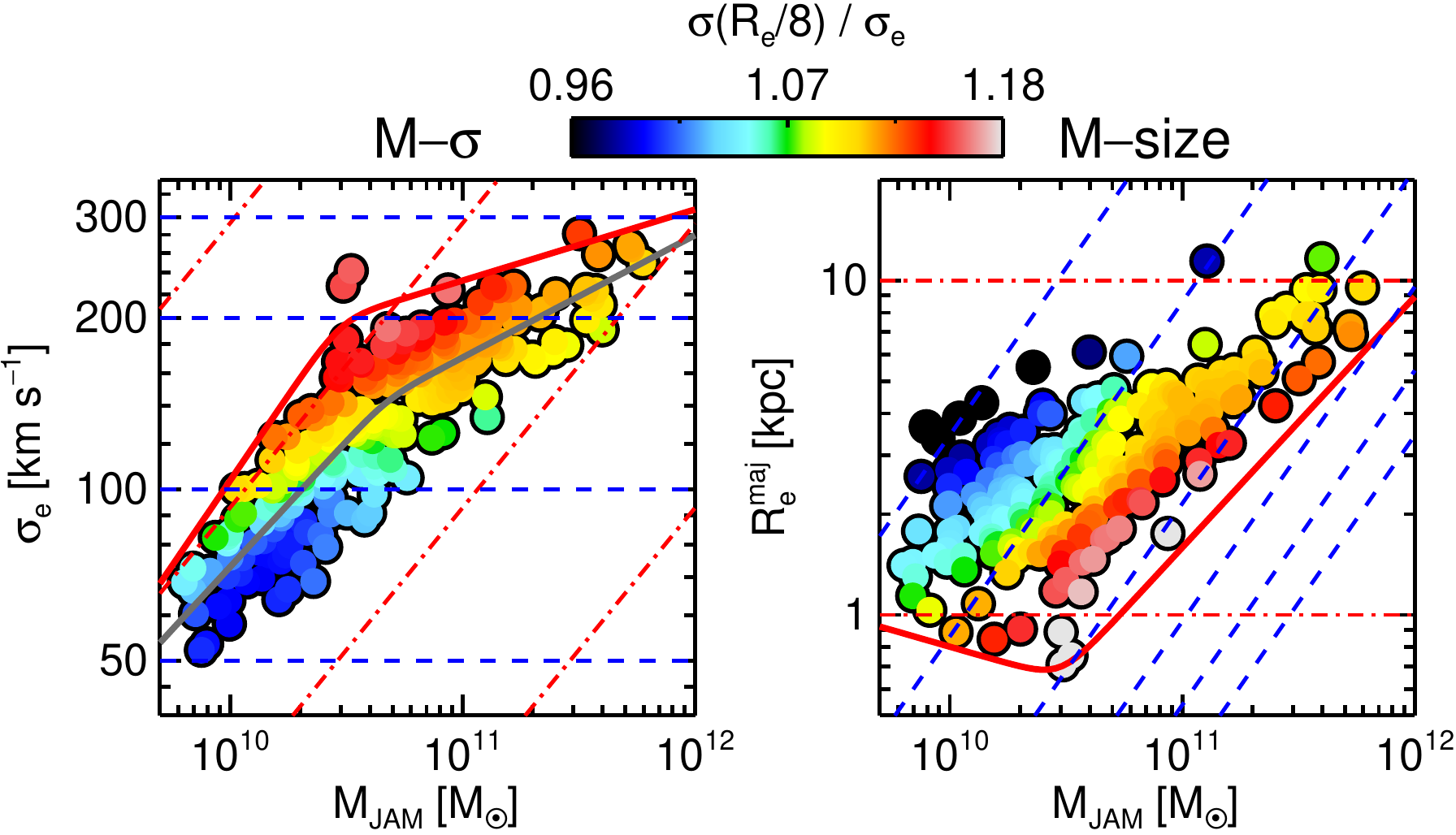}
\includegraphics[width=0.7\textwidth]{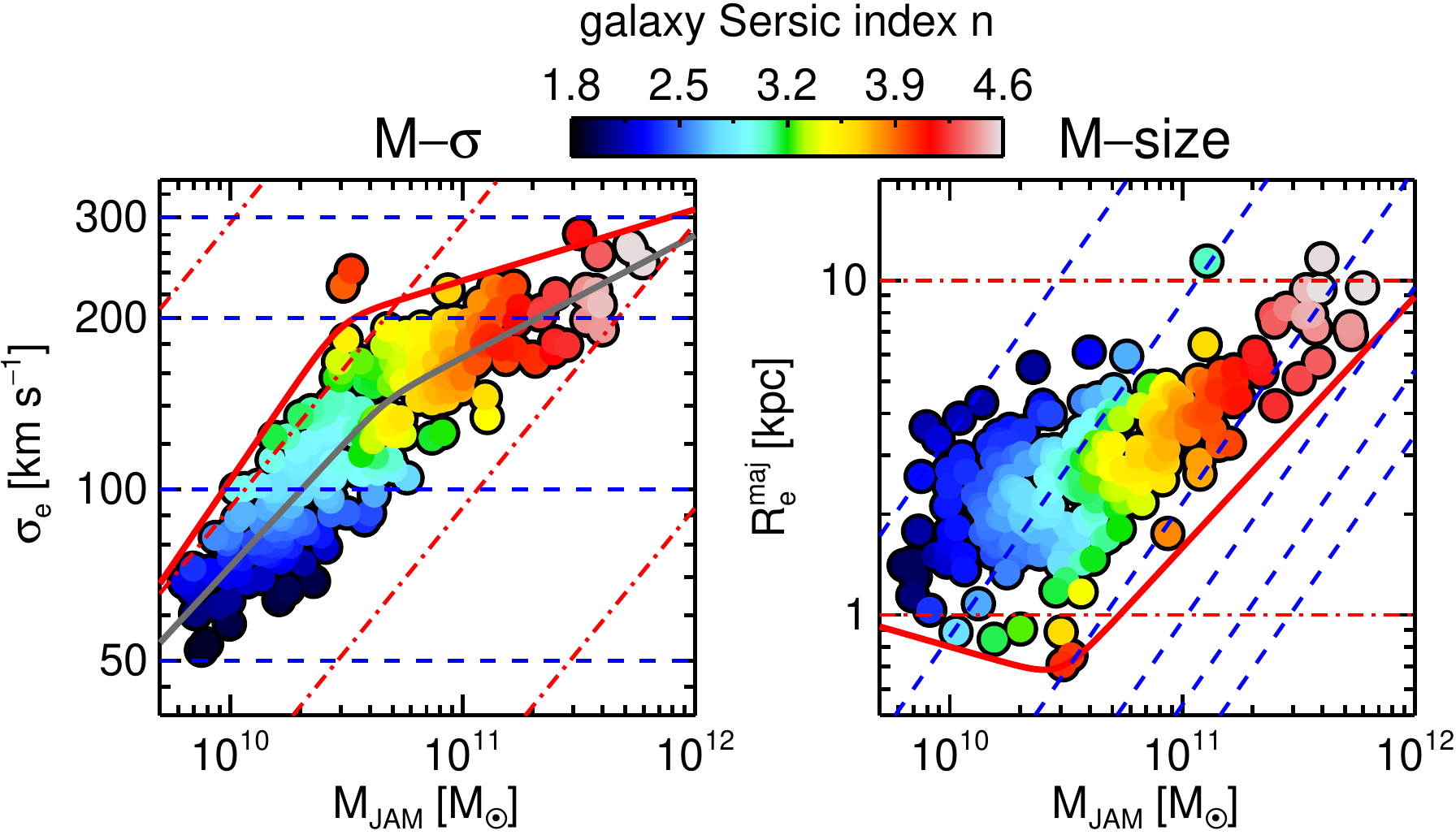}
\caption{Galaxy concentration on the M-size plane. {\em Top Panels:} Same as in \reffig{fig:virial_plane_projections_ml} for the ratio $\sigma(\re/8)/\se$, between the velocity second moment measured within a circular aperture of radius $R=\re/8$ and the second moment \se\ (which includes both rotation and random motions) within the `effective' isophote of semi-major axis \rmaj. As shown in \reffig{fig:bulge_fraction_from_jam_models}, this quantity is related to the bulge fraction, and is weakly sensitive to projection effects. Like for \mljam, the variation of this parameter tends to follow lines of constant \se, or equivalently $\rmaj\propto\mjam$, or even better, it follows lines parallel to the ZOE above its break. Note the slight decrease of $\sigma(\re/8)/\se$ above the characteristic mass $\mjam\ga2\times10^{11}$ \msun.
{\em Bottom Panels:} for comparison this shows the variation of the global Sersic index $n$ for the galaxies, as measured in Paper~XVII. $n$ does not follow lines of constant \se\ and it is not a good empirical predictor of galaxy properties. Above $\mjam\ga2\times10^{11}$ \msun\ galaxies tend to have the value $n\ga4$, which is often associated to pure spheroids.} 
\label{fig:virial_plane_projections_sig8_sige}
\end{figure*}

To try to generalize the qualitative results from the maps of \reffig{fig:vrm_fields_on_mass-size} we need to encode the key information on the shape of the $V_{\rm rms}$ maps into a simple parameter that can be extracted for the full volume-limited \atl\ sample. We found that the following simple approach is sufficient\footnote{For more quantitative results, one could try to break the degeneracy of the photometric bulge-disk decomposition approach by fitting simultaneously both the galaxy images, e.g.\ with a Sersic$+$exponential model, and the $V_{\rm rms}$ maps with the corresponding JAM prediction for different $(i,\beta_z)$. Although interesting and in principle straightforward, this goes beyond the scope of the present paper}. We quantified the variety of shapes in \reffig{fig:bulge_fraction_from_jam_models} by simply measuring the ratio between the central velocity dispersion $\sigma(\re/8)$ (given in Table~1) measured with pPXF (with keyword MOMENTS$=$2) within the commonly used circular aperture of radius $R=\re/8$ and our reference \se\ (given in table~1 of Paper~XV), measured within the `effective' isophote of major axis \rmaj. The distribution of the  $\sigma(\re/8)/\se$ ratio on the MP projections is shown in \reffig{fig:virial_plane_projections_sig8_sige}. It shows that the ratio still approximately follows lines of constant \se\ (or even better lines parallel to the slightly more shallow ZOE $\rmaj\propto M_{\rm JAM}^{0.75}$). This confirms that the trends in the bulge fractions we saw from the maps of the galaxy subset in \reffig{fig:vrm_fields_on_mass-size} are also valid in a statistical sense for the whole \atl\ sample. It demonstrates that at given mass, \se\ traces the bulge fraction, which in turn appear as the main driver for the trends in $M/L$, population parameters and molecular gas content that we observe.

Interestingly the plot also shows a slight decrease in the $\sigma(\re/8)/\se$ ratio above the characteristic mass $M_{\rm JAM}\ga2\times10^{11}  \msun$. This value coincides with the characteristic scale in ETGs above which galaxies start to be rounder \citep{Tremblay1996,vanderWel2009ab}, nearly all have flat (core/deficit) central surface brightness profiles \citep{Faber1997,Graham2003,Ferrarese2006,lauer07prof}, they deviate from power-law colour-magnitude relations \citep{Baldry2004,Ferrarese2006,Bernardi2011mass2e11} and are embedded in massive X-ray halos \citep[hereafter Paper~XIX]{Kormendy2009,Sarzi2013}.

In the bottom panel of \reffig{fig:virial_plane_projections_sig8_sige} we show for comparison the distribution of the \citet{Sersic1968} index $n$ (from Paper~XVII) on the MP projection. These values were obtained by fitting a single Sersic function $I(R)\propto\exp(-k R^{1/n})$ to the entire galaxy profile and do not necessarily provide good descriptions of the galaxies. However this kind of approach is routinely used to characterize sizes and morphologies of large galaxy samples \citep[e.g.][]{Blanton2003,Shen2003} or to study galaxies scaling relations at high-redshift \citep[e.g.][]{Trujillo2007,Cimatti2008,vanDokkum2008}. The plot clarifies that the Sersic index behaves quite differently from the others galaxy indicators and in particular it is weakly sensitive to the bulge fraction. The global Sersic index tends to vary along lines of nearly constant mass rather than \se. Unlike \se, the global Sersic index is not a good predictor of galaxy properties. The plot also clarifies that defining ETGs as having $n>2.5$ \citep[e.g.][]{Shen2003} tends to select ETGs above the characteristic mass $\mjam\approx3\times10^{10}$ \msun, while defining genuine nearby elliptical galaxies as having $n>4$ tends to select objects above the characteristic mass $\mjam\approx2\times10^{11}$ \msun.

\subsection{Variations of galaxy intrinsic flattening and rotation}
\label{sec:rotation}

\begin{figure*}
\includegraphics[width=0.7\textwidth]{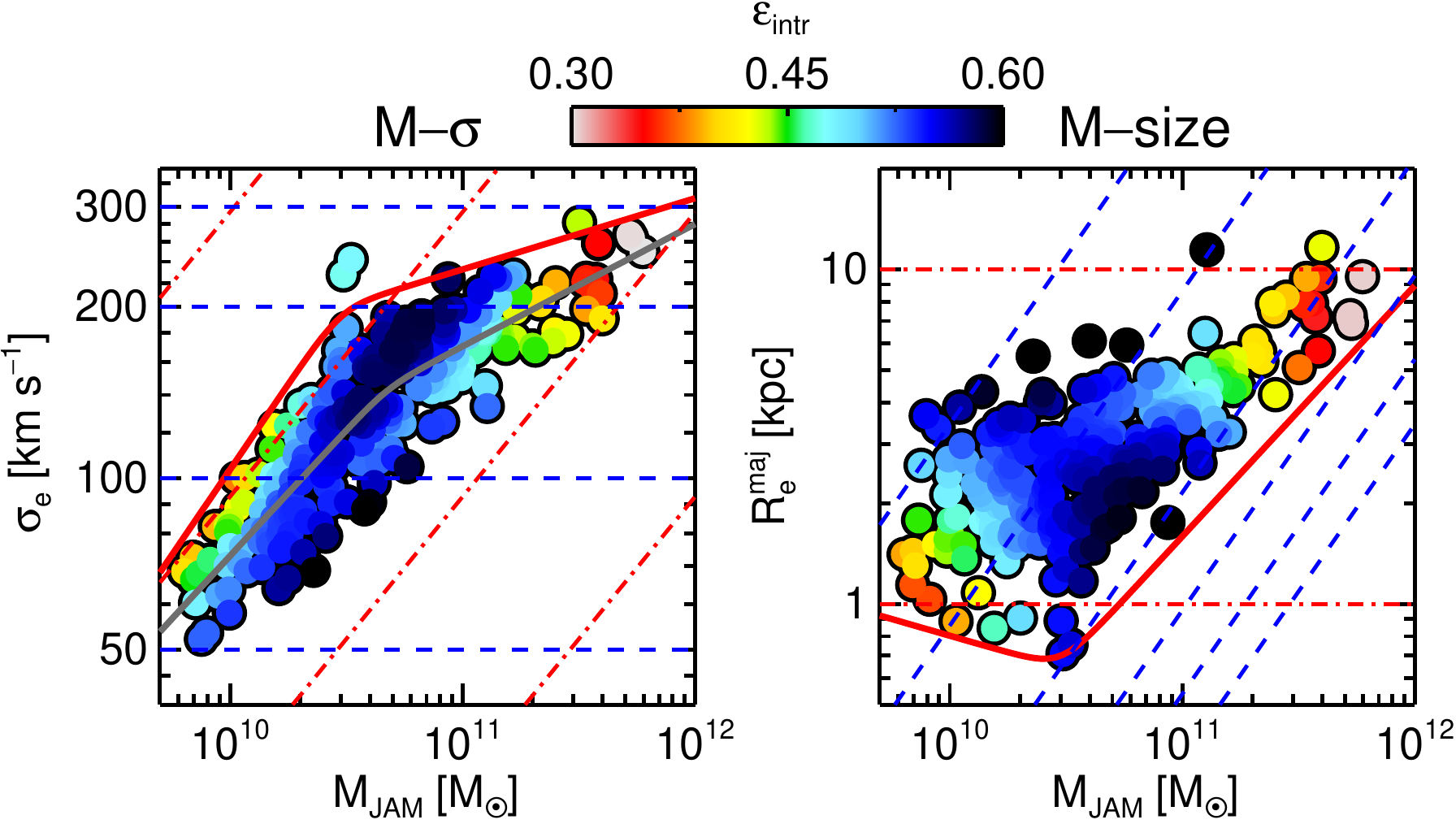}
\includegraphics[width=0.7\textwidth]{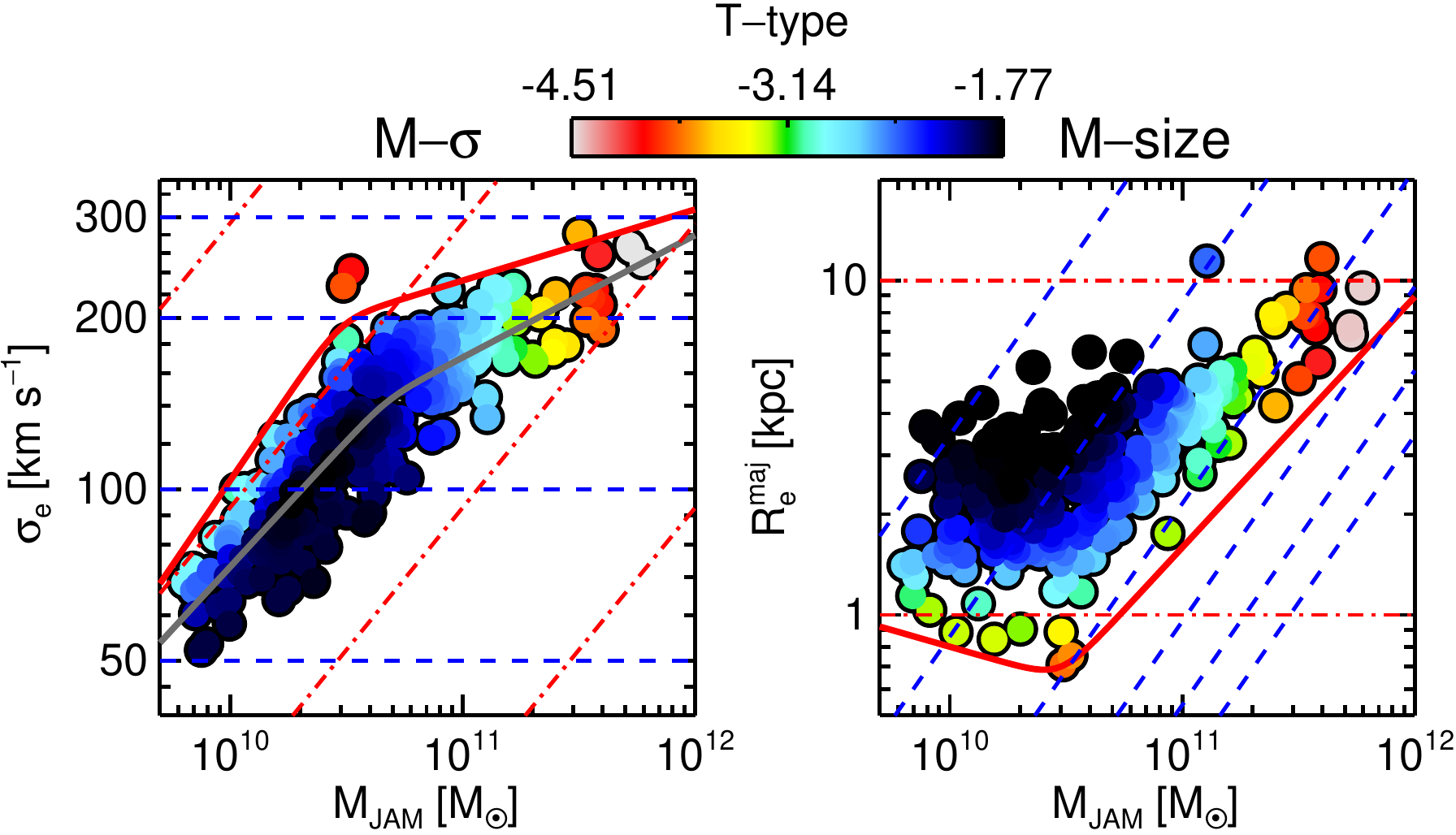}
\caption{Intrinsic shape and morphology. {\em Top Panels:} Same as in \reffig{fig:virial_plane_projections_ml}, with colours indicating the intrinsic ellipticity $\varepsilon_{\rm intr}$ of the galaxies outer regions, computed by de-projecting the observed ellipticity $\epsilon$ at large radii (Paper~II) using the inclination inferred from the JAM models (Paper~XV). Except again for the special region above $M_{\rm JAM}\ga2\times10^{11} \msun$, all the rest of the ETGs have on average the flattening of disks $\varepsilon\approx0.6$ close to the result from the Monte Carlo approach of Paper~III, which suggests fast rotators have a typical intrinsic ellipticity $\varepsilon_{\rm intr}\approx0.7$ and the statistical inversion in Paper~XXIII.
{\em Bottom Panels:} Same as in the top panel, with colours indicating the galaxy morphology as quantified by the T-type parameter from HyperLeda. For $M_{\rm JAM}\ga2\times10^{11} \msun$ galaxies are invariably classified as E ($T<-3.5$), while both E and S0 classifications are present at lower masses.}
\label{fig:virial_plane_projections_concentration}
\end{figure*}

Our dynamical modelling effort provides the unique opportunity of having the galaxy inclination for the entire \atl\ sample of galaxies (from table~1 of Paper~XV). We verified that the JAM inclination agrees with the one inferred from the geometry of the dust, for 26 galaxies with regular dusts disks (see also \citealt{Cappellari2008}). We further verified that we can recover with JAM models the inclination of simulated galaxies that resemble typical objects in our sample (Paper~XII). Although one should not expect the inclination to be reliable for every individual galaxy, it is expected to be accurate for most of our sample. The inclination allows us to recover the intrinsic shape of individual objects. The deprojection is done using the ellipticity $\varepsilon$ of the outer isophote, measured in Paper~II at radii typically around 4\re, at the depth limit of the SDSS photometry. At these radii, when a disk is present, the ellipticity is representative of the outer disks and is insensitive to the possible presence of a bulge. We then computed the intrinsic ellipticity assuming an oblate spheroid geometry as \citep{Binney2008}
\begin{equation}\label{eq:corr_eps}
\varepsilon_{\rm intr}=1-\sqrt{1+\varepsilon (\varepsilon-2)/\sin^2 i},
\end{equation}
The results of the deprojection is shown in the top panels of \reffig{fig:virial_plane_projections_concentration}. Contrary to all previous diagrams presented so far, the distribution of the intrinsic ellipticity for ETGs on the MP shows a completely different trend. Few galaxies stand out in the top-right corner for being nearly round, with average intrinsic ellipticity $\varepsilon_{\rm intr}\approx0.3$. These galaxies are all located above the same characteristic mass $M_{\rm JAM}\ga2\times10^{11}$ \msun, where other observables indicated a transition in galaxy properties. This result confirms previous  statistical studies of ETGs shapes \citep{Tremblay1996,vanderWel2009ab,Bernardi2011mass2e11}. Below this characteristic mass, the mean ellipticity drops dramatically and sharply to about $\varepsilon_{\rm intr}\approx0.6$, characteristics of disks. This galaxy-by-galaxy shape deprojection confirms a similar result on their shape distribution inferred via Monte Carlo simulations from the distribution of fast rotators ETGs on the $(\lambda_R,\varepsilon)$ diagram, suggesting fast rotators have a typical intrinsic ellipticity $\varepsilon_{\rm intr}\approx0.7$  (fig.~15 in Paper~III). It also agrees with a statistical inversion of their shape \citep[hereafter Paper XXIII]{Weijmans2013}. This agreement also seems to confirm the reliability of the inclinations derived via the JAM models. The result strongly confirms our previous conclusions that fast rotators as a class are disk like (Paper~II, III, VII). At the lowest mass range ($\mjam\la10^{10}$ \msun), near the ZOE, there seems to be a marginal decrease of $\varepsilon_{\rm intr}$. The significance or reality of this feature is however unclear.

A similar trend as for the intrinsic shape is seen in the bottom panel of \reffig{fig:virial_plane_projections_concentration} for the optical morphology, as quantified by the T-type parameter from Hyperleda \citep{Paturel2003}. Above $M_{\rm JAM}\ga2\times10^{11}$ \msun\ nearly all galaxies are classified as ellipticals ($T<-3.5$), consistently with the fact that galaxies in this mass range are genuine spheroids. While below this mass both E and S0 classification are present, due to the morphological misclassification of inclined disks for genuine spheroidal ellipticals.

\begin{figure*}
\includegraphics[width=0.7\textwidth]{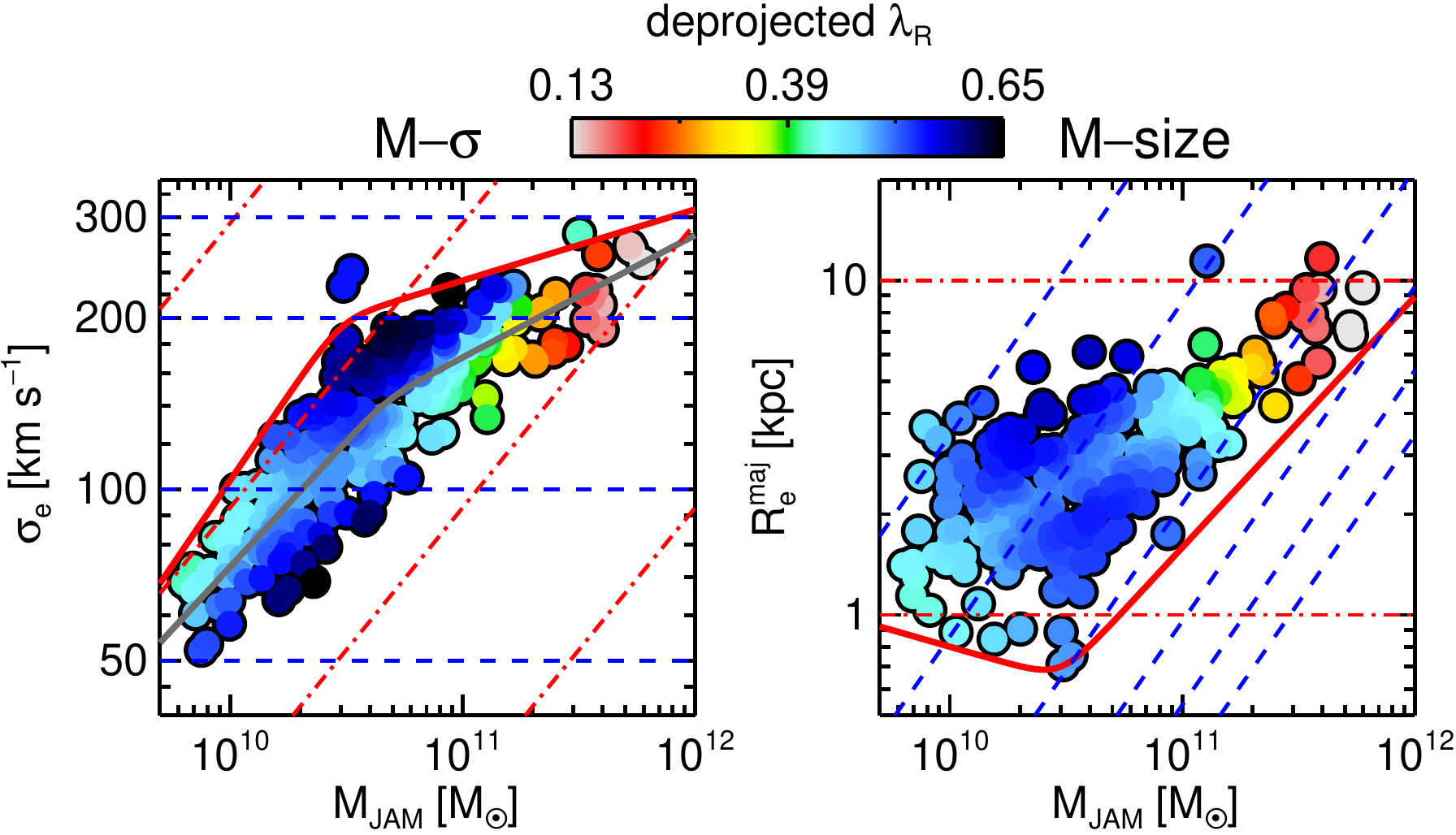}
\includegraphics[width=0.7\textwidth]{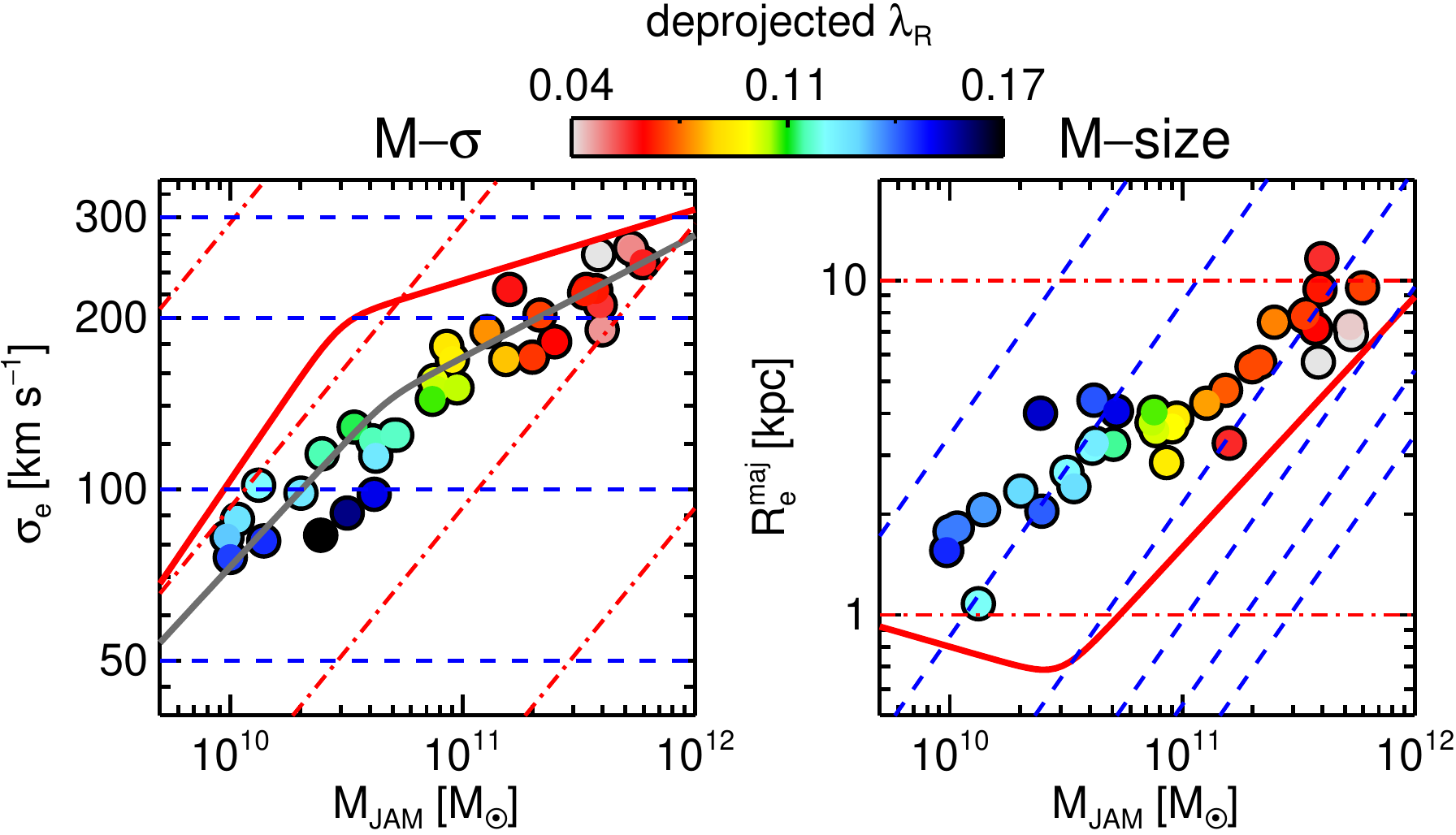}
\caption{Galaxy stellar rotation on the MP projections. Same as in \reffig{fig:virial_plane_projections_ml}, with colours indicating the specific angular momentum parameter $\lambda_R$ of \citet{Emsellem2007}, as given in Paper~III, deprojected to an edge-on view according to the JAM inclination (Paper~XV). {\em Top Panels:} all \atl\ galaxies are included. A clear transition in the mean galaxy rotation is evident above  $M_{\rm JAM}\ga2\times 10^{11}$ \msun, with the population becoming entirely dominated by slow rotators. {\em Bottom Panels:}  Only the slow rotators are shown. They seems to define a sequence in the MP, and they show a decrease in the rotation especially above $M_{\rm JAM}\ga2\times 10^{11}$ \msun, where only the slow rotators with the lowest $\lambda_R$ are present.}
\label{fig:virial_plane_projections_lambda}
\end{figure*}

The round and massive objects that stand out from the shape distribution above $M_{\rm JAM}\ga2\times10^{11}$ \msun\ are the prototypical slow rotators, which also stand out for their extremely low specific stellar angular momentum $\lambda_R$ \citep{Emsellem2007}, as shown in Paper~III. In the top panel of \reffig{fig:virial_plane_projections_lambda} we plot the distribution of the intrinsic $\lambda_R$, which was approximately obtained from the observed one given in Paper~III by deprojecting the observed velocity to an edge-on view, using the best-fitting inclination from the JAM models. The plot shows a clear transition in the average $\lambda_R$ of the galaxy population at the mass $M_{\rm JAM}\ga2\times10^{11}$ \msun. Above this mass essentially all ETGs are slow rotators, while below this mass the population is dominated by fast rotators, although some slow rotators are still present.
The bottom panel of \reffig{fig:virial_plane_projections_lambda} plots the angular momentum parameter $\lambda_R$ for the slow rotator ETGs only. The slow rotators are found to populate a rather narrow sequence on the MP, with a small range of $\Sigma_{\rm e}$. This plot shows that not all slow rotators are the same. In fact there is a clear trend of decreasing rotation with increasing mass as noted in Paper~III, or possibly a transition around $M_{\rm JAM}\approx10^{11} \msun$, which is similar to the characteristic mass defined by other observables. This confirms that the massive slow rotators above our characteristic mass are indeed special and may have formed differently from the rest. Although we tend to cover a smaller field with our kinematics for the largest galaxies, the qualitative difference in the kinematics for the most massive slow rotators is already clearly visible from the velocity fields (Paper~II) and it is not due to the difference in the field coverage. A strong indication that indeed the most massive slow rotators are special also comes from the fact that they all appear to have cores, while the less massive ones are core-less  \citep[hereafter paper~XXIV]{Krajnovic2013}.

\begin{figure*}
\includegraphics[width=0.9\textwidth]{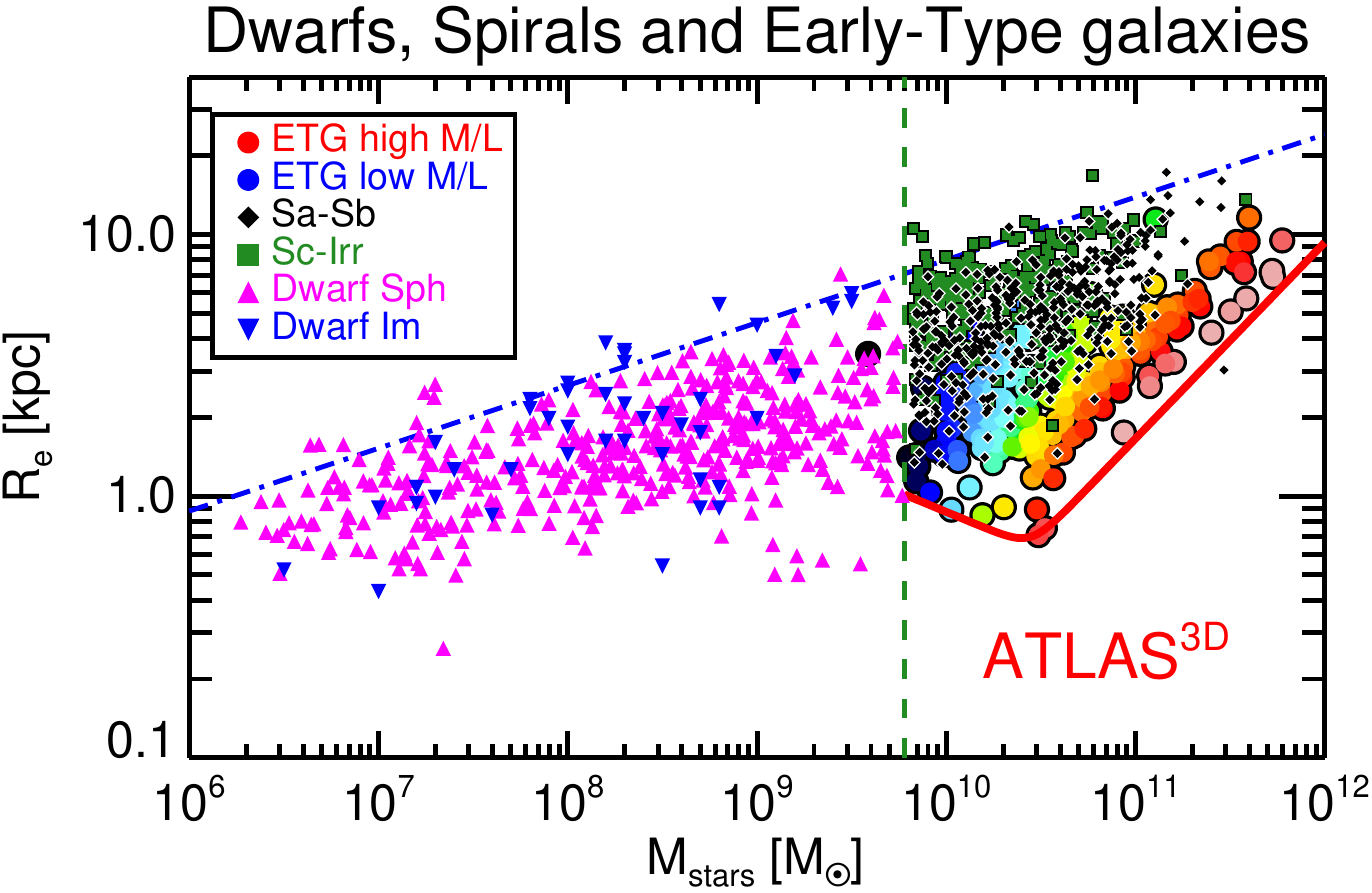}
\caption{The Mass-size distribution for dwarfs, spiral galaxies and ETGs. The ETGs of the \atl\ sample are coloured according to their \mljam\ as in \reffig{fig:virial_plane_projections_ml}. This plot however also includes the approximate location of spiral galaxies in the \atl\ parent sample, where masses are obtained from $K$-band luminosities, scaled to match \mjam\ (a plot using luminosities rather than masses for both ETGs and spirals was shown in fig.~4 of Paper~I). Early spirals (Sa-Sb: $T\le4$) are indicated by the black filled diamonds, while late spirals (Sc-Irr: $T>4$) are shown with the filled green squares. Late spirals are larger than ETGs, while early spirals overlap with the ETGs with low $M/L$. There are essentially no spirals in the region of the oldest and reddest ETGs, which have the largest $M/L$. This illustrates the fact that the increase of the spheroid is required to make a galaxy old and red and consequently produce the largest $M/L$. No spirals are present also above the characteristic mass $\mjam\ga2\times10^{11}$ \msun. Also included in the plot is the approximate location of dwarf early-type galaxies (Sph: magenta filled up triangles. From \citealt{Gavazzi2005}, \citealt{Ferrarese2006acs} \citealt{Misgeld2008} and \citealt{Misgeld2009}) and dwarf irregulars (Im: blue filled down triangles. From \citealt{Kirby2008}). These are only shown below the mass limit of the volume-limited \atl\ survey (vertical dashed line). For reference the dash-dotted blue line shows the relation $(\re/{\rm kpc})=8\times[M_{\rm stars}/(10^{10} M_\odot)]^{0.24}$.}
\label{fig:mass_size_spirals_etgs}
\end{figure*}

In summary the distribution of galaxy properties on the $(M_{\rm JAM},\sigma_{\rm e})$ and $(M_{\rm JAM},R_{\rm e}^{\rm maj})$ projections of the MP can be understood as due to a smooth variation in the bulge mass as already suggested in figure~4 of Paper~I. The connection between bulge mass and galaxy properties, and the close link between ETGs and spiral galaxies is further illustrated in \reffig{fig:mass_size_spirals_etgs}, which includes the location of the spiral galaxies of the \atl\ parent sample (Paper~I) together with the ETGs. The masses of spiral galaxies was approximately estimated assuming a fixed $M/L_K=0.8$ \msun/\lsun, which ensures agreement between $M_{\rm JAM}$ and the $K$-band luminosity at the lowest masses.  The plot shows that ETGs properties vary smoothly with those of spiral galaxies. Galaxies with negligible bulges are almost invariably star forming and classified as late spirals. Galaxies with intermediate bulges can still form stars and be classified as early-spirals, or can be fast rotator ETGs. But the galaxies with the most massive bulges, as indicated by their largest concentration or $\sigma_{\rm e}$ are invariably ETGs. They have the largest $M/L$ (\reffig{fig:virial_plane_projections_ml}), the reddest colours, smallest H$\beta$ and lowest molecular gas fraction (\reffig{fig:virial_plane_projections_hbeta}), but are still flat in their outer parts, indicating they have disks (\reffig{fig:virial_plane_projections_concentration}) and generally still rotate fast (\reffig{fig:virial_plane_projections_lambda}). An exception are galaxies with masses $M_{\rm JAM}\ga2\times10^{11} \msun$, which stand out for being nearly round, and non rotating. Above this characteristic mass essentially no spirals are present.

Also included in \reffig{fig:mass_size_spirals_etgs} is the approximate location of dwarf ETGs (spheroidals) and dwarf irregulars (Im) galaxies in the mass-size diagram. The dwarf ETGs parameters were taken from \citet{Gavazzi2005}, \citet{Ferrarese2006acs}, \citet{Misgeld2008} and \citet{Misgeld2009}, while the Im parameters come from \citet{Kirby2008}. The plot shows that the spheroidal galaxies lie along the continuation of the M-size relation for spiral galaxies and naturally connect to the region of fast rotator ETGs with the lowest $M/L$, young ages and small bulges. One should keep in mind that, while the galaxies parameter space is fully sampled above the mass selection limit $M_{\rm stars}\ga6\times 10^9$ \msun\ of the \atl\ survey, below this limit the data are heterogeneous. A much improved coverage of these parameters will soon be provided by the Next Generation Virgo Cluster Survey \citep{Ferrarese2012} as shown in \citet{Ferrarese2012eso}.

\section{Systematic variation of the IMF}
\label{sec:imf}

Although the IMF variation could have been included among the other galaxy observables described in \refsec{sec:vp_projections}, we keep the IMF in a separate section for a more detailed coverage than for the other observables. In recent months there has been a large amount of interest on the IMF variation. Before describing our new \atl\ results, in the following section we provide a summary of this rapidly evolving field. To clarify the significance and robustness of the various recent claims, including our own, we place particular emphasis on the modelling assumptions that went into the various studies.

\subsection{Summary of previous IMF results}

\subsubsection{IMF of the Milky Way}

The stellar initial mass function (IMF) describes the distribution of stellar masses when the population formed. Nearly every aspect of galaxy formation studies requires a choice of the IMF to calculate predictions for galaxy observables \citep{Kennicutt1998imf}. For this reason the IMF has been subjected to numerous investigations since the first determination more than half a century ago finding it has the form \citep{Salpeter1955}
\begin{equation}
    \zeta(m)\propto m^x=m^{-2.3},
\end{equation}
for $m>0.4 \msun$, where $m$ is the stellar mass.

In the Milky Way the IMF can be measured via direct stellar counts. Various determination in different environments lead to the finding of a remarkable universality in the shape of the IMF \citep{Kroupa2002}, with the IMF being well described by the Salpeter power slope $x=-2.3$ for $m\ge0.5 \msun$ and a shallower one at smaller masses $x=-1.3$  for $m<0.5 \msun$ \citep{Kroupa2001}. The Milky Way IMF can also be described by a log-normal distribution \citep{Chabrier2003}, which has some theoretical justification. However the Kroupa and Chabrier IMFs are essentially indistinguishable from an observational point of view \citep{Kroupa2012}. Recent IMF results were reviewed by \citet{Bastian2010} who concluded no clear evidence existed for a non-universal IMF in our galaxy and among different galaxies.

\subsubsection{IMF from ionized gas emission or redshift evolution}

In external galaxies individual stars cannot yet be resolved down to sufficiently small masses for IMF studies. Some indirect constraints on the slope of the IMF above $\sim1 \msun$ can be obtained by combining observations of the H$\alpha$ equivalent widths, which is related to the number of ionizing photons, with the galaxy colour, which is a function of the galaxy stellar population \citep{Kennicutt1998imf}. Using this technique \citet{Hoversten2008} inferred a variation of the IMF with more massive galaxies having a more top-heavy IMF than Salpeter, a result confirmed under the same assumptions \citep{Gunawardhana2011} by the GAMA survey \citep{Driver2011}. A similar result was found using the ratio between the H$\alpha$ and ultraviolet fluxes by \citet{Meurer2009}, who further pointed out that surface density is more important than mass in driving the IMF trend.

Constraints on the IMF can also be obtained by comparing the local ($z=0$) stellar mass density to the integral of the cosmic star formation history. \citet{Wilkins2008} and \citet{Dave2008} find that the local stellar mass density is lower than the value obtained from integrating the cosmic star formation history (SFH), assuming that all the stars formed with a Salpeter IMF. They propose a time-variable IMF to reconcile the observations.

Alternative constraints on the IMF can be obtained by studying the evolution of the $M/L$ normalization in samples of galaxies as a function of redshift, for example from the evolution of the tilt and normalization of the Fundamental Plane with redshift \citep{Renzini1993}. From these constraints and from the metal abundance in the clusters interstellar medium \citet{Renzini2005} concluded that elliptical galaxies as a class must have an IMF close to Salpeter for stellar masses $m>1 \msun$, while the IMF must flatten at lower masses, similar to the Kroupa IMF.

A related approach was employed by \citet{vanDokkum2008imf} who used the ratio of luminosity evolution to colour evolution of massive galaxies in clusters to constrain the IMF. He originally concluded that the IMF must be top heavy at high redshift in apparent agreement with some theoretical model predictions \citep{Zepf1996,Chiosi1998}. However this result was recently revised with better data and models to conclude that the IMF above $m>1 \msun$ is actually not inconsistent with a Salpeter slope \citep{vanDokkum2012}.

\subsubsection{Lensing or dynamics IMF constraints for external spirals}

A more direct approach to constraining the variation of the IMF in galaxies consists of measuring the mass of the stars, and comparing this with the predictions of stellar population models. It is important to stress however that this method does not measure the shape of the IMF directly, but only its overall mass normalization. The method can only verify whether the stellar mass in a certain region within a galaxy is consistent with that due to a stellar population with a certain assumed IMF.  Moreover these methods do not measure the same IMF one can infer via direct stellar counts in star clusters. They measure instead the stellar mass distribution due to the superposition of the IMFs from a large number of star clusters in the galaxy as well as the IMFs coming from the accretion of satellite galaxies during the galaxy hierarchical growth. As pointed out by \citet{Kroupa2003}, even if the IMF was universal in every cluster in a galaxy, the integrated galactic IMF (IGIMF) would be different from the universal one, due to the non-uniform mass distribution of star clusters \citep[see][for an in-depth discussion]{Kroupa2012}. Hierarchical evolution complicates the picture even further. For these reason, when we state e.~g. that ``a galaxy is consistent with the Salpeter IMF'', we simply imply, consistently with all previous similar studies, that ``the galaxy has the same stellar $M/L$ of a stellar population with the given age and metallicity (and abundances), and the Salpeter IMF''.

The first convincing constraint on the IMF of external galaxies was obtained for a sample of 21 spiral galaxies, using the kinematics of the gas. \citet{Bell2001} concluded that {\em if the IMF is universal}, a sensible assumption at the time, it cannot have the Salpeter form, but it must be lighter, consistently with Kroupa-type. The result was later confirmed, still using gas kinematics, for a sample of 34 bright spiral galaxies by \citet{Kassin2006}. 

The need for a light IMF was inferred for the Einstein Cross spiral galaxy  gravitational lens system  \citep{Huchra1985} by both \citet{vandeVen2010} and \citet{Ferreras2010}, and for another spiral lens galaxy by \citet{Suyu2012}. As part of the DiskMass Survey \citep{Bershady2010}, which obtained integral-field stellar kinematics of a sample of 30 spiral galaxies, a light IMF was indirectly confirmed from the sub-maximality of the disks \citep{Bershady2011}. As part of the SWELLS survey of spiral lens galaxies \citep{Treu2011} the need for a light IMF for most spiral galaxies was also found from the analysis of 20 strong gravitational lens \citep{Brewer2012}.

An inconsistency of the Salpeter IMF normalization for low mass galaxies was also inferred by \citet{Dutton2011imf}, using simple galaxy models with a bulge and a disk, trying to reproduce global trends for a large sample of galaxies extracted from the SDSS.

In summary there is a good agreement on the fact that spiral galaxies as a class must have a normalization lighter than Salpeter and similar to Kroupa/Chabrier. This is a robust result due to the fact that a Salpeter IMF would over-predict the total mass in the galaxy centres. It is unclear whether some spiral galaxies have a Salpeter IMFs, and of course whether the IMF varies within the galaxies themselves, as claimed by \citet{Dutton2012imf_bulges}.

\subsubsection{Lensing or dynamics IMF constraints on early-types}

An early attempt at constraining the IMF of 21 elliptical galaxies, using detailed spherical dynamical models including dark matter, found a general consistency between the stellar $(M/L)_{\rm stars}$ and the one of the population $(M/L)_{\rm pop}$, using the Kroupa IMF, but could not accurately constrain the IMF normalization due to large observational errors \citep{Gerhard2001}. A large study of SDSS elliptical galaxies using fixed spherical Hernquist galaxy models with dark halos \citep{Padmanabhan2004} found a mass excess over the predictions of stellar population models with a fixed IMF, increasing with luminosity. This was interpreted as a increase of the dark matter fraction. A similar conclusion was reached by \citet{Zaritsky2006} while studying the fundamental manifold. A caveat of those studies was the use of homologous stellar profiles or approximate assumptions to study systematic variations in the heterogeneous galaxy population.

To overcome these limitations, as part of the \sauron\ project \citep{deZeeuw2002}, we constructed self-consistent axisymmetric models reproducing in detail both the photometry and the state-of-the-art integral-field stellar kinematics data \citep{Emsellem2004} for 25 early-type galaxies \citep{Cappellari2006}. The assumption that mass follows light is not accurate at large radii, but it is a good assumption for the region where kinematics is available ($\la1\re$), and provides accurate measurements of the total (luminous plus dark) $M/L$ within a sphere of radius $r\approx\re$ (see Paper~XV). The resulting improvement in the $M/L$ accuracy and the removal of systematic biases allowed us to strongly confirm that ``the total and stellar $M/L$ clearly do not follow a one-to-one relation. Dark matter is needed to explain the differences in $M/L$ (if the IMF is not varying)'' \citep[see fig.~17 there]{Cappellari2006}. Although dark matter seemed, at the time, still a more natural explanation of our observations, if the measured trends are re-interpreted as an IMF variation, they would imply an IMF heavier than Salpeter for some of the oldest objects.  That study also concluded that {\em if the IMF is universal} it must have a mass normalization as low as Kroupa IMF, consistently with the results for spiral galaxies, otherwise the stellar mass would overpredict the total one for a number of galaxies. A similar finding was obtained by \citet{Ferreras2008} using gravitational lensing of 9 elliptical galaxies extracted from the SLACS survey \citep{Bolton2006}.

A number of subsequent studies used approximate models reproducing the velocity dispersion of large galaxy samples to study dark matter and IMF variations in galaxies. \citet{Tortora2009} used spherical isotropic models with \citet{Sersic1968} profiles to reproduce the large velocity dispersion compilation by \citet{Prugniel1996fp}. \citet{Graves2010} used the Fundamental Plane of early-type galaxies from SDSS data. \citet{Schulz2010} combined weak lensing measurements in the outer parts to SDSS velocity dispersion determination, using spherical Jeans models with a Hernquist profile. All three studies confirmed the existence of a mass excess which increases with mass. They all preferred a dark matter trend to explain the observations, although they could not exclude the IMF variation alternative.

\citet{Grillo2009} compared stellar population masses, derived from multicolour photometry, to the total masses, inside the Einstein radius of the lenses, published by the SLACS team \citep{Bolton2008slacs5}. He assumed for all galaxies the average dark matter fraction determined for some of the galaxies by \citet{Koopmans2006} and \cite{Gavazzi2007}, which were based on spherical models assuming a single power-law total mass and a \cite{Hernquist1990} (or \citealt{Jaffe1983}) stellar profile. From this comparison \citet{Grillo2009} concluded that elliptical galaxies prefer a Salpeter rather than Kroupa/Chabrier IMF. The same result was found when comparing stellar dynamical masses for galaxies in the Coma cluster by \citet{Thomas2009dm}, which includes dark matter halos with fixed NFW profiles, to stellar population masses \citep{Grillo2010}. However, given that the dynamical masses were determined assuming a fixed NFW profile, the conclusion depended on the correctness of that assumption. A recent reanalysis of the 16 Coma galaxies by \citet{Thomas2011}, still assuming a NFW dark halo profile, interpreted the $M/L$ excess in massive galaxies as more likely due to dark matter, although their results are not inconsistent with a Salpeter IMF instead. The same conclusions were reached by \citet{Wegner2012} for 8 galaxies in Abell~262.

The SLACS team analysed their data assuming all galaxies in their sample can be approximated by homologous spherical and isotropic systems with a \cite{Hernquist1990} (or \citealt{Jaffe1983}) profile. In \citet{Treu2010} they further assumed a spherical NFW profile for the dark matter, with fixed slope and only mass as free parameter. Given that the enclosed total (luminous plus dark) mass inside the Einstein radius is essentially fixed by the lens geometry, under these assumptions the central velocity dispersion is a unique function of the profile of the total mass distribution, which in this case is defined by a single parameter: the ratio between the stellar and dark matter components. Comparing the stellar mass from the lens model to the one from population, based on multicolour photometry \citep{Auger2009}, they found that the data prefer a Salpeter-like normalization of the IMF. \citet{Auger2010imf} used the SLACS data and the same spherical Hernquist models but, instead of allowing for free halos for different galaxies, they assumed the same IMF and the same halos, following the trends predicted by the abundance matching technique \citep{Moster2010}, for the entire ensemble population. They still assumed NFW halos but explored both adiabatically contracted \citep{Gnedin2004} and not contracted halos. They concluded that the data favour a Salpeter-like normalization of the IMF over a lighter Kroupa/Chabrier form.

Although ground-breaking, the analyses of the SLACS data depended on some non-obvious assumptions. \citet{Grillo2009} had assumed a fixed dark matter fraction, as well as a power-law form for the total mass. \citet{Treu2010} result depended on the assumption of a fixed NFW halo slope, as the authors acknowledged concluding that ``the degeneracy between the two [IMF or dark matter] interpretations cannot be broken without additional information, the data imply that massive early-type galaxies cannot have both a universal IMF and universal dark matter halos'', in agreement with previous dynamical analyses. \citet{Auger2010imf} result depended on assuming the halo mass from the abundance matching techniques. This led the authors to conclude that ``better constraints on the star formation efficiency must be obtained from the data in order to draw definitive conclusions about the role of a mass-dependent IMF relative to CDM halo contraction''.

Other independent studies of the same SLACS data did not exclude an IMF variation, but did not confirm the need for a Salpeter IMF to explain the observations. \citet{Tortora2010} reached this conclusion using spherical Hernquist isotropic models like \citet{Treu2010} and \citet{Auger2010imf}, but could not explain the reason for the disagreement. A similar study was also performed by \citet{Deason2012dm}. They showed that the trend in the galaxies enclosed masses can be explained by a toy model similar to the one adopted by \citet{Auger2010imf}, but with a universal Kroupa IMF. \citet{Deason2012dm} study however did not fit the galaxies stellar velocity dispersion, which could be a reason for the difference in the results. \citet{Barnabe2011} re-analysed 16 SLACS galaxies using axisymmetric, rather than spherical, dynamical models, and describing the galaxy images in detail, rather than assuming fixed Hernquist profiles. Their models are fitted to integral-field stellar kinematics, instead of a single velocity dispersion. They find that the data are consistent with both a Kroupa or a Salpeter IMF, in agreement with the early study by \citet{Ferreras2008}.

An additional limitation all previous analyses of the full SLACS sample was the assumption that all galaxies in their sample could be described by homologous spherical Hernquist distributions. These models ignore known systematic variations of galaxy morphology with mass \citep[e.g.][see also \reffig{fig:virial_plane_projections_sig8_sige}]{Caon1993,Kormendy2009}. Moreover real early-type galaxies, even in the mass range of the SLACS survey, are dominated by fast rotating disks (Paper~II, Paper~III), some of which are clearly visible in the SLACS photometry. Disk galaxies are not well described by spherical single-component models. It is unclear whether galaxy models that do not reproduce neither the kinematics nor the photometry of the real galaxies under study can be trusted at the $\sim$10\% level that is required for IMF studies. Possible systematic biases in the results are however difficult to estimate. A reanalysis of the excellent SLACS dataset using more realistic and flexible models would seem the best way to clarify the situation and provide a consistent picture. This approach has already been demonstrated by \citet{Barnabe2012} for a single galaxy using the JAM method \citep{Cappellari2008} in combination with the lensing analysis. One should finally consider that lensing studies measure the mass wihin cylinders along the line-of-sight. The recovered stellar mass depends on both the assumed dark matter profile in the centre, as well as at large radii \citep{Dutton2011swells}.

An simple assumption on the dark matter content of early-type galaxies was made by \citet{Dutton2012}. They selected the most dense galaxies from a large sample of SDSS galaxies and constructed spherical isotropic models to reproduce their stellar velocity dispersion. They assumed the total dynamical $M/L$ of this set of galaxies is the same as the stellar one. Comparing the inferred mass to the one from stellar population they concluded the galaxies require on average a Salpeter IMF normalization.  This study still depends on the spherical approximation. Moreover it is unclear to what accuracy the zero dark matter assumption is verified even for the densest galaxies. Nonetheless this study confirms that, unlike spiral galaxies, the densest early-type galaxies are not inconsistent with a Salpeter IMF.

In summary, the two result on which nearly all previously discussed studies agree are: (i) the {\em total} $M/L$ in the central regions of galaxies ($r\la\re$) does not follow the $M/L$ inferred assuming a universal IMF; (ii) less massive galaxies, especially spiral ones, {\em require} an IMF normalization lighter thank Salpeter and consistent with Kroupa-like, while more massive and dense ones allow for a Salpeter IMF. Indications were found for the Salpeter IMF to be actually {\em required} for massive ellipticals \citep{Grillo2009,Treu2010,Auger2010imf,Dutton2012}. However these conclusions depend on assuming the knowledge of either the dark matter fraction or the slope, or the mass of the dark halo. An exception is the recent work by \citet{Sonnenfeld2012}, which modelled a rare elliptical with two concentric Einstein rings, and concluded for a Salpeter IMF. More similar objects would be needed to draw solid conclusions. Not all studies agreed on the requirement for a Salpeter IMF, even when analysing the same SLACS data \citep{Ferreras2008,Tortora2010,Barnabe2011,Deason2012dm}. It seems that this situation could be resolved by the use of more realistic models.

\subsubsection{Systematic IMF variation in galaxies}

A breakthrough in IMF studies was provided by the work by \citet{vanDokkum2010}, further strengthened in \citet{vanDokkum2011}. Contrary to the previously described set of dynamical and lensing works, they looked for evidences of IMF variation in subtle IMF-sensitive spectral features of the near-infrared region of galaxy spectra, in particular the Wing-Ford band. In this way their study did not suffer from the dependency on the halo assumptions. Although the spectral technique has been around for decades \citep[e.g.][]{Spinrad1971}, only the availability of reliable stellar population models has made the approach sufficiently accurate for IMF studies \citep{Schiavon2000,Cenarro2003}. \citet{vanDokkum2010} analysed the IMF of eight massive ellipticals, from stacked spectra. They used new population models that allow for a variation of the detailed abundance patterns of the stars \citep{Conroy2012models} to distinguish abundance from IMF variations. They concluded that the observed spectra required a bottom-heavy, dwarf-rich IMF. Combining their finding with the previous results on the IMF of spiral galaxies they tentatively concluded that ``Taken at face value, our results imply that the form of the IMF is not universal but depends on the prevailing physical conditions: Kroupa-like in quiet, star-forming disks and dwarf-rich in the progenitors of massive elliptical galaxies''.

Our relatively large and well-selected \atl\ sample and high-quality integral-field stellar kinematics appeared well suited to resolve the halo degeneracies of previous dynamical and lensing analyses and test the claims from the spectral analysis. Contrary to previous large studies, we adopted an axisymmetric modelling method  which  describes in detail both the individual galaxy images using the MGE technique \citep{Emsellem1994,Cappellari2002mge} and the richness of our two-dimensional kinematics \citep{Cappellari2008}, thus avoiding the possible biases of previous more approximate approaches. For the first time, thanks to the tighter constraints to the models provided by the two-dimensional data, our study could leave {\em both} the halo slope and its mass as free parameters. The halo slope is allowed to vary in a range which includes both the flat inner halos predicted by halo expansion models \citep[e.g.][]{Governato2010}, and the steep ones predicted by adiabatic halo contraction \citep[e.g.][]{Gnedin2004}. Our models also explicitly include the galaxy inclination and  anisotropy as free parameters, although the latter is still assumed to be constant in the region where we have data. 
The parameters are estimated in a Bayesian framework with a maximum ignorance (flat) prior on the parameters. 

We showed that even in this relatively general case the models required a variation of the IMF to reproduce the data \citep{Cappellari2012}, unless there are major flaws affecting all available stellar population models. We additionally tested a variety of sensible modelling assumptions on the halo, some of which had already been employed by previous studies. However this time our models accurately described the photometry and kinematics of the real galaxies. We confirmed that a non-universal IMF is required under all those halo assumption. The similarity is due to the fact that in all cases dark matter contributes just about 10-20\% to the total mass within 1\re, so that it cannot explain an observed $M/L$ excess of up to a factor 2-3.

Our \atl\ sample includes a larger range of masses than previous lensing studies. Not only we could quantify the IMF normalization of early-type galaxies as a class, but we showed that the IMF normalization varies systematically within the early-type galaxy population, as a function of the stellar $(M/L)_{\rm stars}$. The IMF was found consistent with Kroupa/Chabrier-like at low $(M/L)_{\rm stars}$ and heavier that Salpeter at large $(M/L)_{\rm stars}$. This finding bridges the gap between the Kroupa-like IMF determinations in spirals and the evidence for a Salpeter-like or heavier in early-type galaxies. This should be expected given the parallelism and continuity in physical parameters between early-type and spiral galaxies as emphasized in our `comb' morphological diagram (Paper~VII). Early-type galaxies at low $(M/L)_{\rm stars}$ in fact completely overlap with the region populated by spiral galaxies on the mass-size projection of the MP (\reffig{fig:mass_size_spirals_etgs}). Our result also reconciled the apparent disagreement between studies claiming that early-type galaxies include cases with Kroupa/Chabrier normalization and those claiming that Salpeter is required.

Although, unsurprisingly, the cleanest IMF trend in \citet{Cappellari2012} was obtained by simply comparing the dynamical and population $M/L$, a trend with velocity dispersion $\sigma_{\rm e}$ was also presented. This is a natural consequence of existence of the $(M/L)-\sigma$ correlation \citep{Cappellari2006,vanderMarel2007}. In particular \citet{Cappellari2012} (their fig.~2) showed an IMF trend, with significant scatter, going from a Kroupa/Chabrier to a Salpeter IMF within the range $\log_{10}(\sigma_{\rm e}/\kms)\approx1.9-2.5$, with a gradual variation in between.

A flurry of papers have appeared in subsequent months from independent groups, all in agreement on the existence of a systematic IMF variation in galaxies. \citet{Spiniello2012} used the models of \citet{Conroy2012models} in combination with SDSS spectra and inferred a variation of the IMF from sodium and titanium-oxide absorptions, which correlate with velocity dispersion (also illustrated by \citealt{Zhu2010}). A similar result was inferred by \citet{Ferreras2012}, also from SDSS spectra, but using the population models by \citet{Vazdekis2012}. \citet{Smith2012} studied the Wing-Ford absorption for galaxies in the Coma Cluster. They find that their galaxy spectra are best reproduced by a Salpeter IMF and detect a weak IMF trend with metallicity, but do not find an IMF trend with velocity dispersion.  \citet{Conroy2012} detected a systematic trend of the IMF, versus either $\sigma$ or metallicity, using both optical and near-infrared spectral features. A key difference from other spectral studies is that they were able to rule out, in a Bayesian framework, a variation of  element abundances as the origin of the empirical trends.

\citet{Dutton2012imf_fp} showed that it is sufficient to assume the knowledge of the halo mass and fit galaxy scaling relations via approximate spherical galaxy models, to infer a systematic variation of the IMF. Similarly \citet{Tortora2012imf} tested a set of different assumption on the dark halo, using approximate spherical \citet{Sersic1968} galaxy models to fit the SDSS velocity dispersion of a large galaxy sample. Some of the halo assumptions were also included in \citet{Cappellari2012} with more detailed dynamical models, producing consistent results. Although these results still depend on simple approximations and on assumption on the halo, they illustrate that many different halo assumptions, motivated by theoretical models, all lead to the same conclusion of a systematic IMF variation. 

In summary, after years of debate on whether the IMF or the dark matter were responsible for the observed disagreement between stellar population and dynamical or lensing indicators, a consensus seems to have emerged, from dynamical, lensing and spectral arguments, on a systematic IMF variation with the IMF becoming heavier with $M/L$, mass or velocity dispersion. It is time to investigate IMF trends against other observables, which we do in the following section.

\subsection{IMF variation on the Mass Plane}

In \citet{Cappellari2012} we presented a systematic trend between the IMF and the stellar $(M/L)_{\rm stars}$ derived using axisymmetric JAM models including a dark halo. The IMF was parametrised by the IMF mismatch parameter $\alpha\equiv(M/L)_{\rm stars}/(M/L)_{\rm Salp}$ (in the notation of \citealt{Treu2010}), where $(M/L)_{\rm Salp}$ was derived from full spectral fitting as summarized in \refsec{sec:global_parameters}. Different population approaches were also tested there, producing less clean trends but consistent results. The $(M/L)_{\rm Salp}$ normalization depends on the adopted lower and upper mass cut-offs for the IMF in the population models. The models we use \citep{Vazdekis2012} adopt standard lower and upper mass cut-offs for the IMF of 0.1 and 100 \msun, respectively. $(M/L)_{\rm Salp}$ further depends on whether the gas lost by the stars during the early stages of their evolution is retained in the central regions we observe or is recycled into stars or expelled at large radii. Evidence suggests the gas is not retained in significant quantities within 1\re, neither in ionized or hot X-ray emission form \citep{Sarzi2010}, nor as a cold component (Paper~IV). If all the gas was retained it would increase $(M/L)_{\rm Salp}$ by up to 30--40\% \citep{Maraston2005}, making population $M/L$ overpredicting total dynamical $M/L$ even for a light Kroupa/Chabrier IMF, for a number of galaxies.

\begin{figure}
\plotone{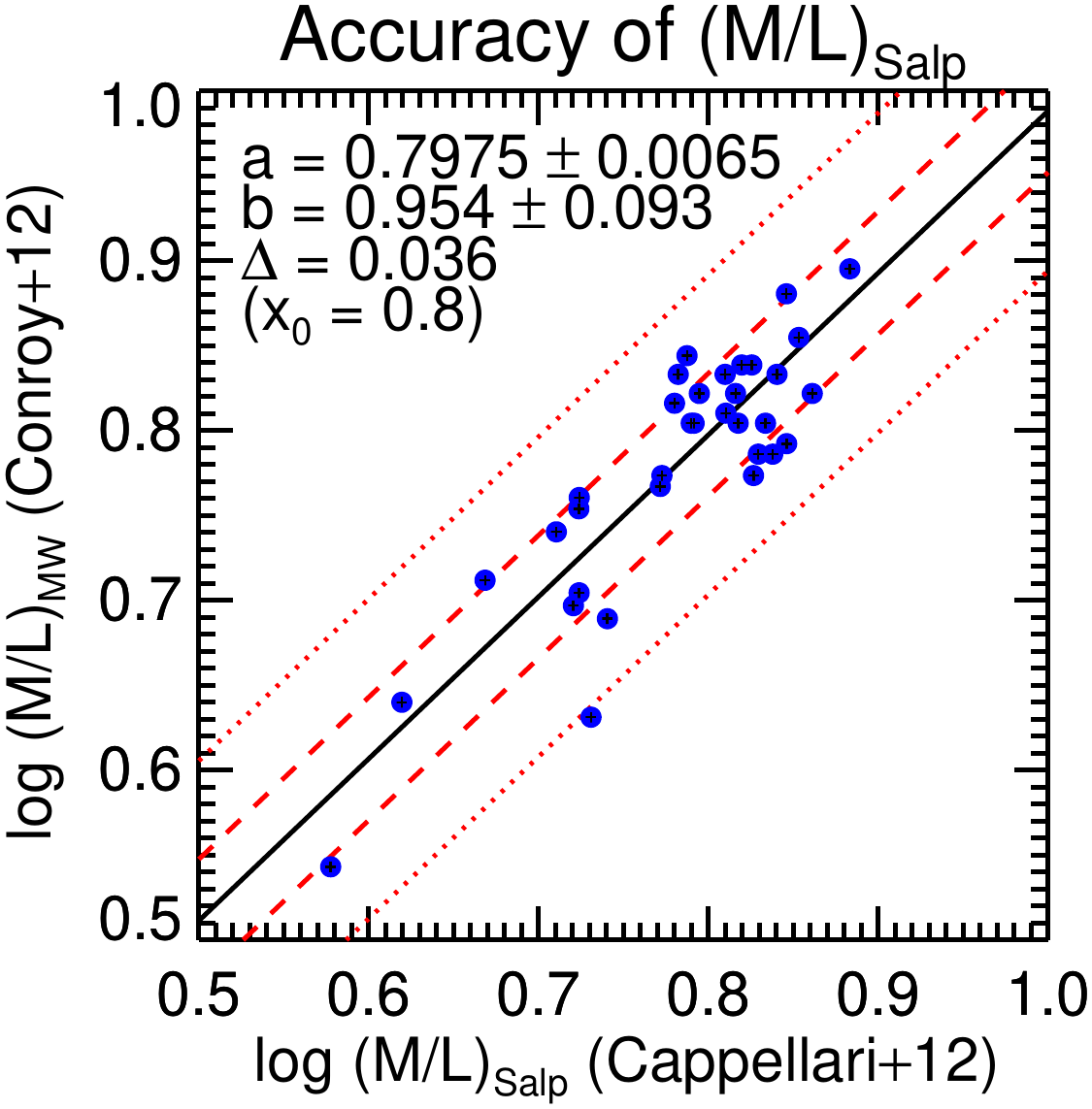}
\caption{Testing the accuracy of $(M/L_r)_{\rm Salp}$ against the completely independent determination of $(M/L_K)_{\rm MW}$ from \citet{Conroy2012}, which use a different population modelling software and different spectra with a longer wavelength range than the \sauron\ ones. Assuming all measurements have comparable errors, the observed scatter between the 35 values is consistent with a $1\sigma$ error $\Delta/\sqrt{2}$ of 6\% in $(M/L_r)_{\rm Salp}$. The $(M/L_K)_{\rm MW}$ values were multiplied by a constant factor to approximately account for the differences in the photometric band and IMF.}
\label{fig:ml_salp_accuracy}
\end{figure}

A careful determination of the population $M/L$ for a fixed IMF was recently provided for 35 galaxies in our sample by \citet{Conroy2012}. They employed different models \citep{Conroy2012models} from the ones we adopted \citep{Vazdekis2012} and used them to fit an independent set of spectra, spanning a larger wavelength range than the \sauron\ ones. In \reffig{fig:ml_salp_accuracy} we compare the two determinations of $M/L$ (allowing for an arbitrary offset in the absolute normalization, which can differ by up to 10\% between different authors). Assuming all measurements have comparable errors, the observed scatter between the 35 values is consistent with an error of just 6\% in $(M/L)_{\rm Salp}$. This error is consistent with the error of 7\% predicted by \citet{Gallazzi2009} when optimal spectra information is available. This gives confidence that our population $(M/L)_{\rm Salp}$ are robust and any trend we observe is not due to the details of our population modelling approach.

\citet{Cappellari2012} showed that consistent IMF trends are found for a variety of different assumptions on the dark halo, as well as for the most general one (model D) which leaves both the halo slope and mass as free parameters, with only an upper limit on the halo slope, derived from model predictions of halo contraction \citep{Gnedin2011} (see Paper~XV for a description of the models). The similarity in $(M/L)_{\rm stars}$ for the different models is due to the fact that in all cases the data allow for a small fraction of dark matter within the region where the kinematics are available ($\sim$1\re), with the most general model implying a median dark matter fractions of just 10\%. Given the similarity of the different approaches, here we adopt as reference the $(M/L)_{\rm stars}$ values obtained with a NFW halo, with halo mass as free parameter (model B). This choice also makes it easy to compare our results with those of others authors that make the same assumption.

\begin{figure}
\plotone{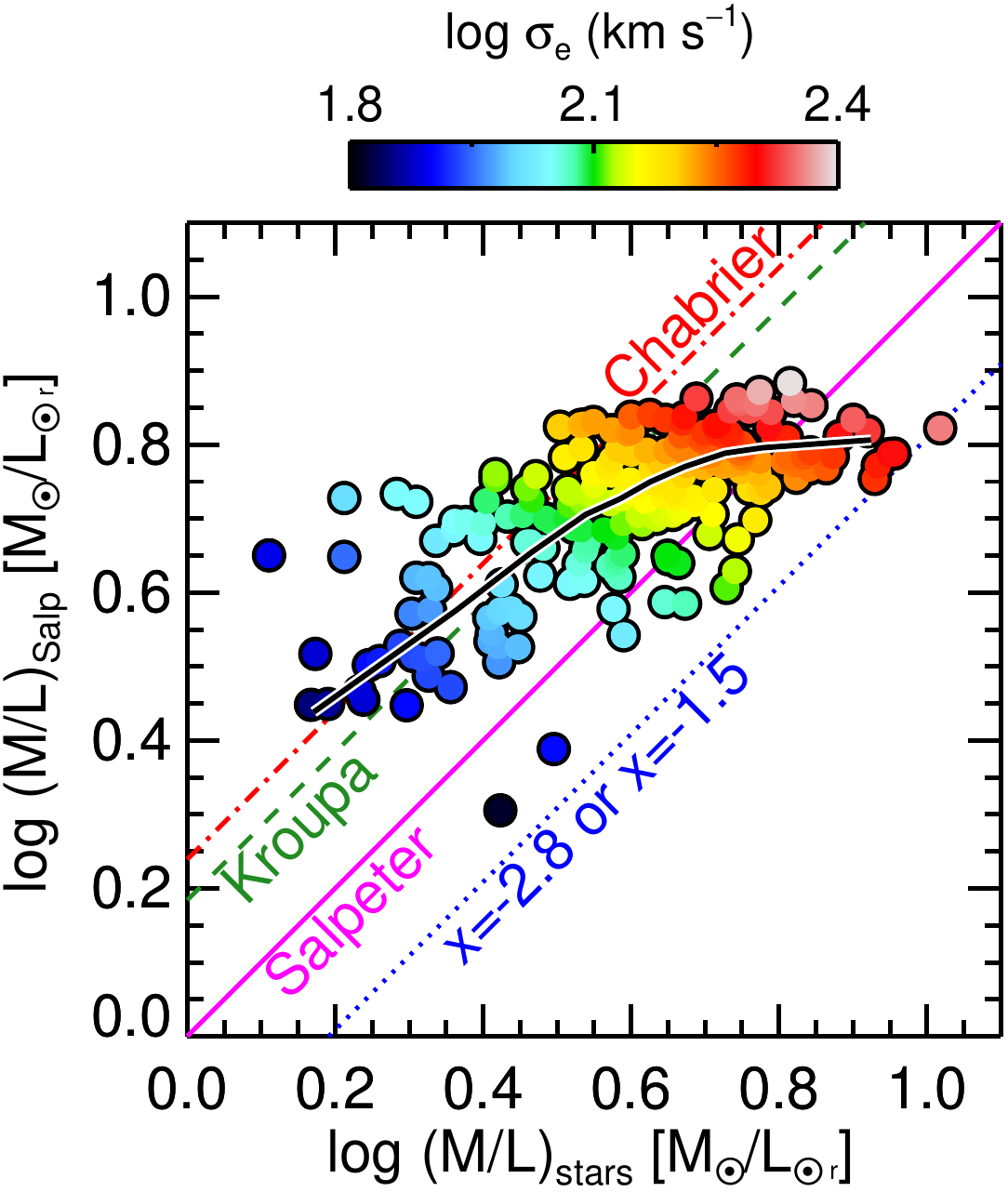}
\caption{The systematic variation of the IMF. The $(M/L_r)_{\rm stars}$ of the stellar component, determined via dynamical models, is compared to the $(M/L_r)_{\rm Salp}$ determined from spectral fitting using stellar population models and assuming for reference a fixed Salpeter IMF. The colours of the symbols show the galaxies $\sigma_{\rm e}$, which was LOESS smoothed in two-dimension to emphasize the trends. The plot only includes the subsets of \atl\ galaxies with H$\beta < 2.3$ \AA, for which $(M/L)_{\rm Salp}$ is sufficiently reliable. Diagonal lines illustrate the expected trends if the IMF was universal with either the Chabrier (red dash-dotted line), Kroupa (green dashed line) or Salpeter ($\zeta(m)\propto m^{-2.3}$, solid magenta line) forms. Also shown is the expected trend for two IMFs heavier than Salpeter (blue dotted line): a top-heavy one, dominated by stellar remnants ($\zeta(m)\propto m^{-1.5}$), and a bottom-heavy one, dominated by dwarfs ($\zeta(m)\propto m^{-2.8}$). The black solid curve is a LOESS smoothed version of the data. A clear systematic trend is evident, with the IMF being closer to Kroupa/Chabrier at the lowest $M/L$, which also have the lowest $\sigma_{\rm e}$, and closer to Salpeter or heavier at the largest $M/L$ or largest $\sigma_{\rm e}$. A different rendition of this plot was presented in fig.~2b of \citet{Cappellari2012}.}
\label{fig:ml_stars_versus_ml_pop}
\end{figure}

In \reffig{fig:ml_stars_versus_ml_pop} we show a different rendition of the similar fig.~2b of \citet{Cappellari2012}. Here we plot $(M/L)_{\rm Salp}$ versus $(M/L)_{\rm stars}$. We still exclude galaxies with very young stellar populations, selected as having stellar absorption line strength index H$\beta>2.3$ \AA. We found that those galaxies have strong gradient in the stellar population and this breaks our approximate assumption of a constant $(M/L)_{\rm stars}$ within the region where we have kinematics, making both $(M/L)_{\rm stars}$ and $(M/L)_{\rm Salp}$ inaccurate and ill-defined.
Different diagonal lines illustrate the expected trends if the IMF was universal or either the Chabrier, Kroupa or Salpeter ($\zeta(m)\propto m^{-2.3}$) forms. Also shown is the expected trend for two IMFs heavier than Salpeter: a top-heavy one, dominated by stellar remnants ($\zeta(m)\propto m^{-1.5}$), and a bottom-heavy one, dominated by dwarfs ($\zeta(m)\propto m^{-2.8}$). This figure clearly illustrates the fact that the two indicators of $M/L$ do not follow a one-to-one relation, but deviate systematically, with the IMF being consistent on average with Kroupa/Chabrier at the lowest $(M/L)_{\rm stars}$ and with Salpeter or heavier at the largest $(M/L)_{\rm stars}$, as already concluded in \citet{Cappellari2012}. Another way to interpret this plot is by noting that galaxies with the largest $(M/L)_{\rm Salp}$, which are characterized by the oldest populations, have a Salpeter or heavier IMF, while those with the lowest $(M/L)_{\rm Salp}$, which have younger populations, have on average a Kroupa/Chabrier IMF. Detailed trends of the IMF with other population indicators will be presented in McDermid et al. (in preparation).

As in the \citet{Cappellari2012} we also show with colours the galaxies velocity dispersion. However, to emphasize the trend, here we apply the same two-dimensional LOESS smoothing approach \citep{cleveland1988locally} introduced in \refsec{sec:vp_projections}, instead of showing the individual $\sigma_{\rm e}$ values. One can still see the trend for the IMF to vary between Kroupa/Chabrier to Salpeter, albeit with large scatter, within the interval $\log_{10}(\sigma_{\rm e}/\kms)\approx1.9-2.5$, with a smooth variation in between. This trend will be more precisely quantified in \refsec{sec:imf_sigma}.

\begin{figure*}
\includegraphics[width=0.7\textwidth]{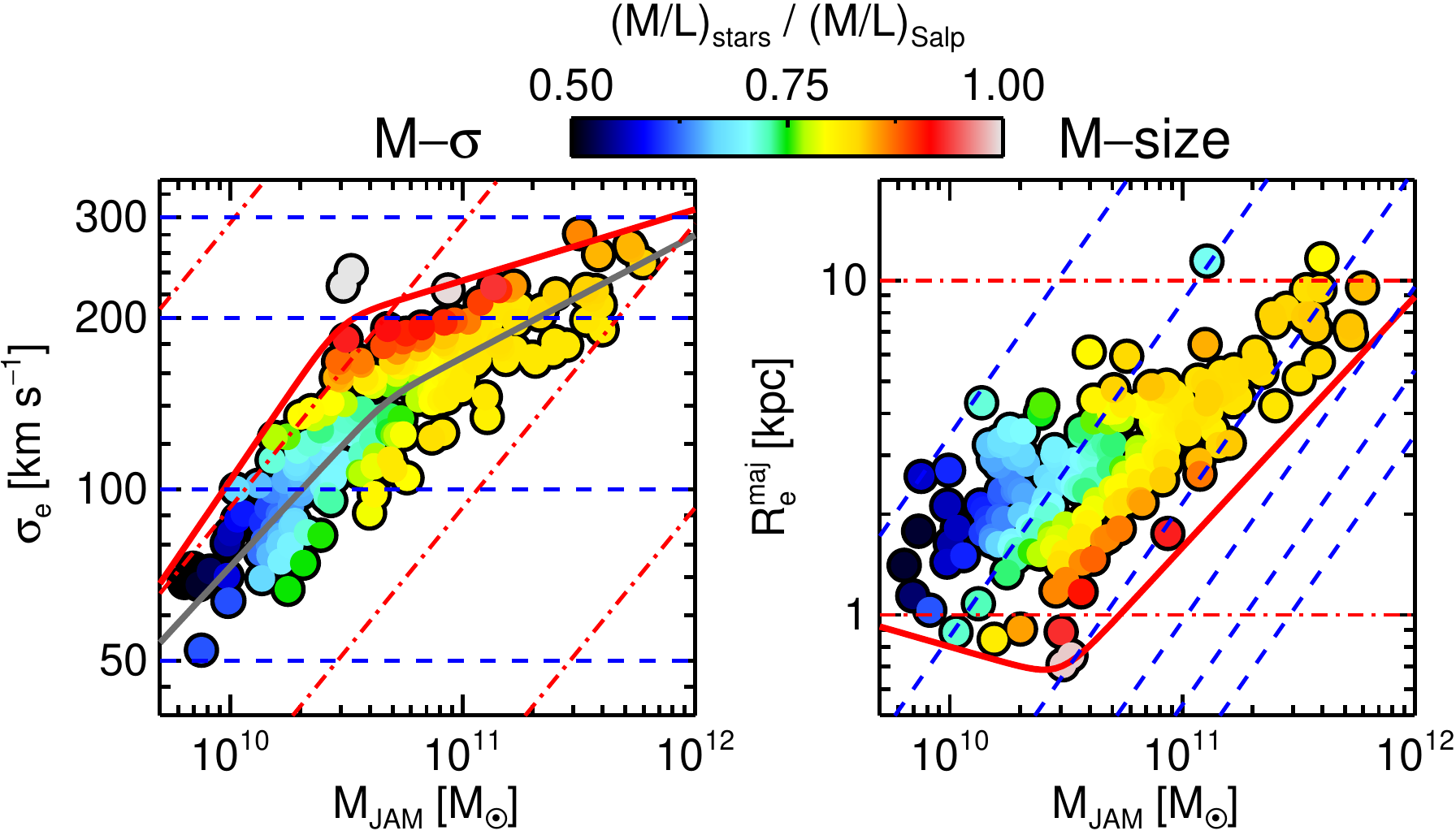}
\caption{Variation of the IMF on the MP projections. Same as in \reffig{fig:virial_plane_projections_ml} for the variation of the IMF mismatch parameter $(M/L)_{\rm stars}/(M/L)_{\rm Salp}$, which measure the ratio between the stellar $M/L$ from dynamical models and the one from population models, with a fixed Salpeter IMF for reference. Like in \reffig{fig:ml_stars_versus_ml_pop} only galaxies with H$\beta < 2.3$ \AA\ are included. The trend follows lines of roughly constant $\sigma_{\rm e}$, or even better, follows lines parallel to the ZOE (thick red line) above its break. Like in other diagrams the trend looks different at large masses $\mjam\ga10^{11}$ \msun, suggesting these objects are different. The heaviest IMF (red colour) is not reached by the most massive galaxies, but by the fast rotators with the biggest bulges.}
\label{fig:vp_projections_imf}
\end{figure*}
 
In \reffig{fig:vp_projections_imf} we show the variation of the IMF mismatch parameter $(M/L)_{\rm stars}/(M/L)_{\rm Salp}$ on the $(\mjam,\se)$ and $(\mjam,\re)$ projections of the MP. The trends in IMF are necessarily noisier that the ones in $(M/L)_{\rm stars}$, as here the errors in both $M/L$ combine. However the structure in this figure closely resembles the one in the previous \reffig{fig:virial_plane_projections_ml}, \ref{fig:virial_plane_projections_hbeta} and the top panel of  \reffig{fig:virial_plane_projections_sig8_sige}. The systematic variation in the IMF follows on average the variations in the total $(M/L)_{\rm JAM}$ and its corresponding stellar population indicators H$\beta$ and galaxy colour, as well as the molecular gas fraction. Like the other quantities, also the IMF appears to roughly follow lines of constant $\sigma_{\rm e}$ on the MP, which we showed is tracing the bulge mass at given galaxy size and mass. In both projections there is indication for some extra substructure, with IMF variation at constant $\sigma_{\rm e}$. Moreover the proximity to the ZOE (thick red line) seems to be an even better indicator of IMF than $\sigma_{\rm e}$ (blue dashed lines). The $(M,\re)$ projection also makes clear why one should expect spirals to have on average light IMF: spirals populate the empty zone above our galaxies (\reffig{fig:mass_size_spirals_etgs} here and fig.~4 in Paper~I) and overlap with the distribution of ETGs with the Kroupa/Chabrier IMF. However our plot suggests that, at fixed galaxy stellar (or $M_{\rm JAM}$) mass and below $\mjam\la10^{11}$ \msun, the bulge mass (as traced by $\sigma_{\rm e}$) is the main driver of the IMF variation, rather than morphological type, and one may expect IMF variations within the spiral population as well. Our finding explains why \citet{Dutton2012} inferred a Salpeter or heavier IMF when selecting the smallest (or densest) galaxies at fixed mass, and assuming no dark matter. Dark matter appears to provide a small contribution to the $M/L$ for our entire sample and not just the smallest ones (Paper~XV), but the densest galaxies are precisely the ones with the heaviest IMF. A recent claim has been made for a Salpeter IMF in the bulge of 5 massive spirals \citep{Dutton2012imf_bulges}, which would go in the suggested direction. However the IMF is currently already difficult to infer when the $M/L$ can be assumed to be spatially constant and the stellar population homogeneous. Relaxing these assumptions makes the results more uncertain unless very good population data from galaxy spectra and accurate non-parametric models for the photometry are employed.

\subsection{IMF versus $\sigma_{\rm e}$ correlation}
\label{sec:imf_sigma}

In the previous section we illustrated the systematic IMF trends already presented in \citet{Cappellari2012} and we additionally presented the variation of the IMF on the projections of the MP. We pointed out that, like the dynamical $M/L$, its population indicators, colour and H$\beta$, the cold gas fraction and the galaxy concentration which traces to the bulge mass, also the IMF broadly follows lines of nearly constant $\sigma_{\rm e}$ on the $(M_{\rm JAM},\sigma_{\rm e})$ MP projection and consequently $\rmaj\propto\mjam$ on the $(M_{\rm JAM},R_{\rm e}^{\rm maj})$ projection. Interestingly a `conspiracy' between the IMF variation and other galaxies properties is required for galaxy scaling relations to still have a small scatter \citep{Renzini1993}. 

\begin{figure}
\plotone{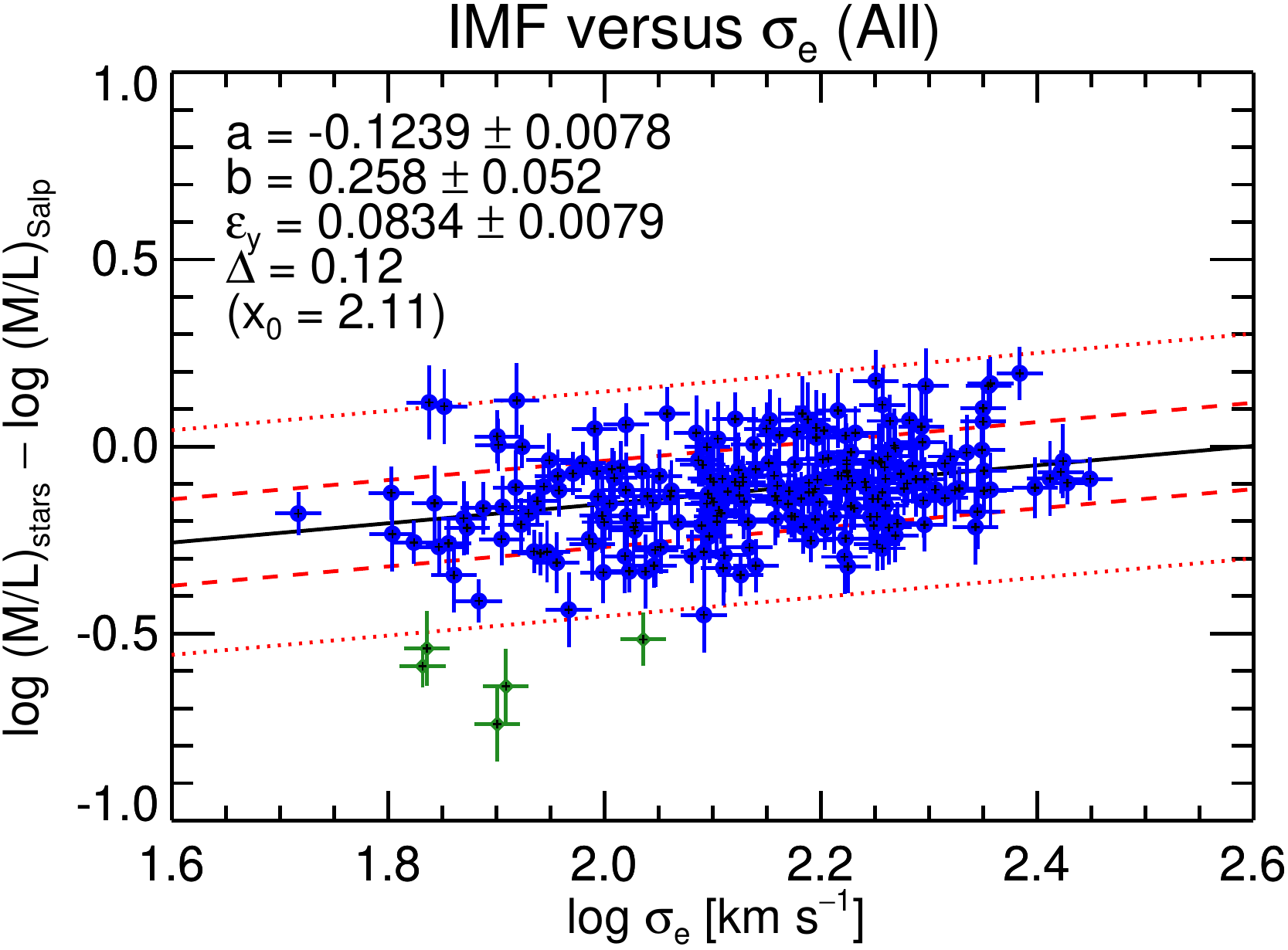}
\plotone{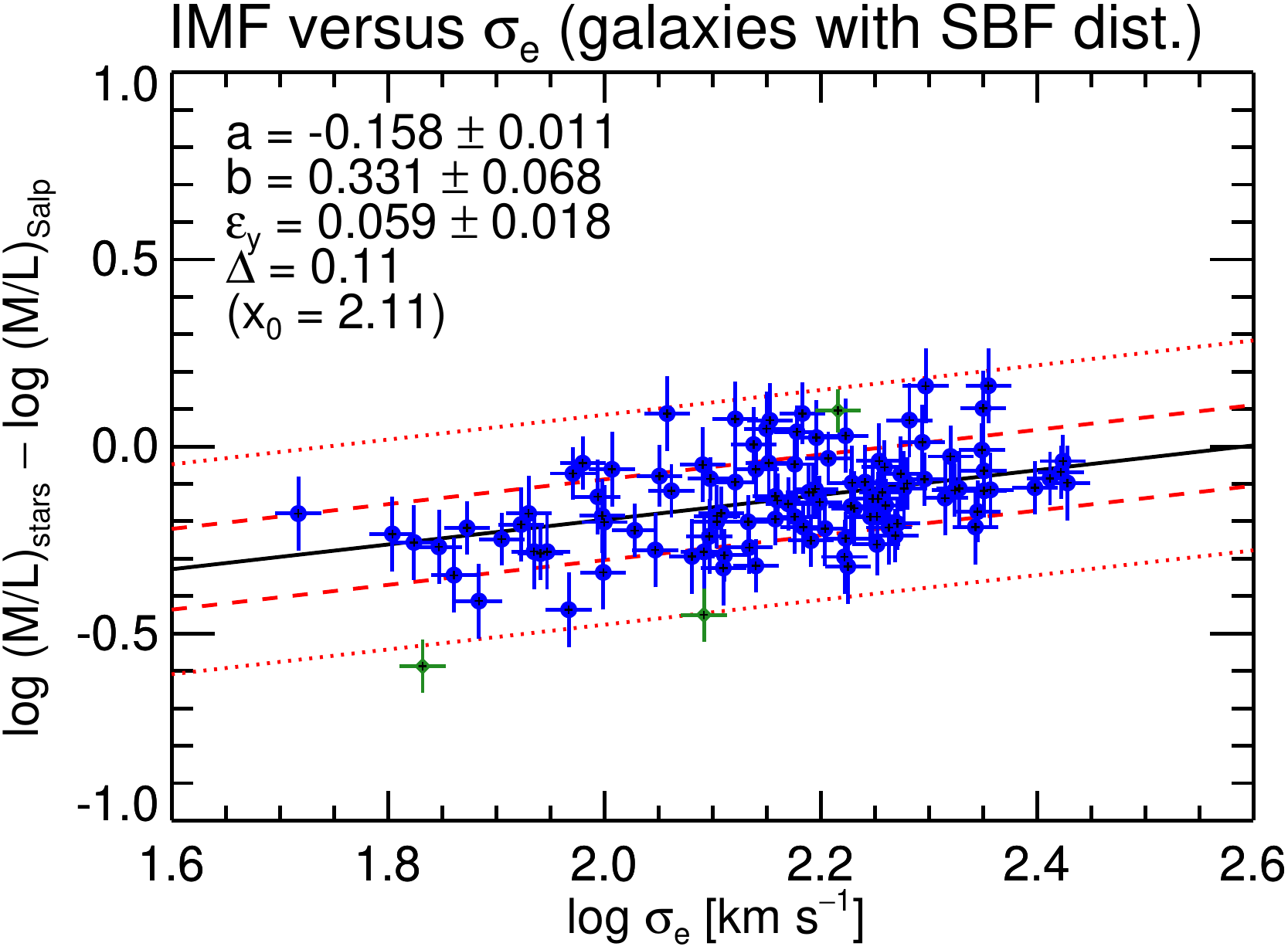}
\plotone{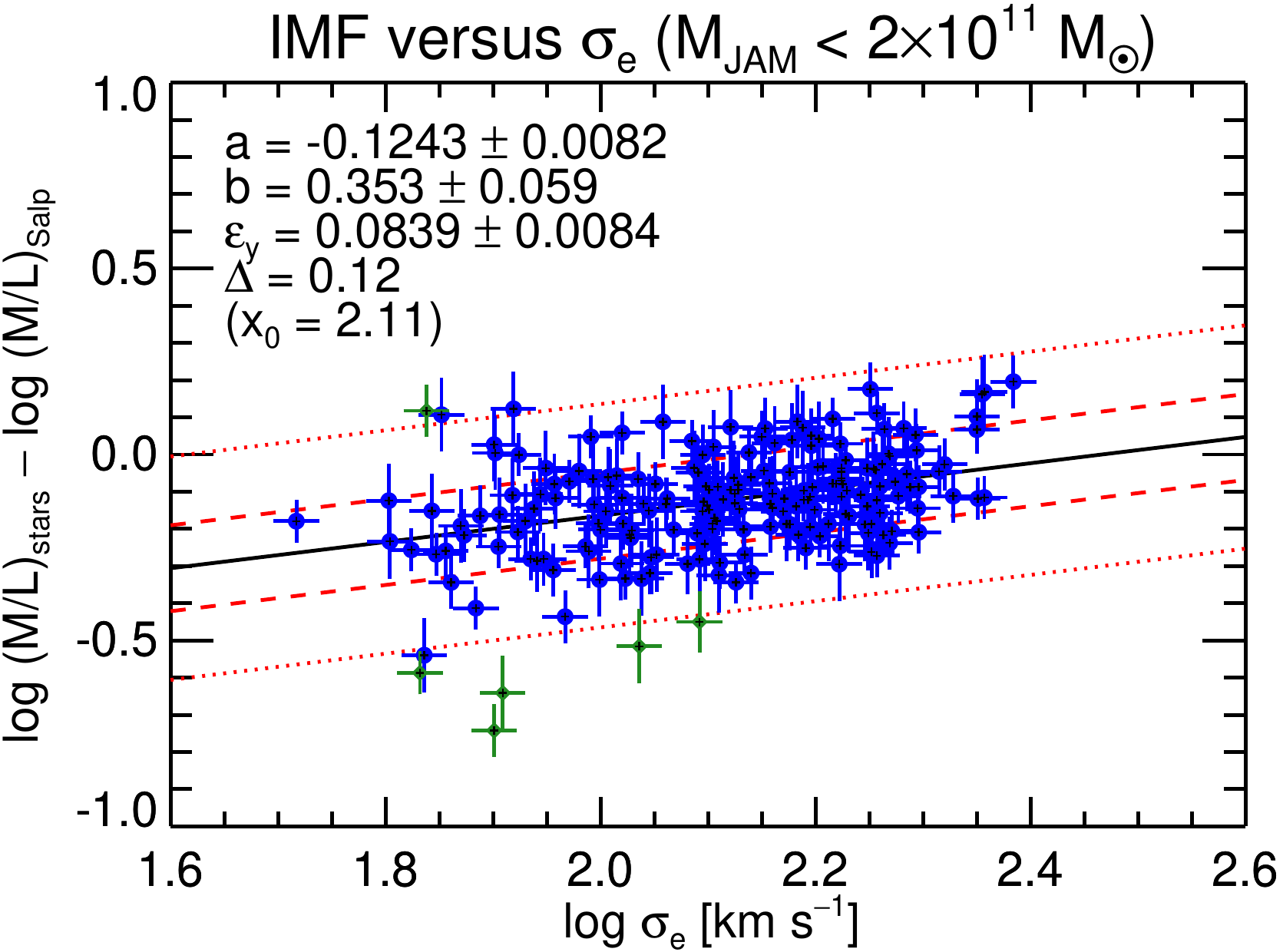}
\caption{IMF versus $\sigma_{\rm e}$ correlations. The plots show as blue filled circles with error bars the logarithm of the IMF mismatch parameter $(M/L)_{\rm stars}/(M/L)_{\rm Salp}$ versus the effective velocity dispersion $\sigma_{\rm e}$, for the subsets of \atl\ galaxies with H$\beta < 2.3$ \AA. Green symbols are outliers automatically removed from the fits. Dashed and dotted red lines indicate the 1$\sigma$ and 2.6$\sigma$ (99\%) observed scatter around the best-fitting relation (black solid line). The top panel includes all galaxies. The middle panel shows galaxies with accurate SBF distances. The bottom panel shows galaxies with $\mjam<2\times10^{11}$ \msun.}
\label{fig:imf_sigma}
\end{figure}

In \reffig{fig:imf_sigma} we present a direct correlation between the logarithm of the IMF mismatch parameter $(M/L)_{\rm stars}/(M/L)_{\rm Salp}$ and the logarithm of $\sigma_{\rm e}$, using the robust linear fitting routine \textsc{lts\_linefit} presented\footnote{Available from http://purl.org/cappellari/idl} in (Paper~XV), which explicitly allows and fits for intrinsic scatter and removes outliers. Like earlier we exclude galaxies with very young stellar populations, selected s having stellar absorption line strength index H$\beta>2.3$ \AA, leading to a sample of 223 out of 260 galaxies. In the fits we quadratically co-added JAM modelling errors of 6\% (Paper~XV) plus distance errors (Paper~I) for $(M/L)_{\rm stars}$ plus population models errors of 6\% for $(M/L)_{\rm Salp}$ (\reffig{fig:ml_salp_accuracy}) plus 10\% errors for our photometry (Paper~XXI). 
We present fits for (i) the full sample, (ii) for the subsample of galaxies with SBF distances (mostly from \citealt{Tonry2001} and \citealt{Mei2007}: see Paper~I) and (iii) for galaxies with $\mjam<2\times10^{11}$ \msun, to eliminate the galaxies above the characteristic mass where the IMF trend, as well as other galaxy properties, seems to change (see \reffig{fig:virial_plane_projections_concentration}, \ref{fig:virial_plane_projections_lambda} and \ref{fig:vp_projections_imf}). All three relations have a comparable observed scatter $\Delta\approx0.10$ dex (26\%) in $(M/L)_{\rm stars}/(M/L)_{\rm Salp}$ and imply a significant intrinsic scatter of about 20\%. When the same outliers are removed, we verified that indistinguishable values for the parameter, their errors and the infered intrinsic scatter are obtained with the Bayesian approach of \citet{Kelly2007}, as implemented in his IDL routine \textsc{linmix\_err}. Our three fits provide consistent values for the best-fitting slopes, within the errors, and nearly consistent normalizations. The fitted relation has the form
\begin{equation}
\log_{10} \frac{(M/L)_{\rm stars}}{(M/L)_{\rm Salp}}=a+b\times\log_{10}\frac{\sigma_{\rm e}}{130\kms},
\label{eq:imf_sigma}
\end{equation}
and our preferred values (bottom panel of \reffig{fig:imf_sigma}) have best fitting parameters and formal errors $a=-0.12\pm0.01$ and $b=0.35\pm0.06$ (parameters and errors for the other fits are given inside the figures). The observed trend of IMF with $\sigma$ appears to account for about half of the total trend in the $(M/L)-\sigma_{\rm e}$ relation $(M/L)_{\rm JAM}\propto\sigma_{\rm e}^{0.72}$ (see Paper~XV), the remaining one being due to stellar population variations.

Our trend implies a transition of the mean IMF from Kroupa to Salpeter in the interval $\log_{10}(\sigma_{\rm e}/\kms)\approx1.9-2.5$ (or $\sigma_{\rm e}\approx90-290$ \kms), with a smooth variation in between, consistently with what can be seen in \citet{Cappellari2012} and in \reffig{fig:ml_stars_versus_ml_pop}. The fact that this trend is slightly weaker than the one implied by \reffig{fig:ml_stars_versus_ml_pop} seems to confirms the intrinsic differences in the IMF of individual galaxies. However, part of this difference could also be explained if the distance errors were underestimated. One way to address this issue would be to repeat our analysis for galaxies in a clusters at intermediate distance like the Coma cluster, for which relative distance errors can be neglected.

Our slope is a factor $\approx3.5$ smaller than the ``tentative'' trend reported in \citet{Treu2010} for the same quantities, and making the same assumption for the dark halo. A reason for this large difference must be due to the fact that their sample only included galaxies with $\sigma\ga200$ \kms\ ($\log_{10}(\sigma/\kms)\ga2.3$). Our sample is too small to reliably study trend in IMF for the galaxies above that $\sigma_{\rm e}$, but their reported trend would exceed the slope of the $(M/L)-\sigma$ relation (Paper~XV), making it difficult to explain. Their steep inferred trend may then be due to their use of spherical and homologous models for all the galaxies, which may introduce systematic trends. Moreover they used single stellar population $M/L$ based on colours is expected to be less reliable than our determinations based on spectra \citep{Gallazzi2009}. A reanalysis of the unique SLACS dataset seems required to clarify this issue. Our trend with $\sigma_{\rm e}$ is also smaller than the one reported by \citet{Ferreras2012} from spectral analysis, who find a rapid change from a Kroupa to a Salpeter IMF in the narrow interval $\sigma\approx150-200$ \kms. Our relation is not inconsistent with the values presented in \citet{Conroy2012} also from spectral analysis, or with the result reported by \citet{Spiniello2012}, with the IMF becoming steeper than Salpeter above $\sigma\ga200$ \kms. We are also broadly consistent with the IMF variation implied by the non-contracted halo spherical dynamical models of \citet{Tortora2012imf}. Overall there is a qualitative agreement between different approaches and the still significant systematic differences in the various methods could account for the differences.

\section{Discussion}

\subsection{Previous relations as seen on the Mass Plane}

We have shown in Paper~XV that the galaxy Fundamental Plane \citep{Dressler1987,Djorgovski1987} can be accurately explained by virial equilibrium combined with a smooth variation of galaxy properties, mainly the total mass-to-light ratio $M/L$, with velocity dispersion, with galaxies lying on a tight MP $(M_{\rm JAM},\sigma_{\rm e},R_{\rm e}^{\rm maj})$, for a large volume-limited sample of ETGs (Paper~I). Once this has been established, the interesting information on galaxy formation is then all contained in the distribution and the physical properties of galaxies on this plane, which we presented in this paper. 

We find that on the MP: (i) galaxies sizes are delimited by a lower-boundary, which has a minimum at a characteristic mass $M_{\rm JAM}\approx3\times10^{10} \msun$; (ii) A number of key galaxy properties: dynamical $M/L$, and its population indicators H$\beta$ and colour, as well as the molecular gas fraction, which are mainly related to age, the normalization of the IMF and the prominence of the bulge, all tend to be constant along lines of constant $\sigma_{\rm e}$, on the MP; (iii) Another characteristic mass for galaxy properties is the value $M_{\rm JAM}\approx2\times10^{11} \msun$ which separates a region dominated by the round or weakly triaxial slow rotators at large masses from one dominated dominated by fast-rotators ETGs, flattened in their outer parts and with embedded exponential disks (Paper~XVII), whose characteristics merge smoothly with the ones of spiral galaxies. A transition in the bulge fraction at this galaxy mass appears required in our models for the formation of fast and slow rotators \citep[hereafter Paper VIII]{Khochfar2011}. 

Although our \atl sample is limited to a minimum mass $M_{\rm JAM}\ga 6\times10^9$ \msun, our picture naturally extends to lower masses. As shown in \reffig{fig:mass_size_spirals_etgs} our trends for fast rotator ETGs continues with the dwarf spheroidal (Sph) sequence at lower masses (see also fig.~7 of \citealt{Binggeli1984}; fig.~38 of \citealt{Kormendy2009}; fig.~12 of \citealt{Chen2010}; fig.~4 of \citealt{Misgeld2011}; fig.~20 of \citealt{Kormendy2012}) , while the spirals sequence continues with a sequence of low-mass late spirals or irregulars (Sc--Irr), as independently noted also by \citet{Kormendy2012}. 
Interestingly the approximate stellar mass $M_\star\approx2\times10^9$ \msun\ where there is a break in the M-size relation of dwarf galaxies and where galaxies with bulge starts to appear (\reffig{fig:mass_size_spirals_etgs}) corresponds to the threshold for quenching of field galaxies recently discovered by \citet{Geha2012}. Below that mass only the cluster or group environment can strip spirals of their gas. Perhaps below that mass bulges cannot grow and star formation cannot be quenched by internal processes.

\begin{figure*}
\includegraphics[width=0.8\textwidth]{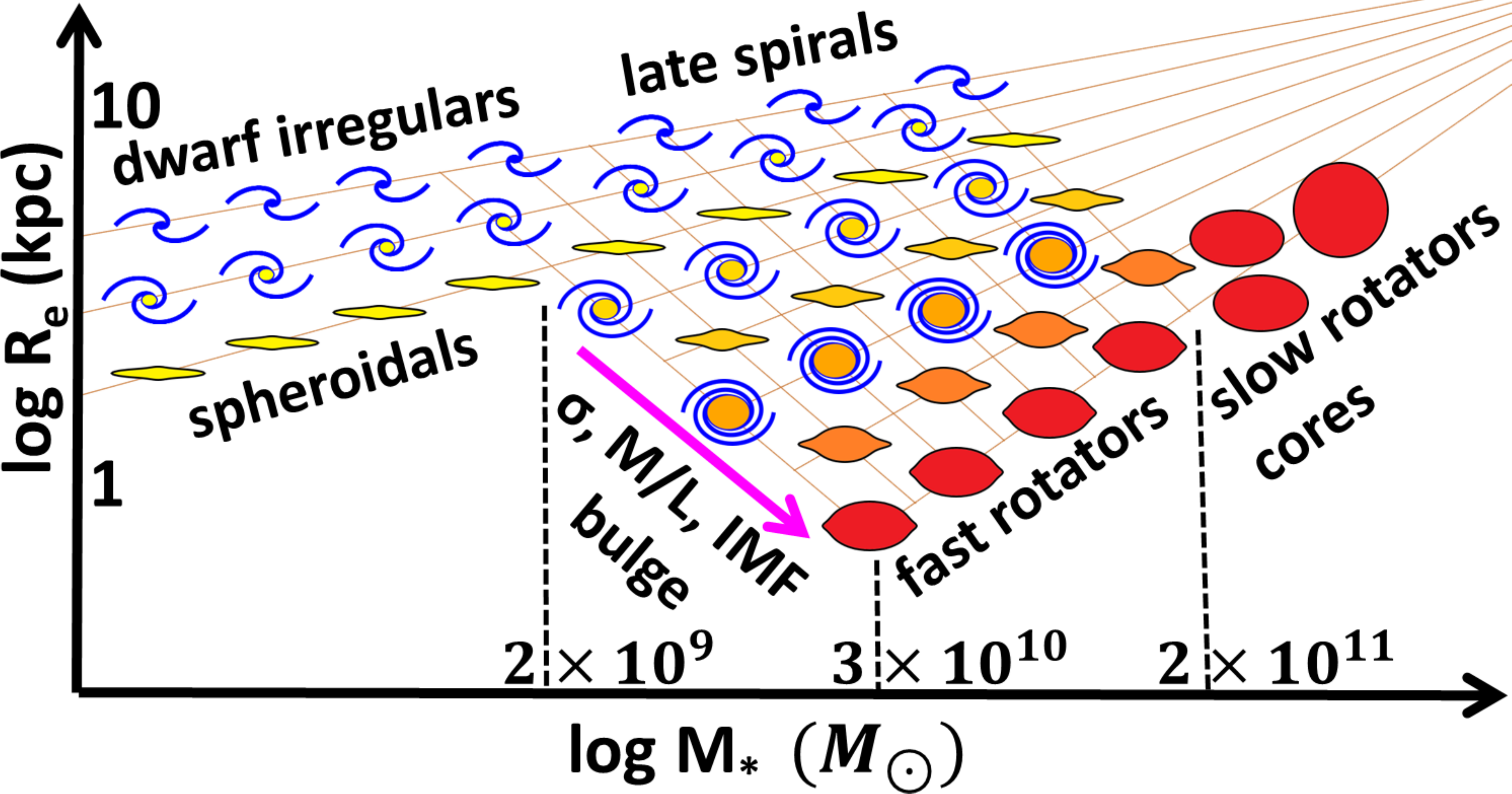}
\caption{Schematic summary of the results presented in \refsec{sec:vp_projections} and \refsec{sec:imf}. ETGs properties, dynamical $M/L$ (\reffig{fig:virial_plane_projections_ml}) or its population proxies, H$\beta$ and galaxy colour, as well as the molecular gas fraction (\reffig{fig:virial_plane_projections_hbeta}), kinematical concentration (\reffig{fig:virial_plane_projections_sig8_sige}), which traces bulge mass, and IMF mass normalization (\reffig{fig:vp_projections_imf}), all tend to vary along lines with roughly $\re\propto M$ (or even better $\re\propto M^{0.75}$), where \se\ is nearly constant. This sequence of ETGs properties merges smoothly with the one of spiral galaxies, with little overlap between late spirals (Sc-Irr) and ETGs, a significant overlap between early spirals (Sa-Sb) and fast-rotator ETGs with low $M/L$ and no overlaps between spirals and fast-rotators with high $M/L$. Three characteristic masses are emphasized in this diagram: (i) below $M_\star\approx2\times10^9$ \msun\ there are no regular ETGs and the mass-size lower boundary is increasing; (ii) $M_\star\approx3\times10^{10}$ \msun\ is the mass at which ETGs reach their minimum size (or maximum stellar density), before a sudden change in slope $\re\propto M^{0.75}$ at larger masses (see also fig.~4 in Paper~I); (iii) Below $M_\star\approx2\times10^{11}$ \msun\ ETGs are dominated by flat fast rotators, showing evidence for disks (Paper~XVII), while slow rotators are rare. Above this mass there are no spirals and the population is dominated by quite round or weakly triaxial slow rotators (paper~III) with flat (core/deficit) central surface brightness profiles (Paper~XXIV). These smooth trends in scaling relations motivated our proposed parallelism between spirals and ETGs. To emphasize this connection we uses the same morphology symbols here as in our `comb' diagram in fig.~2 of Paper~VII.}
\label{fig:etgs_spirals_diagram}
\end{figure*}

The parallelism between fast rotator ETGs and spiral galaxies in scaling relations was one of the driver for our proposed revision (Paper~VII) to the tuning-fork diagram by  \citet{Hubble1936}, in the spirit of \citet{vandenBergh1976} proposed parallelism between S0s and spirals. In \reffig{fig:etgs_spirals_diagram}. we summarize our results on morphology, kinematics, population and scaling relations in a single diagram, using the same galaxy symbols as fig.~2 of Paper~VII, to provide a link between the two papers. This picture allows us to provide a new perspective and a clean empirical view of a number of classic scaling relation and known trends in galaxy properties. 

\subsubsection{$L-\sigma$ trends}

The classic \citet{Faber1976} relation $L\propto\sigma^4$ is well known to be a projection of the FP. We study it here using dynamical mass instead of light. Our study for the first time uses \se\ values integrated within \re, which properly account for both velocity dispersion and stellar rotation, in combination with accurate dynamical masses which, unlike luminosity, can be properly related to the stellar kinematics. We find that the relation is not well described by an approximately linear trend, but it shows a break which is a reflection of the one in the ZOE (equation \ref{eq:zoe}). The slope of the mean $\sigma-M$ relation (equation \ref{eq:m-sigma}) changes from $\mjam\propto\sigma_{\rm e}^{2.3}$ below $\mjam\approx5\times10^{10}$ \msun\ ($\se\approx140$ \kms) to $\mjam\propto\sigma_{\rm e}^{4.7}$ at larger masses. Thanks to our use of dynamical masses and \se\ we find for the first time that the bend in the relation is quantitatively consistent with the one we observe in the M-size relation.

The slope we find at the high-mass range is consistent with the original \citet{Faber1976} relation, while the low-mass slope in our $\sigma-M$ relation is consistent with claims of a change in the slope of the $\sigma-L$ relation of elliptical galaxies at low luminosities, where the relation was reported to become $L\propto\sigma^2$ \citep{Davies1983,Held1992,Matkovic2005,deRijcke2005,Lauer2007,Forbes2008,Tortora2009}. However previous studies differ from ours because of $M/L$ variations, they ignore stellar rotation, suffer from small number statistics and selection biases. This may be the reason why previous studies suggested a value $M_B\approx-20.5$ mag for the break in the $\sigma-L$ relation \citep[see Sec.~3.3.3 of][for a review]{Graham2011}, which corresponds to the characteristic mass $M\approx2\times10^{11}$ \msun, using the $(M/L)-\sigma$ relation \citep{Cappellari2006} and considering a typical galaxy colour $B-I\approx2.2$ mag. The previously reported mass is clearly inconsistent with our accurate value $M_{\rm JAM,b}\approx5\times10^{10}$ \msun. Our new result shows that the break in the relation is associated to the characteristic mass $\mjam\approx3\times10^{10}$ \msun\ of the break in the ZOE and not with the other characteristic mass $\mjam\approx2\times10^{11}$ \msun\ above which round, slow rotator, cored galaxies dominate.

Observations from the much larger SDSS sample have failed to find evidence for a clear break in the $\sigma-L$ relation \citep{Bernardi2003fp,Gallazzi2006,Hyde2009curv}. For this reason the relation is still often assumed to be a power-law. The accuracy and homogeneity of our data, the size of our sample and especially the  quantitative consistency between the $\sigma-M$ and $\re-M$ relation provides an unambiguous confirmation for the break and an accurate determination of the transition mass.

\subsubsection{$L-\re$ trends}

Another well known projection of the FP is the \citet{Kormendy1977} relation. When using mass instead of light it becomes clear it represents the analogue of the \citet{Faber1976}, but this time in the $(M_{\rm JAM},R_{\rm e}^{\rm maj})$ projection of the MP. Also in this case, when samples are morphologically selected to consist of ellipticals, they tend to populate mostly the region of the diagram near the ZOE, defining a relatively narrow sequence \citep{Graham2008curv,Kormendy2009,Chen2010,Misgeld2011}. Although the sequence is useful for a number of studies, it is important to realize that it is not a real sequence in galaxy space on the MP. It is due to the sample selection and it represents essentially one of the contour levels of a continuous trend of galaxy properties, spanning from spiral galaxies, to ETGs, and only terminating on the well defined ZOE (\reffig{fig:mass_size_spirals_etgs} here and figure~4 in Paper~I).

Given that photometry is much easier to obtain than stellar kinematics, a change of slope in the luminosity-size relation has been known for long. It was pointed out by \citet[their fig.~7]{Binggeli1984} when combining photometry measurements of dwarf spheroidals and ordinary ellipticals: the dwarf spheroidals sequence appears to sharply bend from the ellipticals sequence. Similar differences in the slope of the luminosity-size relation of dwarfs and ellipticals were presented by a number of authors \citep[e.g.][]{Kormendy1985,Graham2003,Kormendy2009,Misgeld2011}. The change of slope has been interpreted in different ways.  \citet{Kormendy1985} and \citet{Kormendy2009} interpreted dwarf spheroidal as constituting a separate family, of gas-stripped dwarf spirals/irregulars, while \citet{Graham2003} and \citet{Graham2008curv} explain the change of slope or curvature in the relation as a natural consequence of the variation of the \citet{Sersic1968} index with luminosity \citep[e.g.][]{Young1994,Graham2003} in a homogeneous class of elliptical galaxies with a range of masses (see \citealt{Graham2011} and \citealt{Kormendy2012} for two complementary reviews of this subject). 

Our results cannot be compared to theirs in a statistical sense, as galaxies in their diagrams are, by design, not representative of the population in the nearby Universe, and certain classes of objects (e.g. M32) are over-represented. Our sample is volume-limited and for this reason it gives a statistically representative view of the galaxy population above a certain mass. Still the fact that the sequence of dwarf spheroidals and low-mass spirals/irregulars, lie on the continuation of our trends for fast rotator ETGs and spiral galaxies respectively (\reffig{fig:mass_size_spirals_etgs}), below the $M_\star\la6\times10^{9} \msun$ mass limit of our survey, suggests a continuity between dwarf spheroidals and the low-mass end of our disk-dominated fast-rotator ETGs population, which in turns we showed are closely related to spiral galaxies. For this reason our results reconciles the apparent contrast between the findings of a Sph-E dichotomy \citep{Kormendy1985,deRijcke2005,Janz2008,Kormendy2009} and the ones of a continuity \citep{Graham2003,Gavazzi2005,Ferrarese2006acs,Cote2006,Janz2009,Boselli2008,Graham2008curv,Forbes2011}. Our finding in fact agrees with the proposed common origin of dwarf spheroidal and low-mass spiral galaxies and irregulars \citep{Kormendy1985,Dekel1986}, but also shows an empirical continuity between dwarf spheroidals and a subset of the ellipticals family. The missing link between Sph and E is constituted by disk-dominated fast rotator ETGs \citep{Emsellem2007,Cappellari2007}. In fact the continuity we find is not between ``true'' ellipticals, namely the slow rotator, and dwarf spheroidals, but between ``misclassified'' ellipticals with disks and S0, namely the fast rotators, and dwarf spheroidals. After this text was written and the parallelism between spiral galaxies and early-type galaxies were presented (Paper~VII) to interpret the trends we observed in galaxy scaling relations (Paper~I and \citealt{Cappellari2011dur}), an independent confirmation of this picture, including its extension to low mass systems was also provided by \citet{Kormendy2012}.

\subsubsection{Population trends with $\sigma$ or \re}

The characteristic mass $M_{\rm JAM}\approx3\times10^{10} \msun$ is the same transition mass discovered by \citet{Kauffmann2003mass} who state that ``low-redshift galaxies divide into two distinct families at a stellar mass of $3\times10^{10} \msun$. Lower-mass galaxies have young stellar populations, low surface mass densities and the low concentrations typical of discs. Their star formation histories are more strongly correlated with surface mass density than with stellar mass.'' A similar trend involving colours and also better correlated with surface density (or with the velocity dispersion estimated from the photometry) than with mass, was found to extend to redshift up to $z\approx3$, with red galaxies being systematically small, and blue galaxies being large at a given mass \citep{Franx2008}. This was recently confirmed, still using photometric data alone, by \citet{Bell2012}. The correctness of all these statements can now be easily and accurately verified for the nearby ETGs subset in \reffig{fig:virial_plane_projections_ml} and \ref{fig:virial_plane_projections_hbeta}. 

The novelty of our work, with respect to all previous studies, is that we have unprecedentedly accurate and unbiased dynamical masses, and stellar velocity dispersions \se\ integrated within a large aperture (1\re), instead of $\sigma$ values inferred from photometry. This allow us to conclusively state that neither dynamical mass nor stellar surface density are actually the best descriptor of galaxy properties, the main trend being along the $\sigma_{\rm e}$ direction (which includes rotation in the case of disk galaxies). Our clean \atl\ result was already presented in \citet{Cappellari2011dur} and subsequently confirmed with SDSS data and using virial mass estimates by \citet{Wake2012}. The trend we observed for the ETGs can be extended to spiral galaxies, which fill the region of larger sizes above the ETGs in the $(M_{\rm JAM},R_{\rm e}^{\rm maj})$ projection, smoothly overlapping with the ETGs for the largest spiral bulge fractions (figure~4 of Paper~I and \reffig{fig:mass_size_spirals_etgs}). It has been known for long that in spirals luminosity-weighted ages are lower, star formation is larger and colours bluer on average than the ETGs, essentially by definition \citep[e.g.][]{Hubble1936,vandenBergh1976}.

Similar trends between age and size (or surface brightness), with older objects being smaller at given age, were found to persist in ETGs. \citet{vanderWel2009} state that ``at a given stellar velocity dispersion, SDSS data show that there is no relation between size and age''. The same age-size trend was pointed out by \citet{Shankar2009}, and in different terms by \citet{Graves2009b} who state that ``no stellar population property shows any dependence on \re\ at fixed $\sigma$, suggesting that $\sigma$ and not dynamical mass is the better predictor of past SFH''. These findings are another way of saying that age variations must follow lines of constant $\sigma$ on the MP as we find here for $M/L$, H$\beta$, colour and molecular gas fraction, and confirm using age for our sample in McDermid et al. (in preparation), in agreement with \citet[their figure~15]{Gallazzi2006}.  The age-size trend was also confirmed in different samples of nearby galaxies by \citet{Valentinuzzi2010} and \citet{Napolitano2010}, and a similar trend, with star forming galaxies being larger than passive ones was found in place from redshift $z\sim2.5$ \citep{Williams2010z2,Wuyts2011,Newman2012}. The only contrasting view is the one by \citet{Trujillo2011}, who find a lack of age-size trend both at low and high-redshift.

The same characteristics mass $M_\star\approx3\times10^{10}$ \msun\ of \citet{Kauffmann2003mass}, which constitute the location of the break in our ZOE, was found by \citet{Hyde2009curv} in the mass-size relation of $5\times10^4$ SDSS galaxies. Their trend is significant but quite subtle. The reason for the curvature they find becomes clear from what we find: the average radius of the objects on the MP at constant mass, bends upwards at low masses, due to the cusp in the ZOE, with the strength of the effect being dependent on the specific criterion adopted to select ETGs. 

\subsection{Implications for galaxy formation}

Galaxy formation is the superposition of a number of complex events that happen in parallel. Here we sketch a tentative picture of some of the phenomena that can play a role in explaining what we see.  We refer the reader to \citet[hereafter Paper~VI]{Bois2011}, Paper~VIII and Naab et al. (in preparation) for a more in-depth discussion.

The smoothness and regularity of the trends we observe and the fact that they extend to spirals, appears to indicate a close connection between the formation of the two classes of objects  (Paper~VIII). The same similarity between the fast rotator ETGs and spirals in terms of their morphology and degree of rotation, lead us to propose a revision (Paper~VII) of the classic morphological classification \citep{Hubble1936} to emphasize the parallelism between the fast rotators and spirals, in the same way that \citet{vandenBergh1976} proposed it for S0s and spirals. Only the most massive slow rotators appear to form an empirically separated class. The {\em kinematic} morphology-density relation (Paper~VII), which applies to our kinematic classes the relation discovered by \citet{Dressler1980}, suggest that most spirals are being transformed into fast rotators due to environmental effects \citep{Khochfar2008}, with a mechanism that is sufficiently `gentle' to preserve the near axisymmetry of the disk (Paper~II).

One process which is often mentioned in the context of cluster galaxy populations is stripping by the interstellar medium \citep{Spitzer1951,Abadi1999}, which may impact the global morphology of the galaxy by removing a significant fraction (or most) of its gas content but mostly preserving the stellar disk component. This may explain some of the most flattened galaxies in the fast rotator class (e.g., NGC4762), as emphasised in Paper~VII. This would work by removing some of the mass of the galaxy keeping a relatively constant effective size (of the stellar component), and may contribute to the scatter in the Mass-size plane for fast rotator as well as to their overlap with the more gas-rich spiral galaxies.

As to explain the intermediate to high mass end of the fast rotator class, we would need a process which is able to increase the bulge size, while at the same time removing the gas or shutting off star formation (Paper~VIII). The empirical signatures of this phenomenon are visible in our data as an 'apparent' decrease of the galaxy size, and increase of $\sigma_{\rm e}$, which is actually due to the bulge concentrating more light at smaller radii, accompanied by a increase in $M/L$, which is tracing a decrease of star formation or an age increase. The process appears to generally preserve the intrinsic flatness of the stellar disks at large radii (top panel of \reffig{fig:virial_plane_projections_concentration}). A similar scenario was recently proposed by \citet{Bell2012} to interpret the relationship between rest-frame optical colour, stellar mass, star formation activity, and galaxy structure from $z\approx2$ to the present day.

Intense gas-rich accretion events, mostly via cold streams \citep{Keres2005,Dekel2009}, or major gas rich mergers (Paper~VIII), will increase both the mass and $\sigma_{\rm e}$. During the accretion the gas may sink toward the centre \citep{Mihos1994burst} until it becomes self-gravitating and starts forming stars. It is during this phase of rapid gas accretion that the $(M/L)-\sigma$ \citep{Cappellari2006,vanderMarel2007} and ${\rm IMF}-\sigma$ (\refsec{sec:imf_sigma}) relations, the tilt of the FP \citep{Dressler1987,Faber1987,Djorgovski1987}, the ${\rm Mg}b-\sigma$ \citep{Burstein1988pop,Bender1993pop} or the ${\rm Mg}b-V_{\rm esc}$ relation (\citealt{Davies1993}; \citealt{Scott2009}, Paper~XXI), will be imprinted in the ETGs population \citep{Robertson2006fp,Hopkins2009extralight} and then mostly preserved in the following evolution.

The early progenitors of today's fast rotator ETGs would be high-redshift spirals, which are different from local ones. They are characterized by giant gas clumps \citep{Elmegreen2007,Genzel2011} have high gas fractions \citep{Tacconi2010,Daddi2010}, possess large velocity dispersion and are dominated by turbulent motions \citep{ForsterSchreiber2006,ForsterSchreiber2009,Genzel2006,Genzel2008,Law2012}. In that situation bulges may form naturally as the clumps collide and sink to the centre \citep{Bournaud2007clumps,Dekel2009clumps}, unless they are efficiently destroyed by stellar feedback \citep{Genel2012,Hopkins2011}. Secular effects \citep{Kormendy2004} will also contribute to the bulge growth and $\sigma_{\rm e}$ increase, while keeping the mass unchanged, as will contribute the destabilizing effect of minor mergers. 

\begin{figure}
\plotone{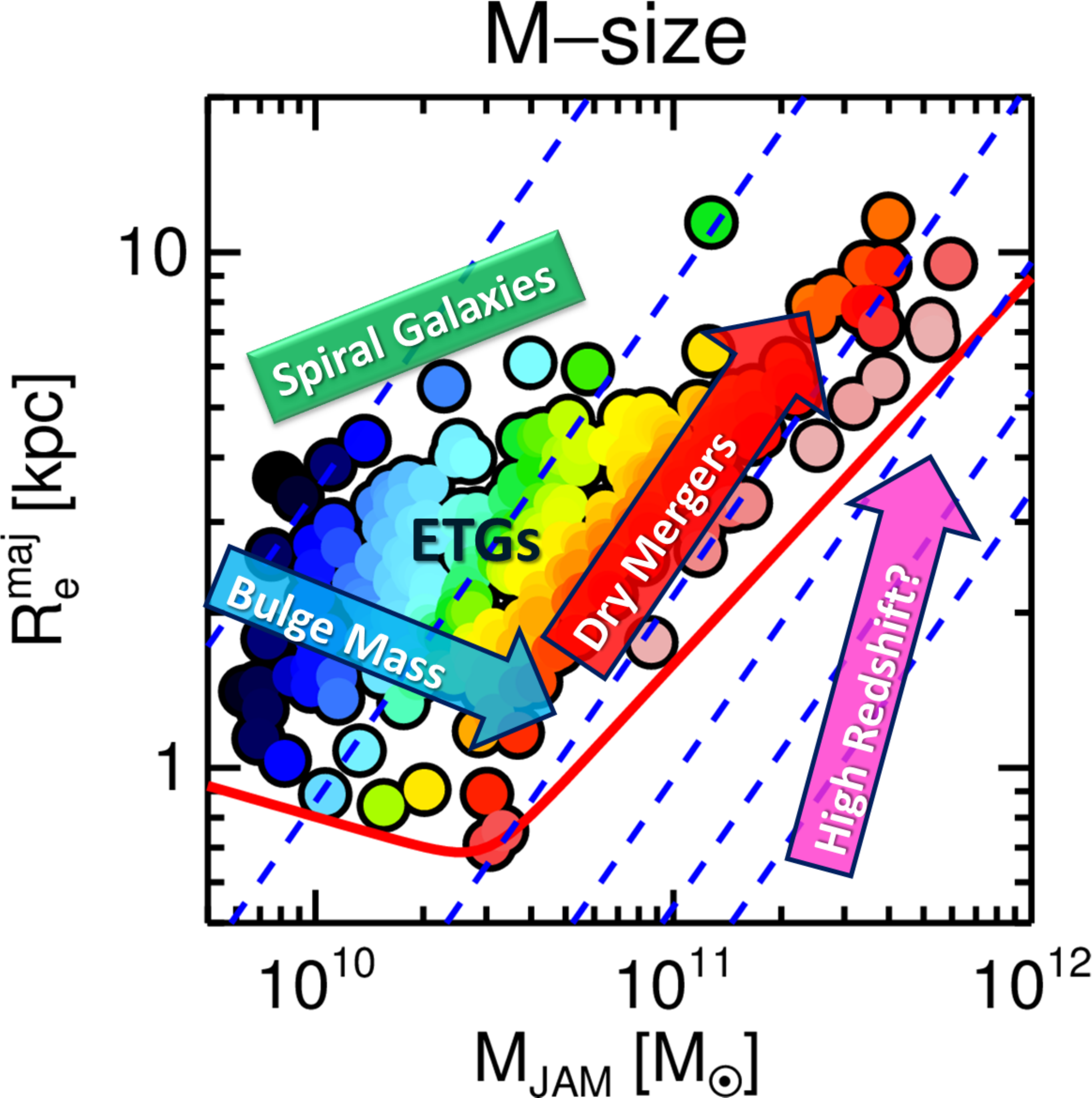}
\caption{Evolution scenario for ETGs. The symbols are the same as in \reffig{fig:virial_plane_projections_ml}, while the large arrows indicate the proposed interpretation of the observed distribution as due to a combination of two processes: (a) in-situ star formation: bulge or spheroid growth, which seems associated to the quenching of star formation, which moves galaxies to the right of towards the bottom, due to the increased concentration (decreasing \re\ and increasing \se), and (b) external accretion: dry mostly minor merging, increasing \re\ by moving galaxies along lines of roughly constant \se\ (or steeper), while leaving the population unchanged. A schematic illustration of these two processes is shown in \reffig{fig:black_holes_growth}.}
\label{fig:etgs_formation_scenario}
\end{figure}

\begin{figure}
\plotone{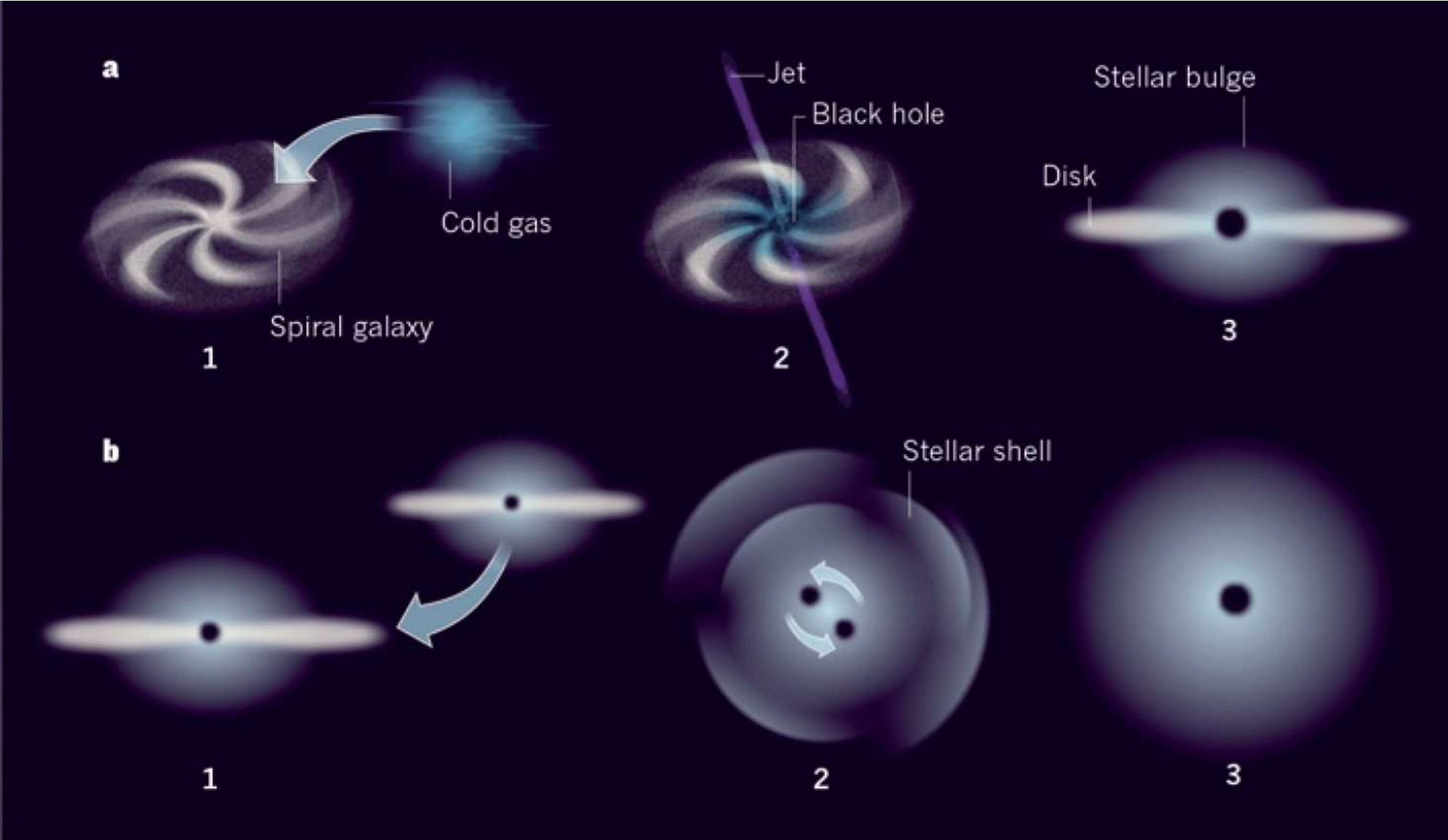}
\caption{Schematic representation of the two main processes responsible for the formation of the observed distribution of galaxies on the MP. (a) In-situ star formation: bulge growth via cold accretion, secular evolution, or minor gas-rich mergers, followed by quenching by AGN or other mechanisms, leaving the galaxy more massive, more compact, and consequently with a larger $\sigma_{\rm e}$, and gas poor (blue arrow in \reffig{fig:etgs_formation_scenario}); (b) external accretion: major or minor dry mergers, increasing galaxy mass and sizes at nearly constant $\sigma_{\rm e}$, or with a possible decrease, leaving the population mostly unchanged (red arrow in \reffig{fig:etgs_formation_scenario}). (taken from \citealt{Cappellari2011nat}).}
\label{fig:black_holes_growth}
\end{figure}

During the bulge growth some process must be able to turn off star formation, without destroying the fast rotating disks that still dominate the local ETGs population (Paper~II; Paper~III) and that dominates the ETGs population already from $z\sim2$ (\citealt{vanderWel2011}; see also \citealt{vanDokkum2011nat}). The reduced efficiency of star formation by morphological quenching \citep{Martig2009,Ceverino2010} may be one of the processes explaining why on average bigger bulges correspond to older ages and larger $M/L$. 
However we showed in \reffig{fig:virial_plane_projections_hbeta} that the cold gas fraction is decreasing with the bulge fraction, which shows that less fuel is available for galaxies with more massive bulges.
Bulge/spheroid growth seems also associated with AGN feedback, which would provide another bulge-related quenching mechanism \citep{Silk1998,Granato2004,Bower2006,Hopkins2006,Croton2006}. During the sequence of bulge/spheroid-growth followed by quenching, the original gas rich spiral will move from the left of the observed  $(M_{\rm JAM},R_{\rm e}^{\rm maj})$ plane towards the right, or possibly the bottom right due to the increased concentration, while intersecting the constant $\sigma$ lines (blue arrow in \reffig{fig:etgs_formation_scenario}). 
At the end of the evens the galaxy will be a fast rotator ETG, generally more massive and with a bigger bulge (smaller \re, larger \se\ and concentration) than the precursor clumpy spiral. 

At any stage during the bulge-growth phase the galaxy may accrete purely stellar satellites \citep{Khochfar2006b}. In the case of these dry mergers the situation is quite different and one can predict the final configuration using energy conservation \citep{Hernquist1993,Cole2000,Boylan-Kolchin2006,Ciotti2007}. The predictions show that sizes will increase as the mass grows. For major  mergers (equal mass) and typical orbital configurations, one can show that the mass and radius double, while $\sigma$ remains constant in agreement with simulations (\citealt{Nipoti2009}; see \citealt{Hilz2012a} and \citealt{Hilz2012b} for more accurate numerical simulations and detailed physical explanations of this process). While in the limit in which the same mass doubling happens via small satellites, as mostly expected from the shape of the \citet{Schechter1976} mass function, the radius will increase by a factor of four and the dispersion will be twice smaller \citep{Naab2009,Bezanson2009,Hopkins2009compact,Hilz2012a}. As a result the galaxy will move along lines that are parallel to the constant $\sigma$ lines, or steeper. During these dry mergers, given that there is little gas involved, the stellar population, colour and $M/L$ will remain unchanged (red arrow in \reffig{fig:etgs_formation_scenario}).

When the galaxy grows more massive than the characteristic mass $M_\star\ga2\times10^{11}$ \msun, a transition seems to happen in their formation, possibly associated to the shock heating of the infalling gas, due to the halo mass \citep{Dekel2006}. Above this mass galaxies are in fact embedded in massive and hot X-ray halos (\citealt{Kormendy2009}, Paper~XIX) which prevents any further cold gas accretion onto the galaxy.

The gas-rich mergers/accretion scenario is generally consistent with the observed correlation between supermassive black holes (BH) and galaxy velocity dispersion or bulge mass \citep{DiMatteo2005}. It is generally believed that the correlations indicate a joint evolution of galaxies and BHs, with BH growth happening at the same time as the bulge growth, and providing a self-regulation via feedback (\citealt{Silk1998}; but see \citealt{Peng2007} and \citealt{Jahnke2011} for a non-causal origin of the correlations). We demonstrate that indeed the $\sigma$ variation directly traces the bulge growth  \citep{Kormendy2001}. In fact $\sigma$ traces the central galaxy concentration and the bulge size as estimated on optical morphology, while the outer galaxy disks remain flat. This implies that, if the BH indeed accretes from the same gas that grows the bulge, BH mass should correlate better with $\sigma$ than with total mass as observed \citep{Gebhardt2000bh,Ferrarese2000}. 

However, when ETGs experience dry mergers, their BH grows in proportion to the mass, but galaxies move along lines of nearly constant $\sigma$. For this reason, an expectation from this picture is that BHs at the high mass end should start to appear too massive with respect to the predictions of the $M_{\rm BH}-\sigma$ relation \citep[see also][]{Nipoti2003,Boylan-Kolchin2006,Ciotti2009}. The high-mass BH end is still not sufficiently populated to reliably test this prediction, but indirect evidence seems to support this possibility \citep{Lauer2007}. Early direct evidences come from the recent detection of two giant black holes at the centre of two bright cluster galaxies, which are clearly more massive than the $M_{\rm BH}-\sigma$ prediction \citep{McConnell2011} or the related observation that non-core ETGs follow a more shallow $M_{\rm BH}-\sigma$ relation than core ETGs \citep{Graham2012}.

In summary we propose that the distribution of galaxy properties on the MP, where $M/L$ and age follows lines of constant $\sigma$ on the MP, could be explained by the combination of two processes, which can happen multiple times during the evolution of a single galaxy: (i) They accrete gas, which grows the bulge and BH, shrinks their \re, and increases $\sigma_{\rm e}$ and concentration, while some process which seems associated with the bulge growth (e.g.\ AGN feedback), quenches star formation; (ii) They experience mostly minor dry mergers that move them along lines of roughly constant $\sigma$ (or steeper).

An open question in the scenario in which ETGs evolve relatively quietly from spirals comes from the comparison of our findings with the empirical scaling relations one observes at larger redshift. In fact at $z\ga1.5$ galaxies are found to be smaller than local ones with the same mass \citep{Daddi2005,Trujillo2006,Trujillo2007,vanDokkum2008,Cimatti2008}. They populate the region below the ZOE of local galaxies, although their $\sigma$ tend to be consistent with our local observations \citep{Cappellari2012iau}. This may indicate that the compact high-redshift ETGs follow a different and more violent evolutionary path than the more quiet majority of local ETGs, as suggested by other high-redshift observations \citep{Barro2012}. A way to reconcile this very mild (or lack of) $\sigma$ evolution is by assuming that the compact primordial ETGs grow mostly by accretion of small satellites in their outer halos, while preserving the central structure \citep{Naab2009,Hopkins2010,Oser2010,Oser2012}. This seems consistent with the shape of the photometric profiles of the early ETGs \citep{Hopkins2009compact,Bezanson2009,vanDokkum2010profiles,Hilz2012b}. A caveat is that significant biases may still exist in the high-redshift photometry \citep{Mancini2010}, considering that systematic differences of up to a factor of two exists even on well observed ETGs in the nearby Universe (\citealt{Kormendy2009}, \citealt{Chen2010}, Paper~I). Moreover comparisons of photometric profiles tend to be made against bona fide ellipticals, while the remnant of the high-redshift ETGs are likely disk-like fast rotators ETG, which have systematically different profiles. Finally the comparisons should ideally be done in mass density, instead of surface density, but kinematic information is available for only a handful of galaxies. This implies that there is perhaps still some room for the compact high-$z$ ETGs to become more consistent with local ones, than currently assumed.

\section{Summary}

In the companion Paper~XV we describe in detail the axisymmetric dynamical models for all the \atl\ ETGs that were introduced in \citet{Cappellari2012}. We found that galaxies lie on a thin Mass Plane (MP) in the three-dimensional parameter space defined by $(M_{\rm JAM},\sigma_{\rm e},R_{\rm e}^{\rm maj})$.

Here we studied the inclined projection of the MP and find that: (i) the location galaxies define a clear zone of exclusion (ZOE), roughly described by two power-laws, joined by a break at a first characteristic mass $M_{\rm JAM}\equiv L\times (M/L)_e\approx3\times10^{10} \msun$, which corresponds to the regime where a number of global galaxy properties were reported to change; (ii) we find a second characteristic mass $\mjam\approx2\times10^{11}$ \msun, which separates a region with only nearly round slow rotator ETGs at large masses from one dominated by fast rotator ETGs with disks and spiral galaxies at lower masses; (iii) The distribution of $(M/L)_e$, as well as population indicators of $(M/L)_{\rm pop}$ like H$\beta$ and colour, and the molecular gas fraction, all tend to be constant along lines of constant $\sigma_{\rm e}$ on the $(\mjam,\se)$ plane, or correspondingly $\rmaj\propto\mjam$ (or even better parallel to the ZOE $\rmaj\propto M_{\rm JAM}^{0.75}$) on the $(\mjam,\rmaj)$ plane; (iv) properties of ETGs merge smoothly and form parallel sequences to those of spiral galaxies; (v) below our survey mass limit, the distribution of fast rotators ETGs continues with that of dwarf spheroidal (Sph), while the sequence of late spirals continues with the dwarf irregulars (Im).

We show that, at fixed mass, \se\ in ETGs traces the galaxy bulge fraction, which appears the main driver for the changes in galaxy properties that we observe. These findings explains why $\sigma_{\rm e}$ (not $L$, $M_\star$, \re\ or $\Sigma_{\rm e}$) has often been the most successful single descriptor of galaxy properties. We discuss a number of previously found galaxy correlation and we find that they are all consistent with what we find. In fact most observed relations turn out to be just special projections of the cleaner and more general view provided by the MP.

The stellar initial mass function was previously shown to vary systematically with $(M/L)_{\rm stars}$ \citep{Cappellari2012} and as expected it also follows the other population indicators. It also tend to vary along lines of nearly constant $\sigma_{\rm e}$ on the MP. This is a necessary expectation if one wants to preserve the tightness of the Fundamental Plane or the $(M/L)-\sigma_{\rm e}$ relation. The trend of IMF with $\sigma$ appears to account for about half of the total trend in the $(M/L)-\sigma_{\rm e}$ relation (see Paper~XV), the remaining one being due to stellar population variations.

The distribution of galaxy properties on the MP can be qualitatively interpreted as due to a combination of two main processes: (i) bulge growth, changing the galaxy population and decreasing \re, while increasing $\sigma_{\rm e}$, and (ii) dry merging, increasing \re\ by moving galaxies along lines of approximately constant $\sigma_{\rm e}$, while leaving the population unchanged. It is unclear where the reported dense ETGs fit in this picture. They may simply follow a different route, which forms the most massive galaxies and is not affecting the fast rotator population, which dominates ETGs in the nearby Universe. Or there may be some remaining biases in the photometric observations, producing an underestimation of the radius. Stellar kinematics of high-redshift ETGs is critically needed to address this question.  

\section*{acknowledgements}

MC acknowledges support from a Royal Society University Research Fellowship.
This work was supported by the rolling grants `Astrophysics at Oxford' PP/E001114/1 and ST/H002456/1 and visitors grants PPA/V/S/2002/00553, PP/E001564/1 and ST/H504862/1 from the UK Research Councils. RLD acknowledges travel and computer grants from Christ Church, Oxford and support from the Royal Society in the form of a Wolfson Merit Award 502011.K502/jd. RLD also acknowledges the support of the ESO Visitor Programme which funded a 3 month stay in 2010.
SK acknowledges support from the Royal Society Joint Projects Grant JP0869822.
RMcD is supported by the Gemini Observatory, which is operated by the Association of Universities for Research in Astronomy, Inc., on behalf of the international Gemini partnership of Argentina, Australia, Brazil, Canada, Chile, the United Kingdom, and the United States of America.
TN and MBois acknowledge support from the DFG Cluster of Excellence `Origin and Structure of the Universe'.
MS acknowledges support from a STFC Advanced Fellowship ST/F009186/1.
PS is a NWO/Veni fellow.
(TAD) The research leading to these results has received funding from the European
Community's Seventh Framework Programme (/FP7/2007-2013/) under grant agreement
No 229517.
MBois has received, during this research, funding from the European Research Council under the Advanced Grant Program Num 267399-Momentum.
The authors acknowledge financial support from ESO.

\footnotesize{{

}

\clearpage

\begin{deluxetable}{lcccc}
\tabletypesize{\small}
\tablewidth{0pt}
\tablecolumns{5}
\tablecaption{Central velocity dispersion, dark matter fraction and $M/L$ for the \atl\ sample of 260 early-type galaxies.\label{tab:atlas3d_parameters}}
\tablehead{
 \colhead{Galaxy} &
 \colhead{$\log\sigma(\re/8)$} &
 \colhead{$f_{\rm DM}(r=\re)$} &
 \colhead{$\log(M/L)_{\rm stars}$} &
 \colhead{$\log(M/L)_{\rm Salp}$} \\
 \colhead{ } &
 \colhead{(\kms)} &
 \colhead{ } &
 \colhead{($M_\odot/L_{\odot r}$)} &
 \colhead{($M_\odot/L_{\odot r}$)} \\
 \colhead{(1)} &
 \colhead{(2)} &
 \colhead{(3)} &
 \colhead{(4)} &
 \colhead{(5)} 
}
\startdata
     IC0560 &       1.920 &        0.12 &       0.429 &       0.381 \\
     IC0598 &       2.001 &        0.24 &       0.268 &       0.364 \\
     IC0676 &       1.825 &        0.20 &       0.481 &       0.261 \\
     IC0719 &       1.934 &        0.11 &       0.833 &       0.518 \\
     IC0782 &       1.818 &        0.57 &       0.318 &       0.510 \\
     IC1024 &       1.772 &        0.15 &       0.570 &       0.194 \\
     IC3631 &       1.733 &        0.00 &      -0.153 &       0.109 \\
    NGC0448 &       1.988 &        0.14 &       0.396 &       0.673 \\
    NGC0474 &       2.223 &        0.00 &       0.620 &       0.724 \\
    NGC0502 &       2.107 &        0.00 &       0.466 &       0.725 \\
    NGC0509 &       1.792 &        0.07 &       0.495 &       0.388 \\
    NGC0516 &       1.743 &        0.16 &       0.421 &       0.573 \\
    NGC0524 &       2.386 &        0.43 &       0.625 &       0.840 \\
    NGC0525 &       1.954 &        0.00 &       0.710 &       0.706 \\
    NGC0661 &       2.279 &        0.08 &       0.929 &       0.754 \\
    NGC0680 &       2.316 &        0.08 &       0.713 &       0.739 \\
    NGC0770 &       2.045 &        0.19 &       0.362 &       0.696 \\
    NGC0821 &       2.300 &        0.18 &       0.734 &       0.773 \\
    NGC0936 &       2.292 &        0.38 &       0.504 &       0.824 \\
    NGC1023 &       2.316 &        0.00 &       0.531 &       0.827 \\
    NGC1121 &       2.235 &        0.13 &       0.706 &       0.786 \\
    NGC1222 &       1.840 &        0.76 &       0.219 &      -0.014 \\
    NGC1248 &       1.965 &        0.00 &       0.326 &       0.488 \\
    NGC1266 &       1.897 &        0.00 &       0.598 &       0.364 \\
    NGC1289 &       2.133 &        0.14 &       0.577 &       0.578 \\
    NGC1665 &       1.999 &        0.12 &       0.450 &       0.567 \\
    NGC2481 &       2.271 &        0.06 &       0.642 &       0.706 \\
    NGC2549 &       2.174 &        0.17 &       0.666 &       0.710 \\
    NGC2577 &       2.320 &        0.06 &       0.843 &       0.790 \\
    NGC2592 &       2.384 &        0.01 &       0.882 &       0.812 \\
    NGC2594 &       2.263 &        0.07 &       0.659 &       0.729 \\
    NGC2679 &       1.983 &        0.17 &       0.423 &       0.576 \\
    NGC2685 &       1.972 &        0.18 &       0.327 &       0.620 \\
    NGC2695 &       2.342 &        0.11 &       0.672 &       0.795 \\
    NGC2698 &       2.355 &        0.00 &       0.749 &       0.802 \\
    NGC2699 &       2.181 &        0.07 &       0.538 &       0.724 \\
    NGC2764 &       1.954 &        0.00 &       0.649 &       0.154 \\
    NGC2768 &       2.306 &        0.00 &       0.933 &       0.772 \\
    NGC2778 &       2.215 &        0.04 &       0.842 &       0.768 \\
    NGC2824 &       2.127 &        0.45 &       0.262 &       0.343 \\
    NGC2852 &       2.306 &        0.00 &       0.834 &       0.784 \\
    NGC2859 &       2.252 &        0.06 &       0.546 &       0.732 \\
    NGC2880 &       2.150 &        0.25 &       0.580 &       0.675 \\
    NGC2950 &       2.230 &        0.11 &       0.539 &       0.661 \\
    NGC2962 &       2.244 &        0.00 &       0.813 &       0.782 \\
    NGC2974 &       2.375 &        0.07 &       0.950 &       0.787 \\
    NGC3032 &       1.983 &        0.42 &       0.164 &      -0.070 \\
    NGC3073 &       1.755 &        0.44 &       0.105 &      -0.145 \\
    NGC3098 &       2.038 &        0.18 &       0.517 &       0.612 \\
    NGC3156 &       1.838 &        0.23 &       0.245 &       0.132 \\
    NGC3182 &       2.060 &        0.48 &       0.336 &       0.606 \\
    NGC3193 &       2.299 &        0.08 &       0.532 &       0.794 \\
    NGC3226 &       2.250 &        0.00 &       0.873 &       0.784 \\
    NGC3230 &       2.303 &        0.04 &       0.754 &       0.795 \\
    NGC3245 &       2.322 &        0.11 &       0.588 &       0.727 \\
    NGC3248 &       1.990 &        0.15 &       0.420 &       0.527 \\
    NGC3301 &       2.096 &        0.21 &       0.297 &       0.448 \\
    NGC3377 &       2.163 &        0.06 &       0.553 &       0.731 \\
    NGC3379 &       2.329 &        0.33 &       0.600 &       0.838 \\
    NGC3384 &       2.208 &        0.22 &       0.404 &       0.724 \\
    NGC3400 &       1.878 &        0.18 &       0.522 &       0.687 \\
    NGC3412 &       2.022 &        0.47 &       0.211 &       0.648 \\
    NGC3414 &       2.365 &        0.14 &       0.714 &       0.826 \\
    NGC3457 &       1.889 &        0.21 &       0.173 &       0.517 \\
    NGC3458 &       2.253 &        0.05 &       0.627 &       0.767 \\
    NGC3489 &       2.028 &        0.24 &       0.113 &       0.394 \\
    NGC3499 &       1.845 &        0.25 &       0.189 &       0.449 \\
    NGC3522 &       2.011 &        0.24 &       0.582 &       0.648 \\
    NGC3530 &       2.049 &        0.23 &       0.493 &       0.695 \\
    NGC3595 &       2.206 &        0.00 &       0.602 &       0.749 \\
    NGC3599 &       1.874 &        0.11 &       0.234 &       0.467 \\
    NGC3605 &       1.966 &        0.00 &       0.459 &       0.668 \\
    NGC3607 &       2.360 &        0.07 &       0.661 &       0.799 \\
    NGC3608 &       2.286 &        0.09 &       0.669 &       0.829 \\
    NGC3610 &       2.244 &        0.12 &       0.409 &       0.566 \\
    NGC3613 &       2.327 &        0.00 &       0.772 &       0.761 \\
    NGC3619 &       2.221 &        0.28 &       0.615 &       0.693 \\
    NGC3626 &       2.134 &        0.23 &       0.332 &       0.261 \\
    NGC3630 &       2.241 &        0.13 &       0.550 &       0.745 \\
    NGC3640 &       2.267 &        0.17 &       0.534 &       0.722 \\
    NGC3641 &       2.241 &        0.34 &       0.821 &       0.773 \\
    NGC3648 &       2.277 &        0.21 &       0.694 &       0.807 \\
    NGC3658 &       2.216 &        0.00 &       0.572 &       0.721 \\
    NGC3665 &       2.352 &        0.14 &       0.748 &       0.764 \\
    NGC3674 &       2.346 &        0.04 &       0.811 &       0.809 \\
    NGC3694 &       1.892 &        0.60 &       0.082 &       0.437 \\
    NGC3757 &       2.170 &        0.20 &       0.610 &       0.705 \\
    NGC3796 &       1.955 &        0.34 &       0.243 &       0.387 \\
    NGC3838 &       2.144 &        0.00 &       0.589 &       0.694 \\
    NGC3941 &       2.144 &        0.00 &       0.400 &       0.693 \\
    NGC3945 &       2.262 &        0.22 &       0.549 &       0.760 \\
    NGC3998 &       2.445 &        0.15 &       0.920 &       0.818 \\
    NGC4026 &       2.279 &        0.08 &       0.633 &       0.747 \\
    NGC4036 &       2.276 &        0.00 &       0.697 &       0.783 \\
    NGC4078 &       2.242 &        0.07 &       0.823 &       0.755 \\
    NGC4111 &       2.201 &        0.00 &       0.651 &       0.616 \\
    NGC4119 &       1.771 &        0.20 &       0.423 &       0.306 \\
    NGC4143 &       2.336 &        0.16 &       0.646 &       0.834 \\
    NGC4150 &       1.945 &        0.20 &       0.336 &       0.358 \\
    NGC4168 &       2.241 &        0.25 &       0.775 &       0.738 \\
    NGC4179 &       2.264 &        0.08 &       0.724 &       0.771 \\
    NGC4191 &       2.127 &        0.22 &       0.533 &       0.685 \\
    NGC4203 &       2.202 &        0.00 &       0.536 &       0.827 \\
    NGC4215 &       2.147 &        0.04 &       0.589 &       0.652 \\
    NGC4233 &       2.337 &        0.22 &       0.724 &       0.812 \\
    NGC4249 &       1.885 &        0.78 &      -0.071 &       0.672 \\
    NGC4251 &       2.137 &        0.08 &       0.360 &       0.684 \\
    NGC4255 &       2.255 &        0.04 &       0.785 &       0.743 \\
    NGC4259 &       2.060 &        0.51 &       0.212 &       0.728 \\
    NGC4261 &       2.469 &        0.28 &       0.820 &       0.859 \\
    NGC4262 &       2.290 &        0.02 &       0.751 &       0.782 \\
    NGC4264 &       2.021 &        0.31 &       0.464 &       0.668 \\
    NGC4267 &       2.193 &        0.17 &       0.548 &       0.831 \\
    NGC4268 &       2.194 &        0.20 &       0.744 &       0.672 \\
    NGC4270 &       2.153 &        0.00 &       0.542 &       0.669 \\
    NGC4278 &       2.395 &        0.26 &       0.734 &       0.846 \\
    NGC4281 &       2.409 &        0.00 &       0.956 &       0.788 \\
    NGC4283 &       2.078 &        0.00 &       0.568 &       0.770 \\
    NGC4324 &       1.969 &        0.32 &       0.310 &       0.620 \\
    NGC4339 &       2.093 &        0.00 &       0.684 &       0.728 \\
    NGC4340 &       2.030 &        0.34 &       0.484 &       0.700 \\
    NGC4342 &       2.403 &        0.01 &       1.018 &       0.822 \\
    NGC4346 &       2.162 &        0.22 &       0.517 &       0.718 \\
    NGC4350 &       2.300 &        0.08 &       0.717 &       0.826 \\
    NGC4365 &       2.408 &        0.18 &       0.688 &       0.862 \\
    NGC4371 &       2.142 &        0.27 &       0.596 &       0.789 \\
    NGC4374 &       2.460 &        0.15 &       0.761 &       0.846 \\
    NGC4377 &       2.164 &        0.53 &       0.284 &       0.733 \\
    NGC4379 &       2.069 &        0.00 &       0.600 &       0.735 \\
    NGC4382 &       2.264 &        0.50 &       0.438 &       0.578 \\
    NGC4387 &       2.024 &        0.00 &       0.574 &       0.759 \\
    NGC4406 &       2.336 &        0.17 &       0.717 &       0.816 \\
    NGC4417 &       2.183 &        0.13 &       0.569 &       0.771 \\
    NGC4425 &       1.921 &        0.14 &       0.529 &       0.638 \\
    NGC4429 &       2.292 &        0.00 &       0.787 &       0.824 \\
    NGC4434 &       2.116 &        0.00 &       0.416 &       0.753 \\
    NGC4435 &       2.214 &        0.13 &       0.518 &       0.734 \\
    NGC4442 &       2.288 &        0.00 &       0.672 &       0.838 \\
    NGC4452 &       1.771 &        0.00 &       0.706 &       0.679 \\
    NGC4458 &       2.032 &        0.15 &       0.460 &       0.740 \\
    NGC4459 &       2.252 &        0.20 &       0.572 &       0.721 \\
    NGC4461 &       2.173 &        0.06 &       0.621 &       0.792 \\
    NGC4472 &       2.460 &        0.00 &       0.746 &       0.856 \\
    NGC4473 &       2.287 &        0.24 &       0.574 &       0.780 \\
    NGC4474 &       1.964 &        0.05 &       0.482 &       0.660 \\
    NGC4476 &       1.854 &        0.20 &       0.324 &       0.382 \\
    NGC4477 &       2.234 &        0.47 &       0.602 &       0.788 \\
    NGC4478 &       2.160 &        0.00 &       0.715 &       0.775 \\
    NGC4483 &       1.966 &        0.30 &       0.472 &       0.757 \\
    NGC4486 &       2.497 &        0.13 &       0.815 &       0.883 \\
   NGC4486A &       2.156 &        0.00 &       0.656 &       0.705 \\
    NGC4489 &       1.812 &        0.77 &      -0.030 &       0.556 \\
    NGC4494 &       2.188 &        0.20 &       0.511 &       0.699 \\
    NGC4503 &       2.170 &        0.00 &       0.733 &       0.815 \\
    NGC4521 &       2.247 &        0.16 &       0.774 &       0.778 \\
    NGC4526 &       2.371 &        0.00 &       0.748 &       0.775 \\
    NGC4528 &       2.044 &        0.01 &       0.580 &       0.641 \\
    NGC4546 &       2.345 &        0.00 &       0.737 &       0.810 \\
    NGC4550 &       1.974 &        0.21 &       0.537 &       0.656 \\
    NGC4551 &       1.998 &        0.00 &       0.697 &       0.769 \\
    NGC4552 &       2.418 &        0.21 &       0.743 &       0.861 \\
    NGC4564 &       2.241 &        0.04 &       0.669 &       0.792 \\
    NGC4570 &       2.281 &        0.15 &       0.574 &       0.820 \\
    NGC4578 &       2.043 &        0.38 &       0.514 &       0.739 \\
    NGC4596 &       2.179 &        0.18 &       0.638 &       0.776 \\
    NGC4608 &       2.139 &        0.00 &       0.633 &       0.766 \\
    NGC4612 &       1.934 &        0.56 &       0.168 &       0.448 \\
    NGC4621 &       2.350 &        0.03 &       0.759 &       0.846 \\
    NGC4623 &       1.835 &        0.50 &       0.309 &       0.723 \\
    NGC4624 &       2.156 &        0.25 &       0.582 &       0.794 \\
    NGC4636 &       2.300 &        0.39 &       0.789 &       0.844 \\
    NGC4638 &       2.080 &        0.05 &       0.433 &       0.703 \\
    NGC4643 &       2.184 &        0.14 &       0.635 &       0.755 \\
    NGC4649 &       2.498 &        0.29 &       0.774 &       0.871 \\
    NGC4660 &       2.338 &        0.10 &       0.601 &       0.817 \\
    NGC4684 &       1.837 &        0.10 &       0.304 &       0.571 \\
    NGC4690 &       2.052 &        0.00 &       0.590 &       0.542 \\
    NGC4694 &       1.792 &        0.00 &       0.150 &      -0.042 \\
    NGC4697 &       2.257 &        0.00 &       0.703 &       0.800 \\
    NGC4710 &       2.021 &        0.00 &       0.645 &       0.587 \\
    NGC4733 &       1.750 &        0.00 &       0.338 &       0.518 \\
    NGC4753 &       2.263 &        0.25 &       0.519 &       0.615 \\
    NGC4754 &       2.254 &        0.21 &       0.596 &       0.815 \\
    NGC4762 &       2.129 &        0.23 &       0.417 &       0.760 \\
    NGC4803 &       2.016 &        0.55 &       0.336 &       0.670 \\
    NGC5103 &       2.061 &        0.22 &       0.376 &       0.695 \\
    NGC5173 &       2.005 &        0.33 &       0.261 &       0.507 \\
    NGC5198 &       2.304 &        0.00 &       0.794 &       0.809 \\
    NGC5273 &       1.870 &        0.55 &       0.247 &       0.502 \\
    NGC5308 &       2.376 &        0.09 &       0.744 &       0.790 \\
    NGC5322 &       2.395 &        0.16 &       0.634 &       0.700 \\
    NGC5342 &       2.254 &        0.21 &       0.676 &       0.773 \\
    NGC5353 &       2.462 &        0.11 &       0.769 &       0.855 \\
    NGC5355 &       1.918 &        0.47 &       0.227 &       0.279 \\
    NGC5358 &       1.886 &        0.27 &       0.477 &       0.623 \\
    NGC5379 &       1.840 &        0.73 &       0.307 &       0.468 \\
    NGC5422 &       2.242 &        0.21 &       0.670 &       0.758 \\
    NGC5473 &       2.332 &        0.49 &       0.511 &       0.783 \\
    NGC5475 &       2.041 &        0.33 &       0.411 &       0.544 \\
    NGC5481 &       2.174 &        0.00 &       0.776 &       0.740 \\
    NGC5485 &       2.253 &        0.09 &       0.806 &       0.777 \\
    NGC5493 &       2.208 &        0.20 &       0.305 &       0.514 \\
    NGC5500 &       1.915 &        0.13 &       0.649 &       0.650 \\
    NGC5507 &       2.320 &        0.17 &       0.708 &       0.786 \\
    NGC5557 &       2.406 &        0.01 &       0.659 &       0.774 \\
    NGC5574 &       1.885 &        0.04 &       0.378 &       0.368 \\
    NGC5576 &       2.269 &        0.00 &       0.454 &       0.705 \\
    NGC5582 &       2.180 &        0.37 &       0.599 &       0.752 \\
    NGC5611 &       2.120 &        0.06 &       0.651 &       0.645 \\
    NGC5631 &       2.207 &        0.10 &       0.580 &       0.627 \\
    NGC5638 &       2.202 &        0.06 &       0.647 &       0.791 \\
    NGC5687 &       2.280 &        0.12 &       0.864 &       0.768 \\
    NGC5770 &       1.988 &        0.12 &       0.328 &       0.576 \\
    NGC5813 &       2.354 &        0.47 &       0.699 &       0.816 \\
    NGC5831 &       2.220 &        0.14 &       0.610 &       0.743 \\
    NGC5838 &       2.441 &        0.00 &       0.901 &       0.834 \\
    NGC5839 &       2.208 &        0.01 &       0.716 &       0.802 \\
    NGC5845 &       2.431 &        0.00 &       0.695 &       0.810 \\
    NGC5846 &       2.365 &        0.14 &       0.844 &       0.853 \\
    NGC5854 &       2.061 &        0.12 &       0.356 &       0.472 \\
    NGC5864 &       2.079 &        0.00 &       0.570 &       0.636 \\
    NGC5866 &       2.205 &        0.00 &       0.663 &       0.640 \\
    NGC5869 &       2.260 &        0.24 &       0.750 &       0.787 \\
    NGC6010 &       2.213 &        0.17 &       0.685 &       0.719 \\
    NGC6014 &       1.953 &        0.48 &       0.429 &       0.518 \\
    NGC6017 &       2.041 &        0.00 &       0.447 &       0.526 \\
    NGC6149 &       2.031 &        0.31 &       0.425 &       0.611 \\
    NGC6278 &       2.338 &        0.16 &       0.667 &       0.812 \\
    NGC6547 &       2.285 &        0.12 &       0.740 &       0.629 \\
    NGC6548 &       2.240 &        0.20 &       0.768 &       0.698 \\
    NGC6703 &       2.260 &        0.00 &       0.775 &       0.736 \\
    NGC6798 &       2.137 &        0.16 &       0.564 &       0.684 \\
    NGC7280 &       2.061 &        0.36 &       0.406 &       0.444 \\
    NGC7332 &       2.143 &        0.07 &       0.288 &       0.528 \\
    NGC7454 &       2.049 &        0.13 &       0.673 &       0.585 \\
    NGC7457 &       1.874 &        0.45 &       0.238 &       0.455 \\
    NGC7465 &       1.971 &        0.05 &       0.333 &       0.348 \\
    NGC7693 &       1.816 &        0.00 &       0.586 &       0.357 \\
    NGC7710 &       2.019 &        0.00 &       0.537 &       0.616 \\
  PGC016060 &       1.992 &        0.00 &       0.651 &       0.509 \\
  PGC028887 &       2.132 &        0.35 &       0.643 &       0.730 \\
  PGC029321 &       1.762 &        0.90 &      -0.541 &       0.362 \\
  PGC035754 &       2.067 &        0.00 &       0.577 &       0.634 \\
  PGC042549 &       2.004 &        0.14 &       0.422 &       0.506 \\
  PGC044433 &       2.123 &        0.10 &       0.688 &       0.724 \\
  PGC050395 &       1.869 &        0.74 &      -0.061 &       0.581 \\
  PGC051753 &       1.892 &        0.15 &       0.584 &       0.621 \\
  PGC054452 &       1.818 &        0.28 &       0.411 &       0.534 \\
  PGC056772 &       1.916 &        0.00 &       0.620 &       0.379 \\
  PGC058114 &       1.902 &        \nodata &    \nodata &       0.474 \\
  PGC061468 &       1.823 &        0.90 &      -0.280 &       0.415 \\
  PGC071531 &       1.969 &        \nodata &    \nodata &       0.479 \\
  PGC170172 &       2.020 &        0.16 &       0.111 &       0.650 \\
   UGC03960 &       1.946 &        0.04 &       0.729 &       0.606 \\
   UGC04551 &       2.265 &        0.04 &       0.658 &       0.737 \\
   UGC05408 &       1.919 &        0.00 &       0.194 &       0.331 \\
   UGC06062 &       2.156 &        0.25 &       0.629 &       0.760 \\
   UGC06176 &       2.045 &        0.50 &       0.331 &       0.328 \\
   UGC08876 &       2.152 &        0.00 &       0.760 &       0.740 \\
   UGC09519 &       1.992 &        0.10 &       0.513 &       0.345 \\
\enddata
\tablecomments{
Column (1): The Name is the principal designation from LEDA \citep{Paturel2003}, which is used as standard designation for our project.
Column (2): Central stellar velocity dispersion ($1\sigma$ random error of 5\% or 0.021 dex) within a circular aperture of radius $R=\re/8$, where \re\ is the value given in table~1 of Paper~XV. This $\sigma(\re/8)$ is measured by co-adding all \sauron\ spectra contained within the aperture. The velocity dispersion is measured on that single spectrum using pPXF and adopting a Gaussian line-of-sight velocity distribution (keyword MOMENTS$=$2). 
Column (3): fraction of dark matter contained within an iso-surface of volume $V=4\pi R_{\rm e}^3/3$ (a sphere of radius $r=\re$ for spherical galaxies).
Column (4): Mass-to-light ratio ($1\sigma$ random error of 6\% or 0.027 dex) of the stellar component (approximately within a sphere of radius $r\sim\re$) in the SDSS $r$-band, for the best-fitting JAM model with NFW halo (model B in Paper~XV) for the assumed distance and extinction of table~3 of Paper~I. A description of the quality of the model fits is given in table~1 of Paper~XV. The $(M/L)_{\rm stars}$ and $f_{\rm DM}$ of models with ${\rm qual}<2$ should be used with caution.
Column (5): Mass-to-light ratio of the stellar population ($1\sigma$ random error of 6\% or 0.027 dex)  within 1\re, in the SDSS $r$-band, assuming a Salpeter IMF slope from 0.1 to 100 \msun\ and using the models of \citet{Vazdekis2012}. This value is mass weighted over the multiple populations fitted to the \sauron\ spectra with pPXF.
Table 1 is also available from our project website http://purl.org/atlas3d.
}
\end{deluxetable}

\label{lastpage}

\begin{thebibliography}{295}
\expandafter\ifx\csname natexlab\endcsname\relax\def\natexlab#1{#1}\fi

\bibitem[{{Abadi}, {Moore} \& {Bower}(1999){Abadi} M.~G., {Moore} B., {Bower}
  R.~G.}]{Abadi1999}
{Abadi} M.~G., {Moore} B., {Bower} R.~G., 1999, \mnras, 308, 947

\bibitem[{{Aihara} {et~al}\mbox{.}(2011){Aihara} H. {et~al.}}]{Aihara2011}
{Aihara} H. {et~al.}, 2011, \apjs, 193, 29

\bibitem[{{Auger} {et~al}\mbox{.}(2009){Auger} M.~W. {et~al.}}]{Auger2009}
{Auger} M.~W., {Treu} T., {Bolton} A.~S., {Gavazzi} R., {Koopmans} L.~V.~E.,
  {Marshall} P.~J., {Bundy} K., {Moustakas} L.~A., 2009, \apj, 705, 1099

\bibitem[{{Auger} {et~al}\mbox{.}(2010{\natexlab{a}}){Auger} M.~W.
  {et~al.}}]{Auger2010}
{Auger} M.~W., {Treu} T., {Bolton} A.~S., {Gavazzi} R., {Koopmans} L.~V.~E.,
  {Marshall} P.~J., {Moustakas} L.~A., {Burles} S., 2010{\natexlab{a}}, \apj,
  724, 511

\bibitem[{{Auger} {et~al}\mbox{.}(2010{\natexlab{b}}){Auger} M.~W.
  {et~al.}}]{Auger2010imf}
{Auger} M.~W., {Treu} T., {Gavazzi} R., {Bolton} A.~S., {Koopmans} L.~V.~E.,
  {Marshall} P.~J., 2010{\natexlab{b}}, \apjl, 721, L163

\bibitem[{{Bacon} {et~al}\mbox{.}(2001){Bacon} R. {et~al.}}]{Bacon2001}
{Bacon} R. {et~al.}, 2001, \mnras, 326, 23

\bibitem[{{Baldry} {et~al}\mbox{.}(2004){Baldry} I.~K. {et~al.}}]{Baldry2004}
{Baldry} I.~K., {Glazebrook} K., {Brinkmann} J., {Ivezi{\'c}} {\v Z}., {Lupton}
  R.~H., {Nichol} R.~C., {Szalay} A.~S., 2004, \apj, 600, 681

\bibitem[{{Barnab{\`e}} {et~al}\mbox{.}(2011){Barnab{\`e}} M.
  {et~al.}}]{Barnabe2011}
{Barnab{\`e}} M., {Czoske} O., {Koopmans} L.~V.~E., {Treu} T., {Bolton} A.~S.,
  2011, \mnras, 415, 2215

\bibitem[{{Barnab{\`e}} {et~al}\mbox{.}(2012){Barnab{\`e}} M.
  {et~al.}}]{Barnabe2012}
{Barnab{\`e}} M. {et~al.}, 2012, \mnras, 423, 1073

\bibitem[{{Barro} {et~al}\mbox{.}(2013){Barro} G. {et~al.}}]{Barro2012}
{Barro} G. {et~al.}, 2013, \apj, 765, 104

\bibitem[{{Bastian}, {Covey} \& {Meyer}(2010){Bastian} N., {Covey} K.~R.,
  {Meyer} M.~R.}]{Bastian2010}
{Bastian} N., {Covey} K.~R., {Meyer} M.~R., 2010, \araa, 48, 339

\bibitem[{{Bell} \& {de Jong}(2001){Bell} E.~F., {de Jong} R.~S.}]{Bell2001}
{Bell} E.~F., {de Jong} R.~S., 2001, \apj, 550, 212

\bibitem[{{Bell} {et~al}\mbox{.}(2012){Bell} E.~F. {et~al.}}]{Bell2012}
{Bell} E.~F. {et~al.}, 2012, \apj, 753, 167

\bibitem[{{Bender}, {Burstein} \& {Faber}(1992){Bender} R., {Burstein} D.,
  {Faber} S.~M.}]{Bender1992}
{Bender} R., {Burstein} D., {Faber} S.~M., 1992, \apj, 399, 462

\bibitem[{{Bender}, {Burstein} \& {Faber}(1993){Bender} R., {Burstein} D.,
  {Faber} S.~M.}]{Bender1993pop}
{Bender} R., {Burstein} D., {Faber} S.~M., 1993, \apj, 411, 153

\bibitem[{{Bernardi} {et~al}\mbox{.}(2011){Bernardi} M.
  {et~al.}}]{Bernardi2011mass2e11}
{Bernardi} M., {Roche} N., {Shankar} F., {Sheth} R.~K., 2011, \mnras, 412, L6

\bibitem[{{Bernardi} {et~al}\mbox{.}(2003){Bernardi} M.
  {et~al.}}]{Bernardi2003fp}
{Bernardi} M. {et~al.}, 2003, \aj, 125, 1866

\bibitem[{{Bershady} {et~al}\mbox{.}(2011){Bershady} M.~A.
  {et~al.}}]{Bershady2011}
{Bershady} M.~A., {Martinsson} T.~P.~K., {Verheijen} M.~A.~W., {Westfall}
  K.~B., {Andersen} D.~R., {Swaters} R.~A., 2011, \apjl, 739, L47

\bibitem[{{Bershady} {et~al}\mbox{.}(2010){Bershady} M.~A.
  {et~al.}}]{Bershady2010}
{Bershady} M.~A., {Verheijen} M.~A.~W., {Swaters} R.~A., {Andersen} D.~R.,
  {Westfall} K.~B., {Martinsson} T., 2010, \apj, 716, 198

\bibitem[{{Bezanson} {et~al}\mbox{.}(2009){Bezanson} R.
  {et~al.}}]{Bezanson2009}
{Bezanson} R., {van Dokkum} P.~G., {Tal} T., {Marchesini} D., {Kriek} M.,
  {Franx} M., {Coppi} P., 2009, \apj, 697, 1290

\bibitem[{{Binggeli}, {Sandage} \& {Tarenghi}(1984){Binggeli} B., {Sandage} A.,
  {Tarenghi} M.}]{Binggeli1984}
{Binggeli} B., {Sandage} A., {Tarenghi} M., 1984, \aj, 89, 64

\bibitem[{{Binney} \& {Tremaine}(2008){Binney} J., {Tremaine} S.}]{Binney2008}
{Binney} J., {Tremaine} S., 2008, Galactic Dynamics: Second Edition. Princeton
  University Press

\bibitem[{{Blanton} {et~al}\mbox{.}(2003){Blanton} M.~R.
  {et~al.}}]{Blanton2003}
{Blanton} M.~R. {et~al.}, 2003, \apj, 592, 819

\bibitem[{{Bois} {et~al}\mbox{.}(2011){Bois} M. {et~al.}}]{Bois2011}
{Bois} M. {et~al.}, 2011, \mnras, 416, 1654 (Paper~VI)

\bibitem[{{Bolton} {et~al}\mbox{.}(2008{\natexlab{a}}){Bolton} A.~S.
  {et~al.}}]{Bolton2008slacs5}
{Bolton} A.~S., {Burles} S., {Koopmans} L.~V.~E., {Treu} T., {Gavazzi} R.,
  {Moustakas} L.~A., {Wayth} R., {Schlegel} D.~J., 2008{\natexlab{a}}, \apj,
  682, 964

\bibitem[{{Bolton} {et~al}\mbox{.}(2006){Bolton} A.~S. {et~al.}}]{Bolton2006}
{Bolton} A.~S., {Burles} S., {Koopmans} L.~V.~E., {Treu} T., {Moustakas} L.~A.,
  2006, \apj, 638, 703

\bibitem[{{Bolton} {et~al}\mbox{.}(2008{\natexlab{b}}){Bolton} A.~S.
  {et~al.}}]{Bolton2008}
{Bolton} A.~S., {Treu} T., {Koopmans} L.~V.~E., {Gavazzi} R., {Moustakas}
  L.~A., {Burles} S., {Schlegel} D.~J., {Wayth} R., 2008{\natexlab{b}}, \apj,
  684, 248

\bibitem[{{Boselli} {et~al}\mbox{.}(2008){Boselli} A. {et~al.}}]{Boselli2008}
{Boselli} A., {Boissier} S., {Cortese} L., {Gavazzi} G., 2008, \apj, 674, 742

\bibitem[{{Bournaud}, {Elmegreen} \& {Elmegreen}(2007){Bournaud} F.,
  {Elmegreen} B.~G., {Elmegreen} D.~M.}]{Bournaud2007clumps}
{Bournaud} F., {Elmegreen} B.~G., {Elmegreen} D.~M., 2007, \apj, 670, 237

\bibitem[{{Bower} {et~al}\mbox{.}(2006){Bower} R.~G. {et~al.}}]{Bower2006}
{Bower} R.~G., {Benson} A.~J., {Malbon} R., {Helly} J.~C., {Frenk} C.~S.,
  {Baugh} C.~M., {Cole} S., {Lacey} C.~G., 2006, \mnras, 370, 645

\bibitem[{{Boylan-Kolchin}, {Ma} \& {Quataert}(2006){Boylan-Kolchin} M., {Ma}
  C.-P., {Quataert} E.}]{Boylan-Kolchin2006}
{Boylan-Kolchin} M., {Ma} C.-P., {Quataert} E., 2006, \mnras, 369, 1081

\bibitem[{{Brewer} {et~al}\mbox{.}(2012){Brewer} B.~J. {et~al.}}]{Brewer2012}
{Brewer} B.~J. {et~al.}, 2012, \mnras, 422, 3574

\bibitem[{{Burstein} {et~al}\mbox{.}(1997){Burstein} D.
  {et~al.}}]{Burstein1997}
{Burstein} D., {Bender} R., {Faber} S., {Nolthenius} R., 1997, \aj, 114, 1365

\bibitem[{{Burstein} {et~al}\mbox{.}(1988){Burstein} D.
  {et~al.}}]{Burstein1988pop}
{Burstein} D., {Davies} R.~L., {Dressler} A., {Faber} S.~M., {Lynden-Bell} D.,
  1988, in Astrophysics and Space Science Library, Vol. 141, Towards
  Understanding Galaxies at Large Redshift, Renzini R.~G.~K. .~A., ed., pp.
  17--21

\bibitem[{{Caon}, {Capaccioli} \& {D'Onofrio}(1993){Caon} N., {Capaccioli} M.,
  {D'Onofrio} M.}]{Caon1993}
{Caon} N., {Capaccioli} M., {D'Onofrio} M., 1993, \mnras, 265, 1013

\bibitem[{{Cappellari}(2002){Cappellari} M.}]{Cappellari2002mge}
{Cappellari} M., 2002, \mnras, 333, 400

\bibitem[{{Cappellari}(2008){Cappellari} M.}]{Cappellari2008}
{Cappellari} M., 2008, \mnras, 390, 71

\bibitem[{{Cappellari}(2011{\natexlab{a}}){Cappellari} M.}]{Cappellari2011dur}
{Cappellari} M., 2011{\natexlab{a}}, in Paper presented at the conference on
  Galaxy Formation held 18--22 July, 2011 at Durham University, Durham, UK.
  Online at http://astro.dur.ac.uk/Gal2011/talks.php

\bibitem[{{Cappellari}(2011{\natexlab{b}}){Cappellari} M.}]{Cappellari2011nat}
{Cappellari} M., 2011{\natexlab{b}}, \nat, 480, 187

\bibitem[{{Cappellari}(2013){Cappellari} M.}]{Cappellari2012iau}
{Cappellari} M., 2013, in IAU Symposium, Vol. 295, The Intriguing Life of
  Massive Galaxies, {Thomas} D., {Pasquali} A., {Ferreras} I., eds., p.
  (arXiv:1210.7742)

\bibitem[{{Cappellari} {et~al}\mbox{.}(2006){Cappellari} M.
  {et~al.}}]{Cappellari2006}
{Cappellari} M. {et~al.}, 2006, \mnras, 366, 1126

\bibitem[{{Cappellari} \& {Emsellem}(2004){Cappellari} M., {Emsellem}
  E.}]{Cappellari2004}
{Cappellari} M., {Emsellem} E., 2004, \pasp, 116, 138

\bibitem[{{Cappellari} {et~al}\mbox{.}(2007){Cappellari} M.
  {et~al.}}]{Cappellari2007}
{Cappellari} M. {et~al.}, 2007, \mnras, 379, 418

\bibitem[{{Cappellari} {et~al}\mbox{.}(2011{\natexlab{a}}){Cappellari} M.
  {et~al.}}]{Cappellari2011a}
{Cappellari} M. {et~al.}, 2011{\natexlab{a}}, \mnras, 413, 813 (Paper~I)

\bibitem[{{Cappellari} {et~al}\mbox{.}(2011{\natexlab{b}}){Cappellari} M.
  {et~al.}}]{Cappellari2011b}
{Cappellari} M. {et~al.}, 2011{\natexlab{b}}, \mnras, 416, 1680 (Paper~VII)

\bibitem[{{Cappellari} {et~al}\mbox{.}(2012){Cappellari} M.
  {et~al.}}]{Cappellari2012}
{Cappellari} M. {et~al.}, 2012, \nat, 484, 485

\bibitem[{{Cappellari} {et~al}\mbox{.}(2013){Cappellari} M.
  {et~al.}}]{Cappellari2012p15}
{Cappellari} M. {et~al.}, 2013, \mnras\ in press (ArXiv:1208.3522), (Paper~XV)

\bibitem[{{Carlberg}(1986){Carlberg} R.~G.}]{Carlberg1986}
{Carlberg} R.~G., 1986, \apj, 310, 593

\bibitem[{{Cenarro} {et~al}\mbox{.}(2003){Cenarro} A.~J.
  {et~al.}}]{Cenarro2003}
{Cenarro} A.~J., {Gorgas} J., {Vazdekis} A., {Cardiel} N., {Peletier} R.~F.,
  2003, \mnras, 339, L12

\bibitem[{{Ceverino}, {Dekel} \& {Bournaud}(2010){Ceverino} D., {Dekel} A.,
  {Bournaud} F.}]{Ceverino2010}
{Ceverino} D., {Dekel} A., {Bournaud} F., 2010, \mnras, 404, 2151

\bibitem[{{Chabrier}(2003){Chabrier} G.}]{Chabrier2003}
{Chabrier} G., 2003, \pasp, 115, 763

\bibitem[{{Chen} {et~al}\mbox{.}(2010){Chen} C. {et~al.}}]{Chen2010}
{Chen} C., {C{\^o}t{\'e}} P., {West} A.~A., {Peng} E.~W., {Ferrarese} L., 2010,
  \apjs, 191, 1

\bibitem[{{Chiosi} {et~al}\mbox{.}(1998){Chiosi} C. {et~al.}}]{Chiosi1998}
{Chiosi} C., {Bressan} A., {Portinari} L., {Tantalo} R., 1998, \aap, 339, 355

\bibitem[{{Cimatti} {et~al}\mbox{.}(2008){Cimatti} A. {et~al.}}]{Cimatti2008}
{Cimatti} A. {et~al.}, 2008, \aap, 482, 21

\bibitem[{{Ciotti}(1991){Ciotti} L.}]{Ciotti1991}
{Ciotti} L., 1991, \aap, 249, 99

\bibitem[{{Ciotti}(2009){Ciotti} L.}]{Ciotti2009}
{Ciotti} L., 2009, Nuovo Cimento Rivista Serie, 32, 1

\bibitem[{{Ciotti}, {Lanzoni} \& {Volonteri}(2007){Ciotti} L., {Lanzoni} B.,
  {Volonteri} M.}]{Ciotti2007}
{Ciotti} L., {Lanzoni} B., {Volonteri} M., 2007, \apj, 658, 65

\bibitem[{Cleveland(1979)Cleveland W.}]{cleveland1979robust}
Cleveland W., 1979, Journal of the American statistical association, 829

\bibitem[{Cleveland \& Devlin(1988)Cleveland W., Devlin
  S.}]{cleveland1988locally}
Cleveland W., Devlin S., 1988, Journal of the American Statistical Association,
  596

\bibitem[{{Cole} {et~al}\mbox{.}(2000){Cole} S. {et~al.}}]{Cole2000}
{Cole} S., {Lacey} C.~G., {Baugh} C.~M., {Frenk} C.~S., 2000, \mnras, 319, 168

\bibitem[{{Conroy}, {Gunn} \& {White}(2009){Conroy} C., {Gunn} J.~E., {White}
  M.}]{Conroy2009}
{Conroy} C., {Gunn} J.~E., {White} M., 2009, \apj, 699, 486

\bibitem[{{Conroy} \& {van Dokkum}(2012{\natexlab{a}}){Conroy} C., {van Dokkum}
  P.}]{Conroy2012models}
{Conroy} C., {van Dokkum} P., 2012{\natexlab{a}}, \apj, 747, 69

\bibitem[{{Conroy} \& {van Dokkum}(2012{\natexlab{b}}){Conroy} C., {van Dokkum}
  P.~G.}]{Conroy2012}
{Conroy} C., {van Dokkum} P.~G., 2012{\natexlab{b}}, \apj, 760, 71

\bibitem[{{C{\^o}t{\'e}} {et~al}\mbox{.}(2006){C{\^o}t{\'e}} P.
  {et~al.}}]{Cote2006}
{C{\^o}t{\'e}} P. {et~al.}, 2006, \apjs, 165, 57

\bibitem[{{Croton} {et~al}\mbox{.}(2006){Croton} D.~J. {et~al.}}]{Croton2006}
{Croton} D.~J. {et~al.}, 2006, \mnras, 365, 11

\bibitem[{{Daddi} {et~al}\mbox{.}(2010){Daddi} E. {et~al.}}]{Daddi2010}
{Daddi} E. {et~al.}, 2010, \apj, 713, 686

\bibitem[{{Daddi} {et~al}\mbox{.}(2005){Daddi} E. {et~al.}}]{Daddi2005}
{Daddi} E. {et~al.}, 2005, \apj, 626, 680

\bibitem[{{Dav{\'e}}(2008){Dav{\'e}} R.}]{Dave2008}
{Dav{\'e}} R., 2008, \mnras, 385, 147

\bibitem[{{Davies} {et~al}\mbox{.}(1983){Davies} R.~L. {et~al.}}]{Davies1983}
{Davies} R.~L., {Efstathiou} G., {Fall} S.~M., {Illingworth} G., {Schechter}
  P.~L., 1983, \apj, 266, 41

\bibitem[{{Davies}, {Sadler} \& {Peletier}(1993){Davies} R.~L., {Sadler} E.~M.,
  {Peletier} R.~F.}]{Davies1993}
{Davies} R.~L., {Sadler} E.~M., {Peletier} R.~F., 1993, \mnras, 262, 650

\bibitem[{{de Rijcke} {et~al}\mbox{.}(2005){de Rijcke} S.
  {et~al.}}]{deRijcke2005}
{de Rijcke} S., {Michielsen} D., {Dejonghe} H., {Zeilinger} W.~W., {Hau}
  G.~K.~T., 2005, \aap, 438, 491

\bibitem[{{de Vaucouleurs}(1959){de Vaucouleurs} G.}]{deVaucouleurs1959}
{de Vaucouleurs} G., 1959, Handbuch der Physik, 53, 311

\bibitem[{{de Zeeuw} {et~al}\mbox{.}(2002){de Zeeuw} P.~T.
  {et~al.}}]{deZeeuw2002}
{de Zeeuw} P.~T. {et~al.}, 2002, \mnras, 329, 513

\bibitem[{{Deason} {et~al}\mbox{.}(2012){Deason} A.~J. {et~al.}}]{Deason2012dm}
{Deason} A.~J., {Belokurov} V., {Evans} N.~W., {McCarthy} I.~G., 2012, \apj,
  748, 2

\bibitem[{{Dekel} \& {Birnboim}(2006){Dekel} A., {Birnboim} Y.}]{Dekel2006}
{Dekel} A., {Birnboim} Y., 2006, \mnras, 368, 2

\bibitem[{{Dekel} {et~al}\mbox{.}(2009){Dekel} A. {et~al.}}]{Dekel2009}
{Dekel} A. {et~al.}, 2009, \nat, 457, 451

\bibitem[{{Dekel}, {Sari} \& {Ceverino}(2009){Dekel} A., {Sari} R., {Ceverino}
  D.}]{Dekel2009clumps}
{Dekel} A., {Sari} R., {Ceverino} D., 2009, \apj, 703, 785

\bibitem[{{Dekel} \& {Silk}(1986){Dekel} A., {Silk} J.}]{Dekel1986}
{Dekel} A., {Silk} J., 1986, \apj, 303, 39

\bibitem[{{Di Matteo}, {Springel} \& {Hernquist}(2005){Di Matteo} T.,
  {Springel} V., {Hernquist} L.}]{DiMatteo2005}
{Di Matteo} T., {Springel} V., {Hernquist} L., 2005, \nat, 433, 604

\bibitem[{{Djorgovski} \& {Davis}(1987){Djorgovski} S., {Davis}
  M.}]{Djorgovski1987}
{Djorgovski} S., {Davis} M., 1987, \apj, 313, 59

\bibitem[{{D'Onofrio} {et~al}\mbox{.}(2008){D'Onofrio} M.
  {et~al.}}]{dOnofrio2008}
{D'Onofrio} M. {et~al.}, 2008, \apj, 685, 875

\bibitem[{{Dressler}(1980){Dressler} A.}]{Dressler1980}
{Dressler} A., 1980, \apj, 236, 351

\bibitem[{{Dressler} {et~al}\mbox{.}(1987){Dressler} A.
  {et~al.}}]{Dressler1987}
{Dressler} A., {Lynden-Bell} D., {Burstein} D., {Davies} R.~L., {Faber} S.~M.,
  {Terlevich} R., {Wegner} G., 1987, \apj, 313, 42

\bibitem[{{Driver} {et~al}\mbox{.}(2011){Driver} S.~P. {et~al.}}]{Driver2011}
{Driver} S.~P. {et~al.}, 2011, \mnras, 413, 971

\bibitem[{{Dutton} {et~al}\mbox{.}(2011{\natexlab{a}}){Dutton} A.~A.
  {et~al.}}]{Dutton2011swells}
{Dutton} A.~A. {et~al.}, 2011{\natexlab{a}}, \mnras, 417, 1621

\bibitem[{{Dutton} {et~al}\mbox{.}(2011{\natexlab{b}}){Dutton} A.~A.
  {et~al.}}]{Dutton2011imf}
{Dutton} A.~A. {et~al.}, 2011{\natexlab{b}}, \mnras, 416, 322

\bibitem[{{Dutton} {et~al}\mbox{.}(2012){Dutton} A.~A.
  {et~al.}}]{Dutton2012imf_fp}
{Dutton} A.~A., {Maccio'} A.~V., {Mendel} J.~T., {Simard} L., 2012, \mnras\
  submitted (ArXiv:1204.2825)

\bibitem[{{Dutton}, {Mendel} \& {Simard}(2012){Dutton} A.~A., {Mendel} J.~T.,
  {Simard} L.}]{Dutton2012}
{Dutton} A.~A., {Mendel} J.~T., {Simard} L., 2012, \mnras, 422, L33

\bibitem[{{Dutton} {et~al}\mbox{.}(2013){Dutton} A.~A.
  {et~al.}}]{Dutton2012imf_bulges}
{Dutton} A.~A. {et~al.}, 2013, \mnras, 428, 3183

\bibitem[{{Elmegreen} {et~al}\mbox{.}(2007){Elmegreen} D.~M.
  {et~al.}}]{Elmegreen2007}
{Elmegreen} D.~M., {Elmegreen} B.~G., {Ravindranath} S., {Coe} D.~A., 2007,
  \apj, 658, 763

\bibitem[{{Emsellem} {et~al}\mbox{.}(2007){Emsellem} E.
  {et~al.}}]{Emsellem2007}
{Emsellem} E. {et~al.}, 2007, \mnras, 379, 401

\bibitem[{{Emsellem} {et~al}\mbox{.}(2004){Emsellem} E.
  {et~al.}}]{Emsellem2004}
{Emsellem} E. {et~al.}, 2004, \mnras, 352, 721

\bibitem[{{Emsellem}, {Monnet} \& {Bacon}(1994){Emsellem} E., {Monnet} G.,
  {Bacon} R.}]{Emsellem1994}
{Emsellem} E., {Monnet} G., {Bacon} R., 1994, \aap, 285, 723

\bibitem[{{Faber} {et~al}\mbox{.}(1997){Faber} S.~M. {et~al.}}]{Faber1997}
{Faber} S.~M. {et~al.}, 1997, \aj, 114, 1771

\bibitem[{{Faber} {et~al}\mbox{.}(1987){Faber} S.~M. {et~al.}}]{Faber1987}
{Faber} S.~M., {Dressler} A., {Davies} R.~L., {Burstein} D., {Lynden-Bell} D.,
  1987, in Nearly Normal Galaxies. From the Planck Time to the Present, Faber
  S.~M., ed., pp. 175--183

\bibitem[{{Faber} \& {Jackson}(1976){Faber} S.~M., {Jackson} R.~E.}]{Faber1976}
{Faber} S.~M., {Jackson} R.~E., 1976, \apj, 204, 668

\bibitem[{{Falc{\'o}n-Barroso} {et~al}\mbox{.}(2011){Falc{\'o}n-Barroso} J.
  {et~al.}}]{FalconBarroso2011miles}
{Falc{\'o}n-Barroso} J., {S{\'a}nchez-Bl{\'a}zquez} P., {Vazdekis} A.,
  {Ricciardelli} E., {Cardiel} N., {Cenarro} A.~J., {Gorgas} J., {Peletier}
  R.~F., 2011, \aap, 532, A95

\bibitem[{{Ferrarese}(2012){Ferrarese} L.}]{Ferrarese2012eso}
{Ferrarese} L., 2012, in Paper presented at the conference on Science from the
  Next Generation Imaging and Spectroscopic Surveys held 15-18 October, 2012 at
  ESO Garching, Germany. Online at
  http://www.eso.org/sci/meetings/2012/surveys2012/program.html

\bibitem[{{Ferrarese} {et~al}\mbox{.}(2012){Ferrarese} L.
  {et~al.}}]{Ferrarese2012}
{Ferrarese} L. {et~al.}, 2012, \apjs, 200, 4

\bibitem[{{Ferrarese} {et~al}\mbox{.}(2006{\natexlab{a}}){Ferrarese} L.
  {et~al.}}]{Ferrarese2006}
{Ferrarese} L. {et~al.}, 2006{\natexlab{a}}, \apjl, 644, L21

\bibitem[{{Ferrarese} {et~al}\mbox{.}(2006{\natexlab{b}}){Ferrarese} L.
  {et~al.}}]{Ferrarese2006acs}
{Ferrarese} L. {et~al.}, 2006{\natexlab{b}}, \apjs, 164, 334

\bibitem[{{Ferrarese} \& {Merritt}(2000){Ferrarese} L., {Merritt}
  D.}]{Ferrarese2000}
{Ferrarese} L., {Merritt} D., 2000, \apjl, 539, L9

\bibitem[{{Ferreras} {et~al}\mbox{.}(2013){Ferreras} I.
  {et~al.}}]{Ferreras2012}
{Ferreras} I., {La Barbera} F., {de la Rosa} I.~G., {Vazdekis} A., {de
  Carvalho} R.~R., {Falc{\'o}n-Barroso} J., {Ricciardelli} E., 2013, \mnras,
  429, L15

\bibitem[{{Ferreras}, {Saha} \& {Burles}(2008){Ferreras} I., {Saha} P.,
  {Burles} S.}]{Ferreras2008}
{Ferreras} I., {Saha} P., {Burles} S., 2008, \mnras, 383, 857

\bibitem[{{Ferreras} {et~al}\mbox{.}(2010){Ferreras} I.
  {et~al.}}]{Ferreras2010}
{Ferreras} I., {Saha} P., {Leier} D., {Courbin} F., {Falco} E.~E., 2010,
  \mnras, 409, L30

\bibitem[{{Forbes} {et~al}\mbox{.}(2008){Forbes} D.~A. {et~al.}}]{Forbes2008}
{Forbes} D.~A., {Lasky} P., {Graham} A.~W., {Spitler} L., 2008, \mnras, 389,
  1924

\bibitem[{{Forbes} {et~al}\mbox{.}(2011){Forbes} D.~A. {et~al.}}]{Forbes2011}
{Forbes} D.~A., {Spitler} L.~R., {Graham} A.~W., {Foster} C., {Hau} G.~K.~T.,
  {Benson} A., 2011, \mnras, 413, 2665

\bibitem[{{F{\"o}rster Schreiber} {et~al}\mbox{.}(2009){F{\"o}rster Schreiber}
  N.~M. {et~al.}}]{ForsterSchreiber2009}
{F{\"o}rster Schreiber} N.~M. {et~al.}, 2009, \apj, 706, 1364

\bibitem[{{F{\"o}rster Schreiber} {et~al}\mbox{.}(2006){F{\"o}rster Schreiber}
  N.~M. {et~al.}}]{ForsterSchreiber2006}
{F{\"o}rster Schreiber} N.~M. {et~al.}, 2006, \apj, 645, 1062

\bibitem[{{Franx} {et~al}\mbox{.}(2008){Franx} M. {et~al.}}]{Franx2008}
{Franx} M., {van Dokkum} P.~G., {Schreiber} N.~M.~F., {Wuyts} S., {Labb{\'e}}
  I., {Toft} S., 2008, \apj, 688, 770

\bibitem[{{Freeman}(1970){Freeman} K.~C.}]{Freeman1970}
{Freeman} K.~C., 1970, \apj, 160, 811

\bibitem[{{Gallazzi} \& {Bell}(2009){Gallazzi} A., {Bell} E.~F.}]{Gallazzi2009}
{Gallazzi} A., {Bell} E.~F., 2009, \apjs, 185, 253

\bibitem[{{Gallazzi} {et~al}\mbox{.}(2006){Gallazzi} A.
  {et~al.}}]{Gallazzi2006}
{Gallazzi} A., {Charlot} S., {Brinchmann} J., {White} S.~D.~M., 2006, \mnras,
  370, 1106

\bibitem[{{Gargiulo} {et~al}\mbox{.}(2009){Gargiulo} A.
  {et~al.}}]{Gargiulo2009}
{Gargiulo} A. {et~al.}, 2009, \mnras, 397, 75

\bibitem[{{Gavazzi} {et~al}\mbox{.}(2005){Gavazzi} G. {et~al.}}]{Gavazzi2005}
{Gavazzi} G., {Donati} A., {Cucciati} O., {Sabatini} S., {Boselli} A., {Davies}
  J., {Zibetti} S., 2005, \aap, 430, 411

\bibitem[{{Gavazzi} {et~al}\mbox{.}(2007){Gavazzi} R. {et~al.}}]{Gavazzi2007}
{Gavazzi} R., {Treu} T., {Rhodes} J.~D., {Koopmans} L.~V.~E., {Bolton} A.~S.,
  {Burles} S., {Massey} R.~J., {Moustakas} L.~A., 2007, \apj, 667, 176

\bibitem[{{Gebhardt} {et~al}\mbox{.}(2000){Gebhardt} K.
  {et~al.}}]{Gebhardt2000bh}
{Gebhardt} K. {et~al.}, 2000, \apjl, 539, L13

\bibitem[{{Geha} {et~al}\mbox{.}(2012){Geha} M. {et~al.}}]{Geha2012}
{Geha} M., {Blanton} M.~R., {Yan} R., {Tinker} J.~L., 2012, \apj, 757, 85

\bibitem[{Gelman {et~al}\mbox{.}(2004)Gelman A. {et~al.}}]{gelman2004bayesian}
Gelman A., Carlin J., Stern H., Rubin D., 2004, Bayesian data analysis. CRC
  press

\bibitem[{{Genel} {et~al}\mbox{.}(2012){Genel} S. {et~al.}}]{Genel2012}
{Genel} S. {et~al.}, 2012, \apj, 745, 11

\bibitem[{{Genzel} {et~al}\mbox{.}(2008){Genzel} R. {et~al.}}]{Genzel2008}
{Genzel} R. {et~al.}, 2008, \apj, 687, 59

\bibitem[{{Genzel} {et~al}\mbox{.}(2011){Genzel} R. {et~al.}}]{Genzel2011}
{Genzel} R. {et~al.}, 2011, \apj, 733, 101

\bibitem[{{Genzel} {et~al}\mbox{.}(2006){Genzel} R. {et~al.}}]{Genzel2006}
{Genzel} R. {et~al.}, 2006, \nat, 442, 786

\bibitem[{{Gerhard} {et~al}\mbox{.}(2001){Gerhard} O. {et~al.}}]{Gerhard2001}
{Gerhard} O., {Kronawitter} A., {Saglia} R.~P., {Bender} R., 2001, \aj, 121,
  1936

\bibitem[{{Gerhard} \& {Binney}(1996){Gerhard} O.~E., {Binney}
  J.~J.}]{Gerhard1996}
{Gerhard} O.~E., {Binney} J.~J., 1996, \mnras, 279, 993

\bibitem[{{Gnedin} {et~al}\mbox{.}(2011){Gnedin} O.~Y. {et~al.}}]{Gnedin2011}
{Gnedin} O.~Y., {Ceverino} D., {Gnedin} N.~Y., {Klypin} A.~A., {Kravtsov}
  A.~V., {Levine} R., {Nagai} D., {Yepes} G., 2011, \apj\ submitted
  (ArXiv:1108.5736)

\bibitem[{{Gnedin} {et~al}\mbox{.}(2004){Gnedin} O.~Y. {et~al.}}]{Gnedin2004}
{Gnedin} O.~Y., {Kravtsov} A.~V., {Klypin} A.~A., {Nagai} D., 2004, \apj, 616,
  16

\bibitem[{{Governato} {et~al}\mbox{.}(2010){Governato} F.
  {et~al.}}]{Governato2010}
{Governato} F. {et~al.}, 2010, \nat, 463, 203

\bibitem[{{Graham}(2012){Graham} A.~W.}]{Graham2012}
{Graham} A.~W., 2012, \apj, 746, 113

\bibitem[{{Graham}(2013){Graham} A.~W.}]{Graham2011}
{Graham} A.~W., 2013, in Planets, stars and stellar systems, {Oswalt} T.~D.,
  {Keel} W.~C., eds., Vol.~6, (Berlin: Springer), pp. 91--139 (arXiv:1108.0997)

\bibitem[{{Graham} \& {Guzm{\'a}n}(2003){Graham} A.~W., {Guzm{\'a}n}
  R.}]{Graham2003}
{Graham} A.~W., {Guzm{\'a}n} R., 2003, \aj, 125, 2936

\bibitem[{{Graham} \& {Worley}(2008){Graham} A.~W., {Worley}
  C.~C.}]{Graham2008curv}
{Graham} A.~W., {Worley} C.~C., 2008, \mnras, 388, 1708

\bibitem[{{Granato} {et~al}\mbox{.}(2004){Granato} G.~L.
  {et~al.}}]{Granato2004}
{Granato} G.~L., {De Zotti} G., {Silva} L., {Bressan} A., {Danese} L., 2004,
  \apj, 600, 580

\bibitem[{{Graves} \& {Faber}(2010){Graves} G.~J., {Faber} S.~M.}]{Graves2010}
{Graves} G.~J., {Faber} S.~M., 2010, \apj, 717, 803

\bibitem[{{Graves}, {Faber} \& {Schiavon}(2009){Graves} G.~J., {Faber} S.~M.,
  {Schiavon} R.~P.}]{Graves2009b}
{Graves} G.~J., {Faber} S.~M., {Schiavon} R.~P., 2009, \apj, 698, 1590

\bibitem[{{Grillo} \& {Gobat}(2010){Grillo} C., {Gobat} R.}]{Grillo2010}
{Grillo} C., {Gobat} R., 2010, \mnras, 402, L67

\bibitem[{{Grillo} {et~al}\mbox{.}(2009){Grillo} C. {et~al.}}]{Grillo2009}
{Grillo} C., {Gobat} R., {Lombardi} M., {Rosati} P., 2009, \aap, 501, 461

\bibitem[{{Gunawardhana} {et~al}\mbox{.}(2011){Gunawardhana} M.~L.~P.
  {et~al.}}]{Gunawardhana2011}
{Gunawardhana} M.~L.~P. {et~al.}, 2011, \mnras, 415, 1647

\bibitem[{{Heavens} {et~al}\mbox{.}(2004){Heavens} A. {et~al.}}]{Heavens2004}
{Heavens} A., {Panter} B., {Jimenez} R., {Dunlop} J., 2004, \nat, 428, 625

\bibitem[{{Held} {et~al}\mbox{.}(1992){Held} E.~V. {et~al.}}]{Held1992}
{Held} E.~V., {de Zeeuw} T., {Mould} J., {Picard} A., 1992, \aj, 103, 851

\bibitem[{{Hernquist}(1990){Hernquist} L.}]{Hernquist1990}
{Hernquist} L., 1990, \apj, 356, 359

\bibitem[{{Hernquist}, {Spergel} \& {Heyl}(1993){Hernquist} L., {Spergel}
  D.~N., {Heyl} J.~S.}]{Hernquist1993}
{Hernquist} L., {Spergel} D.~N., {Heyl} J.~S., 1993, \apj, 416, 415

\bibitem[{{Hilz}, {Naab} \& {Ostriker}(2013){Hilz} M., {Naab} T., {Ostriker}
  J.~P.}]{Hilz2012b}
{Hilz} M., {Naab} T., {Ostriker} J.~P., 2013, \mnras, 429, 2924

\bibitem[{{Hilz} {et~al}\mbox{.}(2012){Hilz} M. {et~al.}}]{Hilz2012a}
{Hilz} M., {Naab} T., {Ostriker} J.~P., {Thomas} J., {Burkert} A., {Jesseit}
  R., 2012, \mnras, 425, 3119

\bibitem[{{Hopkins} {et~al}\mbox{.}(2010){Hopkins} P.~F.
  {et~al.}}]{Hopkins2010}
{Hopkins} P.~F., {Bundy} K., {Hernquist} L., {Wuyts} S., {Cox} T.~J., 2010,
  \mnras, 401, 1099

\bibitem[{{Hopkins} {et~al}\mbox{.}(2009{\natexlab{a}}){Hopkins} P.~F.
  {et~al.}}]{Hopkins2009compact}
{Hopkins} P.~F., {Bundy} K., {Murray} N., {Quataert} E., {Lauer} T.~R., {Ma}
  C.-P., 2009{\natexlab{a}}, \mnras, 398, 898

\bibitem[{{Hopkins} {et~al}\mbox{.}(2006){Hopkins} P.~F.
  {et~al.}}]{Hopkins2006}
{Hopkins} P.~F., {Hernquist} L., {Cox} T.~J., {Di Matteo} T., {Robertson} B.,
  {Springel} V., 2006, \apjs, 163, 1

\bibitem[{{Hopkins} {et~al}\mbox{.}(2009{\natexlab{b}}){Hopkins} P.~F.
  {et~al.}}]{Hopkins2009extralight}
{Hopkins} P.~F., {Hernquist} L., {Cox} T.~J., {Keres} D., {Wuyts} S.,
  2009{\natexlab{b}}, \apj, 691, 1424

\bibitem[{{Hopkins} {et~al}\mbox{.}(2012){Hopkins} P.~F.
  {et~al.}}]{Hopkins2011}
{Hopkins} P.~F., {Kere{\v s}} D., {Murray} N., {Quataert} E., {Hernquist} L.,
  2012, \mnras, 427, 968

\bibitem[{{Hoversten} \& {Glazebrook}(2008){Hoversten} E.~A., {Glazebrook}
  K.}]{Hoversten2008}
{Hoversten} E.~A., {Glazebrook} K., 2008, \apj, 675, 163

\bibitem[{{Hubble}(1936){Hubble} E.~P.}]{Hubble1936}
{Hubble} E.~P., 1936, Realm of the Nebulae. Yale Univ. Press, New Haven

\bibitem[{{Huchra} {et~al}\mbox{.}(1985){Huchra} J. {et~al.}}]{Huchra1985}
{Huchra} J., {Gorenstein} M., {Kent} S., {Shapiro} I., {Smith} G., {Horine} E.,
  {Perley} R., 1985, \aj, 90, 691

\bibitem[{{Huchra} {et~al}\mbox{.}(2012){Huchra} J.~P. {et~al.}}]{Huchra2012}
{Huchra} J.~P. {et~al.}, 2012, \apjs, 199, 26

\bibitem[{{Hyde} \& {Bernardi}(2009{\natexlab{a}}){Hyde} J.~B., {Bernardi}
  M.}]{Hyde2009curv}
{Hyde} J.~B., {Bernardi} M., 2009{\natexlab{a}}, \mnras, 394, 1978

\bibitem[{{Hyde} \& {Bernardi}(2009{\natexlab{b}}){Hyde} J.~B., {Bernardi}
  M.}]{Hyde2009fp}
{Hyde} J.~B., {Bernardi} M., 2009{\natexlab{b}}, \mnras, 396, 1171

\bibitem[{{Jaffe}(1983){Jaffe} W.}]{Jaffe1983}
{Jaffe} W., 1983, \mnras, 202, 995

\bibitem[{{Jahnke} \& {Macci{\`o}}(2011){Jahnke} K., {Macci{\`o}}
  A.~V.}]{Jahnke2011}
{Jahnke} K., {Macci{\`o}} A.~V., 2011, \apj, 734, 92

\bibitem[{{Janz} \& {Lisker}(2008){Janz} J., {Lisker} T.}]{Janz2008}
{Janz} J., {Lisker} T., 2008, \apjl, 689, L25

\bibitem[{{Janz} \& {Lisker}(2009){Janz} J., {Lisker} T.}]{Janz2009}
{Janz} J., {Lisker} T., 2009, \apjl, 696, L102

\bibitem[{{Jiang} {et~al}\mbox{.}(2012){Jiang} F. {et~al.}}]{Jiang2012}
{Jiang} F., {van Dokkum} P., {Bezanson} R., {Franx} M., 2012, \apjl, 749, L10

\bibitem[{{Kassin}, {de Jong} \& {Weiner}(2006){Kassin} S.~A., {de Jong} R.~S.,
  {Weiner} B.~J.}]{Kassin2006}
{Kassin} S.~A., {de Jong} R.~S., {Weiner} B.~J., 2006, \apj, 643, 804

\bibitem[{{Kauffmann} {et~al}\mbox{.}(2003{\natexlab{a}}){Kauffmann} G.
  {et~al.}}]{Kauffmann2003mass}
{Kauffmann} G. {et~al.}, 2003{\natexlab{a}}, \mnras, 341, 33

\bibitem[{{Kauffmann} {et~al}\mbox{.}(2003{\natexlab{b}}){Kauffmann} G.
  {et~al.}}]{Kauffmann2003sfh}
{Kauffmann} G. {et~al.}, 2003{\natexlab{b}}, \mnras, 341, 54

\bibitem[{{Kelly}(2007){Kelly} B.~C.}]{Kelly2007}
{Kelly} B.~C., 2007, \apj, 665, 1489

\bibitem[{{Kelson} {et~al}\mbox{.}(1997){Kelson} D.~D. {et~al.}}]{Kelson1997}
{Kelson} D.~D., {van Dokkum} P.~G., {Franx} M., {Illingworth} G.~D.,
  {Fabricant} D., 1997, \apjl, 478, L13

\bibitem[{{Kennicutt}(1998){Kennicutt}, Jr. R.~C.}]{Kennicutt1998imf}
{Kennicutt}, Jr. R.~C., 1998, in Astronomical Society of the Pacific Conference
  Series, Vol. 142, The Stellar Initial Mass Function (38th Herstmonceux
  Conference), Howell G.~G. .~D., ed., p.~1

\bibitem[{{Kere{\v s}} {et~al}\mbox{.}(2005){Kere{\v s}} D.
  {et~al.}}]{Keres2005}
{Kere{\v s}} D., {Katz} N., {Weinberg} D.~H., {Dav{\'e}} R., 2005, \mnras, 363,
  2

\bibitem[{{Khochfar} {et~al}\mbox{.}(2011){Khochfar} S.
  {et~al.}}]{Khochfar2011}
{Khochfar} S. {et~al.}, 2011, \mnras, 417, 845 (Paper~VIII)

\bibitem[{{Khochfar} \& {Ostriker}(2008){Khochfar} S., {Ostriker}
  J.~P.}]{Khochfar2008}
{Khochfar} S., {Ostriker} J.~P., 2008, \apj, 680, 54

\bibitem[{{Khochfar} \& {Silk}(2006){Khochfar} S., {Silk} J.}]{Khochfar2006b}
{Khochfar} S., {Silk} J., 2006, \mnras, 370, 902

\bibitem[{{Kirby} {et~al}\mbox{.}(2008){Kirby} E.~M. {et~al.}}]{Kirby2008}
{Kirby} E.~M., {Jerjen} H., {Ryder} S.~D., {Driver} S.~P., 2008, \aj, 136, 1866

\bibitem[{{Koopmans} {et~al}\mbox{.}(2006){Koopmans} L.~V.~E.
  {et~al.}}]{Koopmans2006}
{Koopmans} L.~V.~E., {Treu} T., {Bolton} A.~S., {Burles} S., {Moustakas} L.~A.,
  2006, \apj, 649, 599

\bibitem[{{Kormendy}(1977){Kormendy} J.}]{Kormendy1977}
{Kormendy} J., 1977, \apj, 218, 333

\bibitem[{{Kormendy}(1985){Kormendy} J.}]{Kormendy1985}
{Kormendy} J., 1985, \apj, 295, 73

\bibitem[{{Kormendy} \& {Bender}(2012){Kormendy} J., {Bender}
  R.}]{Kormendy2012}
{Kormendy} J., {Bender} R., 2012, \apjs, 198, 2

\bibitem[{{Kormendy} {et~al}\mbox{.}(2009){Kormendy} J.
  {et~al.}}]{Kormendy2009}
{Kormendy} J., {Fisher} D.~B., {Cornell} M.~E., {Bender} R., 2009, \apjs, 182,
  216

\bibitem[{{Kormendy} \& {Gebhardt}(2001){Kormendy} J., {Gebhardt}
  K.}]{Kormendy2001}
{Kormendy} J., {Gebhardt} K., 2001, in American Institute of Physics Conference
  Series, Vol. 586, 20th Texas Symposium on relativistic astrophysics,
  {Wheeler} J.~C., {Martel} H., eds., pp. 363--+

\bibitem[{{Kormendy} \& {Kennicutt}(2004){Kormendy} J., {Kennicutt}, Jr.
  R.~C.}]{Kormendy2004}
{Kormendy} J., {Kennicutt}, Jr. R.~C., 2004, \araa, 42, 603

\bibitem[{{Krajnovi{\'c}} {et~al}\mbox{.}(2013){Krajnovi{\'c}} D.
  {et~al.}}]{Krajnovic2012}
{Krajnovi{\'c}} D. {et~al.}, 2013, \mnras\ in press (arXiv:1210.8167),
  (Paper~XVII)

\bibitem[{{Krajnovi\'c} {et~al}\mbox{.}(2013){Krajnovi\'c} D.
  {et~al.}}]{Krajnovic2013}
{Krajnovi\'c} D. {et~al.}, 2013, \mnras\ submitted, (Paper~XXIV)

\bibitem[{{Kroupa}(2001){Kroupa} P.}]{Kroupa2001}
{Kroupa} P., 2001, \mnras, 322, 231

\bibitem[{{Kroupa}(2002){Kroupa} P.}]{Kroupa2002}
{Kroupa} P., 2002, Science, 295, 82

\bibitem[{{Kroupa} \& {Weidner}(2003){Kroupa} P., {Weidner} C.}]{Kroupa2003}
{Kroupa} P., {Weidner} C., 2003, \apj, 598, 1076

\bibitem[{Kroupa {et~al}\mbox{.}(2013)Kroupa P. {et~al.}}]{Kroupa2012}
Kroupa P., Weidner C., Pflamm-Altenburg J., Thies I., Dabringhausen J., Marks
  M., Maschberger T., 2013, in Planets, stars and stellar systems, {Oswalt}
  T.~D., {Gilmore} G., eds., Vol.~5, (Berlin: Springer), pp. 115--242
  (arXiv:1112.3340)

\bibitem[{{Kuntschner} {et~al}\mbox{.}(2006){Kuntschner} H.
  {et~al.}}]{Kuntschner2006}
{Kuntschner} H. {et~al.}, 2006, \mnras, 369, 497

\bibitem[{{Kuntschner}, {Emsellem} \& {et al.}(2010){Kuntschner} H., {Emsellem}
  E., {et al.}}]{Kuntschner2010}
{Kuntschner} H., {Emsellem} E., {et al.}, 2010, \mnras, 408, 97

\bibitem[{{Lablanche} {et~al}\mbox{.}(2012){Lablanche} P.-Y.
  {et~al.}}]{Lablanche2012}
{Lablanche} P.-Y. {et~al.}, 2012, \mnras, 424, 1495 (Paper~XII)

\bibitem[{{Lauer} {et~al}\mbox{.}(2007{\natexlab{a}}){Lauer} T.~R.
  {et~al.}}]{lauer07prof}
{Lauer} T.~R. {et~al.}, 2007{\natexlab{a}}, \apj, 664, 226

\bibitem[{{Lauer} {et~al}\mbox{.}(2007{\natexlab{b}}){Lauer} T.~R.
  {et~al.}}]{Lauer2007}
{Lauer} T.~R. {et~al.}, 2007{\natexlab{b}}, \apj, 662, 808

\bibitem[{{Law} {et~al}\mbox{.}(2012){Law} D.~R. {et~al.}}]{Law2012}
{Law} D.~R., {Shapley} A.~E., {Steidel} C.~C., {Reddy} N.~A., {Christensen}
  C.~R., {Erb} D.~K., 2012, \nat, 487, 338

\bibitem[{{Longhetti} \& {Saracco}(2009){Longhetti} M., {Saracco}
  P.}]{Longhetti2009}
{Longhetti} M., {Saracco} P., 2009, \mnras, 394, 774

\bibitem[{{Magorrian} {et~al}\mbox{.}(1998){Magorrian} J.
  {et~al.}}]{Magorrian1998}
{Magorrian} J. {et~al.}, 1998, \aj, 115, 2285

\bibitem[{{Mancini} {et~al}\mbox{.}(2010){Mancini} C. {et~al.}}]{Mancini2010}
{Mancini} C. {et~al.}, 2010, \mnras, 401, 933

\bibitem[{{Maraston}(2005){Maraston} C.}]{Maraston2005}
{Maraston} C., 2005, \mnras, 362, 799

\bibitem[{{Maraston} {et~al}\mbox{.}(2006){Maraston} C.
  {et~al.}}]{Maraston2006}
{Maraston} C., {Daddi} E., {Renzini} A., {Cimatti} A., {Dickinson} M.,
  {Papovich} C., {Pasquali} A., {Pirzkal} N., 2006, \apj, 652, 85

\bibitem[{{Martig} {et~al}\mbox{.}(2009){Martig} M. {et~al.}}]{Martig2009}
{Martig} M., {Bournaud} F., {Teyssier} R., {Dekel} A., 2009, \apj, 707, 250

\bibitem[{{Matkovi{\'c}} \& {Guzm{\'a}n}(2005){Matkovi{\'c}} A., {Guzm{\'a}n}
  R.}]{Matkovic2005}
{Matkovi{\'c}} A., {Guzm{\'a}n} R., 2005, \mnras, 362, 289

\bibitem[{{McConnell} {et~al}\mbox{.}(2011){McConnell} N.~J.
  {et~al.}}]{McConnell2011}
{McConnell} N.~J., {Ma} C.-P., {Gebhardt} K., {Wright} S.~A., {Murphy} J.~D.,
  {Lauer} T.~R., {Graham} J.~R., {Richstone} D.~O., 2011, \nat, 480, 215

\bibitem[{{Mei} {et~al}\mbox{.}(2007){Mei} S. {et~al.}}]{Mei2007}
{Mei} S. {et~al.}, 2007, \apj, 655, 144

\bibitem[{{Meurer} {et~al}\mbox{.}(2009){Meurer} G.~R. {et~al.}}]{Meurer2009}
{Meurer} G.~R. {et~al.}, 2009, \apj, 695, 765

\bibitem[{{Mihos} \& {Hernquist}(1994){Mihos} J.~C., {Hernquist}
  L.}]{Mihos1994burst}
{Mihos} J.~C., {Hernquist} L., 1994, \apjl, 431, L9

\bibitem[{{Misgeld} \& {Hilker}(2011){Misgeld} I., {Hilker} M.}]{Misgeld2011}
{Misgeld} I., {Hilker} M., 2011, \mnras, 414, 3699

\bibitem[{{Misgeld}, {Hilker} \& {Mieske}(2009){Misgeld} I., {Hilker} M.,
  {Mieske} S.}]{Misgeld2009}
{Misgeld} I., {Hilker} M., {Mieske} S., 2009, \aap, 496, 683

\bibitem[{{Misgeld}, {Mieske} \& {Hilker}(2008){Misgeld} I., {Mieske} S.,
  {Hilker} M.}]{Misgeld2008}
{Misgeld} I., {Mieske} S., {Hilker} M., 2008, \aap, 486, 697

\bibitem[{{Moster} {et~al}\mbox{.}(2010){Moster} B.~P. {et~al.}}]{Moster2010}
{Moster} B.~P., {Somerville} R.~S., {Maulbetsch} C., {van den Bosch} F.~C.,
  {Macci{\`o}} A.~V., {Naab} T., {Oser} L., 2010, \apj, 710, 903

\bibitem[{{Naab}, {Johansson} \& {Ostriker}(2009){Naab} T., {Johansson} P.~H.,
  {Ostriker} J.~P.}]{Naab2009}
{Naab} T., {Johansson} P.~H., {Ostriker} J.~P., 2009, \apjl, 699, L178

\bibitem[{{Napolitano}, {Romanowsky} \& {Tortora}(2010){Napolitano} N.~R.,
  {Romanowsky} A.~J., {Tortora} C.}]{Napolitano2010}
{Napolitano} N.~R., {Romanowsky} A.~J., {Tortora} C., 2010, \mnras, 405, 2351

\bibitem[{{Navarro}, {Frenk} \& {White}(1996){Navarro} J.~F., {Frenk} C.~S.,
  {White} S.~D.~M.}]{navarro96}
{Navarro} J.~F., {Frenk} C.~S., {White} S.~D.~M., 1996, \apj, 462, 563

\bibitem[{{Newman} {et~al}\mbox{.}(2012){Newman} A.~B. {et~al.}}]{Newman2012}
{Newman} A.~B., {Ellis} R.~S., {Bundy} K., {Treu} T., 2012, \apj, 746, 162

\bibitem[{{Nipoti}, {Londrillo} \& {Ciotti}(2003){Nipoti} C., {Londrillo} P.,
  {Ciotti} L.}]{Nipoti2003}
{Nipoti} C., {Londrillo} P., {Ciotti} L., 2003, \mnras, 342, 501

\bibitem[{{Nipoti} {et~al}\mbox{.}(2009){Nipoti} C. {et~al.}}]{Nipoti2009}
{Nipoti} C., {Treu} T., {Auger} M.~W., {Bolton} A.~S., 2009, \apjl, 706, L86

\bibitem[{{Onodera} {et~al}\mbox{.}(2012){Onodera} M. {et~al.}}]{Onodera2012}
{Onodera} M. {et~al.}, 2012, \apj, 755, 26

\bibitem[{{Oser} {et~al}\mbox{.}(2012){Oser} L. {et~al.}}]{Oser2012}
{Oser} L., {Naab} T., {Ostriker} J.~P., {Johansson} P.~H., 2012, \apj, 744, 63

\bibitem[{{Oser} {et~al}\mbox{.}(2010){Oser} L. {et~al.}}]{Oser2010}
{Oser} L., {Ostriker} J.~P., {Naab} T., {Johansson} P.~H., {Burkert} A., 2010,
  \apj, 725, 2312

\bibitem[{{Padmanabhan} {et~al}\mbox{.}(2004){Padmanabhan} N.
  {et~al.}}]{Padmanabhan2004}
{Padmanabhan} N. {et~al.}, 2004, \na, 9, 329

\bibitem[{{Paturel} {et~al}\mbox{.}(2003){Paturel} G. {et~al.}}]{Paturel2003}
{Paturel} G., {Petit} C., {Prugniel} P., {Theureau} G., {Rousseau} J., {Brouty}
  M., {Dubois} P., {Cambr{\'e}sy} L., 2003, \aap, 412, 45

\bibitem[{{Peng}(2007){Peng} C.~Y.}]{Peng2007}
{Peng} C.~Y., 2007, \apj, 671, 1098

\bibitem[{{Press} {et~al}\mbox{.}(1992){Press} W.~H. {et~al.}}]{Press1992}
{Press} W.~H., {Teukolsky} S.~A., {Vetterling} W.~T., {Flannery} B.~P., 1992,
  Numerical recipes in FORTRAN. The art of scientific computing. Cambridge:
  University Press, |c1992, 2nd ed.

\bibitem[{{Prugniel} \& {Simien}(1996){Prugniel} P., {Simien}
  F.}]{Prugniel1996fp}
{Prugniel} P., {Simien} F., 1996, \aap, 309, 749

\bibitem[{{Renzini}(2005){Renzini} A.}]{Renzini2005}
{Renzini} A., 2005, in Astrophysics and Space Science Library, Vol. 327, The
  Initial Mass Function 50 Years Later, {Corbelli} E., {Palla} F., {Zinnecker}
  H., eds., p. 221

\bibitem[{{Renzini} \& {Ciotti}(1993){Renzini} A., {Ciotti} L.}]{Renzini1993}
{Renzini} A., {Ciotti} L., 1993, \apjl, 416, L49+

\bibitem[{{Robertson} {et~al}\mbox{.}(2006){Robertson} B.
  {et~al.}}]{Robertson2006fp}
{Robertson} B., {Cox} T.~J., {Hernquist} L., {Franx} M., {Hopkins} P.~F.,
  {Martini} P., {Springel} V., 2006, \apj, 641, 21

\bibitem[{{Rybicki}(1987){Rybicki} G.~B.}]{Rybicki1987}
{Rybicki} G.~B., 1987, in IAU Symposium, Vol. 127, Structure and Dynamics of
  Elliptical Galaxies, {de Zeeuw} P.~T., ed., p. 397

\bibitem[{{Salpeter}(1955){Salpeter} E.~E.}]{Salpeter1955}
{Salpeter} E.~E., 1955, \apj, 121, 161

\bibitem[{{S{\'a}nchez-Bl{\'a}zquez}
  {et~al}\mbox{.}(2006){S{\'a}nchez-Bl{\'a}zquez} P.
  {et~al.}}]{Sanchez-Blazquez2006}
{S{\'a}nchez-Bl{\'a}zquez} P. {et~al.}, 2006, \mnras, 371, 703

\bibitem[{{Sandage}(1961){Sandage} A.}]{Sandage1961}
{Sandage} A., 1961, The Hubble Atlas. Carnegie Institution, Washington

\bibitem[{{Sarzi} {et~al}\mbox{.}(2013){Sarzi} M. {et~al.}}]{Sarzi2013}
{Sarzi} M. {et~al.}, 2013, \mnras\ in press (arXiv:1301.2589), (Paper~XIX)

\bibitem[{{Sarzi} {et~al}\mbox{.}(2010){Sarzi} M. {et~al.}}]{Sarzi2010}
{Sarzi} M. {et~al.}, 2010, \mnras, 402, 2187

\bibitem[{{Schechter}(1976){Schechter} P.}]{Schechter1976}
{Schechter} P., 1976, \apj, 203, 297

\bibitem[{{Schiavon}, {Barbuy} \& {Bruzual A.}(2000){Schiavon} R.~P., {Barbuy}
  B., {Bruzual A.} G.}]{Schiavon2000}
{Schiavon} R.~P., {Barbuy} B., {Bruzual A.} G., 2000, \apj, 532, 453

\bibitem[{{Schulz}, {Mandelbaum} \& {Padmanabhan}(2010){Schulz} A.~E.,
  {Mandelbaum} R., {Padmanabhan} N.}]{Schulz2010}
{Schulz} A.~E., {Mandelbaum} R., {Padmanabhan} N., 2010, \mnras, 408, 1463

\bibitem[{{Scott} {et~al}\mbox{.}(2009){Scott} N. {et~al.}}]{Scott2009}
{Scott} N. {et~al.}, 2009, \mnras, 398, 1835

\bibitem[{{Scott} {et~al}\mbox{.}(2013){Scott} N. {et~al.}}]{Scott2011}
{Scott} N. {et~al.}, 2013, \mnras\ in press (arXiv:1211.4615), (Paper~XXI)

\bibitem[{{Sersic}(1968){Sersic} J.~L.}]{Sersic1968}
{Sersic} J.~L., 1968, Atlas de galaxias australes. Cordoba, Argentina:
  Observatorio Astronomico, 1968

\bibitem[{{Shankar} \& {Bernardi}(2009){Shankar} F., {Bernardi}
  M.}]{Shankar2009}
{Shankar} F., {Bernardi} M., 2009, \mnras, 396, L76

\bibitem[{{Shen} {et~al}\mbox{.}(2003){Shen} S. {et~al.}}]{Shen2003}
{Shen} S. {et~al.}, 2003, \mnras, 343, 978

\bibitem[{{Silk} \& {Rees}(1998){Silk} J., {Rees} M.~J.}]{Silk1998}
{Silk} J., {Rees} M.~J., 1998, \aap, 331, L1

\bibitem[{Silverman(1986)Silverman B.~W.}]{silverman1986density}
Silverman B.~W., 1986, Density estimation for statistics and data analysis,
  Vol.~26. Chapman \& Hall/CRC, Boca Raton

\bibitem[{{Skrutskie} {et~al}\mbox{.}(2006){Skrutskie} M.~F.
  {et~al.}}]{Skrutskie2006}
{Skrutskie} M.~F. {et~al.}, 2006, \aj, 131, 1163

\bibitem[{{Smith}, {Lucey} \& {Carter}(2012){Smith} R.~J., {Lucey} J.~R.,
  {Carter} D.}]{Smith2012}
{Smith} R.~J., {Lucey} J.~R., {Carter} D., 2012, \mnras, 426, 2994

\bibitem[{{Sonnenfeld} {et~al}\mbox{.}(2012){Sonnenfeld} A.
  {et~al.}}]{Sonnenfeld2012}
{Sonnenfeld} A., {Treu} T., {Gavazzi} R., {Marshall} P.~J., {Auger} M.~W.,
  {Suyu} S.~H., {Koopmans} L.~V.~E., {Bolton} A.~S., 2012, \apj, 752, 163

\bibitem[{{Spiniello} {et~al}\mbox{.}(2012){Spiniello} C.
  {et~al.}}]{Spiniello2012}
{Spiniello} C., {Trager} S.~C., {Koopmans} L.~V.~E., {Chen} Y.~P., 2012, \apjl,
  753, L32

\bibitem[{{Spinrad} \& {Taylor}(1971){Spinrad} H., {Taylor}
  B.~J.}]{Spinrad1971}
{Spinrad} H., {Taylor} B.~J., 1971, \apjs, 22, 445

\bibitem[{{Spitzer} \& {Baade}(1951){Spitzer}, Jr. L., {Baade}
  W.}]{Spitzer1951}
{Spitzer}, Jr. L., {Baade} W., 1951, \apj, 113, 413

\bibitem[{{Suyu} {et~al}\mbox{.}(2012){Suyu} S.~H. {et~al.}}]{Suyu2012}
{Suyu} S.~H. {et~al.}, 2012, \apj, 750, 10

\bibitem[{{Tacconi} {et~al}\mbox{.}(2010){Tacconi} L.~J.
  {et~al.}}]{Tacconi2010}
{Tacconi} L.~J. {et~al.}, 2010, \nat, 463, 781

\bibitem[{{Thomas} {et~al}\mbox{.}(2005){Thomas} D.
  {et~al.}}]{Thomas2005daniel}
{Thomas} D., {Maraston} C., {Bender} R., {Mendes de Oliveira} C., 2005, \apj,
  621, 673

\bibitem[{{Thomas} {et~al}\mbox{.}(2009){Thomas} J. {et~al.}}]{Thomas2009dm}
{Thomas} J., {Saglia} R.~P., {Bender} R., {Thomas} D., {Gebhardt} K.,
  {Magorrian} J., {Corsini} E.~M., {Wegner} G., 2009, \apj, 691, 770

\bibitem[{{Thomas} {et~al}\mbox{.}(2011){Thomas} J. {et~al.}}]{Thomas2011}
{Thomas} J. {et~al.}, 2011, \mnras, 415, 545

\bibitem[{{Tonry} {et~al}\mbox{.}(2001){Tonry} J.~L. {et~al.}}]{Tonry2001}
{Tonry} J.~L., {Dressler} A., {Blakeslee} J.~P., {Ajhar} E.~A., {Fletcher}
  A.~B., {Luppino} G.~A., {Metzger} M.~R., {Moore} C.~B., 2001, \apj, 546, 681

\bibitem[{{Tortora} {et~al}\mbox{.}(2009){Tortora} C. {et~al.}}]{Tortora2009}
{Tortora} C., {Napolitano} N.~R., {Romanowsky} A.~J., {Capaccioli} M., {Covone}
  G., 2009, \mnras, 396, 1132

\bibitem[{{Tortora} {et~al}\mbox{.}(2010){Tortora} C. {et~al.}}]{Tortora2010}
{Tortora} C., {Napolitano} N.~R., {Romanowsky} A.~J., {Jetzer} P., 2010, \apjl,
  721, L1

\bibitem[{{Tortora}, {Romanowsky} \& {Napolitano}(2013){Tortora} C.,
  {Romanowsky} A.~J., {Napolitano} N.~R.}]{Tortora2012imf}
{Tortora} C., {Romanowsky} A.~J., {Napolitano} N.~R., 2013, \apj, 765, 8

\bibitem[{{Tremblay} \& {Merritt}(1996){Tremblay} B., {Merritt}
  D.}]{Tremblay1996}
{Tremblay} B., {Merritt} D., 1996, \aj, 111, 2243

\bibitem[{{Treu} {et~al}\mbox{.}(2010){Treu} T. {et~al.}}]{Treu2010}
{Treu} T., {Auger} M.~W., {Koopmans} L.~V.~E., {Gavazzi} R., {Marshall} P.~J.,
  {Bolton} A.~S., 2010, \apj, 709, 1195

\bibitem[{{Treu} {et~al}\mbox{.}(2011){Treu} T. {et~al.}}]{Treu2011}
{Treu} T., {Dutton} A.~A., {Auger} M.~W., {Marshall} P.~J., {Bolton} A.~S.,
  {Brewer} B.~J., {Koo} D.~C., {Koopmans} L.~V.~E., 2011, \mnras, 417, 1601

\bibitem[{{Treu} {et~al}\mbox{.}(2005){Treu} T. {et~al.}}]{Treu2005}
{Treu} T. {et~al.}, 2005, \apj, 633, 174

\bibitem[{{Trujillo} {et~al}\mbox{.}(2007){Trujillo} I.
  {et~al.}}]{Trujillo2007}
{Trujillo} I., {Conselice} C.~J., {Bundy} K., {Cooper} M.~C., {Eisenhardt} P.,
  {Ellis} R.~S., 2007, \mnras, 382, 109

\bibitem[{{Trujillo}, {Ferreras} \& {de La Rosa}(2011){Trujillo} I., {Ferreras}
  I., {de La Rosa} I.~G.}]{Trujillo2011}
{Trujillo} I., {Ferreras} I., {de La Rosa} I.~G., 2011, \mnras, 938

\bibitem[{{Trujillo} {et~al}\mbox{.}(2006){Trujillo} I.
  {et~al.}}]{Trujillo2006}
{Trujillo} I. {et~al.}, 2006, \mnras, 373, L36

\bibitem[{{Valentinuzzi} {et~al}\mbox{.}(2010){Valentinuzzi} T.
  {et~al.}}]{Valentinuzzi2010}
{Valentinuzzi} T. {et~al.}, 2010, \apj, 712, 226

\bibitem[{{van de Ven} {et~al}\mbox{.}(2010){van de Ven} G.
  {et~al.}}]{vandeVen2010}
{van de Ven} G., {Falc{\'o}n-Barroso} J., {McDermid} R.~M., {Cappellari} M.,
  {Miller} B.~W., {de Zeeuw} P.~T., 2010, \apj, 719, 1481

\bibitem[{{van den Bergh}(1976){van den Bergh} S.}]{vandenBergh1976}
{van den Bergh} S., 1976, \apj, 206, 883

\bibitem[{{van den Bosch}(1997){van den Bosch} F.~C.}]{vandenBosch1997}
{van den Bosch} F.~C., 1997, \mnras, 287, 543

\bibitem[{{van der Marel} \& {van Dokkum}(2007){van der Marel} R.~P., {van
  Dokkum} P.~G.}]{vanderMarel2007}
{van der Marel} R.~P., {van Dokkum} P.~G., 2007, \apj, 668, 756

\bibitem[{{van der Wel} {et~al}\mbox{.}(2009{\natexlab{a}}){van der Wel} A.
  {et~al.}}]{vanderWel2009}
{van der Wel} A., {Bell} E.~F., {van den Bosch} F.~C., {Gallazzi} A., {Rix}
  H.-W., 2009{\natexlab{a}}, \apj, 698, 1232

\bibitem[{{van der Wel} {et~al}\mbox{.}(2009{\natexlab{b}}){van der Wel} A.
  {et~al.}}]{vanderWel2009ab}
{van der Wel} A., {Rix} H.-W., {Holden} B.~P., {Bell} E.~F., {Robaina} A.~R.,
  2009{\natexlab{b}}, \apjl, 706, L120

\bibitem[{{van der Wel} {et~al}\mbox{.}(2011){van der Wel} A.
  {et~al.}}]{vanderWel2011}
{van der Wel} A. {et~al.}, 2011, \apj, 730, 38

\bibitem[{{van Dokkum}(2011){van Dokkum} P.}]{vanDokkum2011nat}
{van Dokkum} P., 2011, \nat, 473, 160

\bibitem[{{van Dokkum}(2008){van Dokkum} P.~G.}]{vanDokkum2008imf}
{van Dokkum} P.~G., 2008, \apj, 674, 29

\bibitem[{{van Dokkum} \& {Conroy}(2010){van Dokkum} P.~G., {Conroy}
  C.}]{vanDokkum2010}
{van Dokkum} P.~G., {Conroy} C., 2010, \nat, 468, 940

\bibitem[{{van Dokkum} \& {Conroy}(2011){van Dokkum} P.~G., {Conroy}
  C.}]{vanDokkum2011}
{van Dokkum} P.~G., {Conroy} C., 2011, \apjl, 735, L13

\bibitem[{{van Dokkum} \& {Conroy}(2012){van Dokkum} P.~G., {Conroy}
  C.}]{vanDokkum2012}
{van Dokkum} P.~G., {Conroy} C., 2012, \apj, 760, 70

\bibitem[{{van Dokkum} \& {Franx}(1996){van Dokkum} P.~G., {Franx}
  M.}]{vanDokkum1996}
{van Dokkum} P.~G., {Franx} M., 1996, \mnras, 281, 985

\bibitem[{{van Dokkum} {et~al}\mbox{.}(1998){van Dokkum} P.~G.
  {et~al.}}]{vanDokkum1998}
{van Dokkum} P.~G., {Franx} M., {Kelson} D.~D., {Illingworth} G.~D., 1998,
  \apjl, 504, L17

\bibitem[{{van Dokkum} {et~al}\mbox{.}(2008){van Dokkum} P.~G.
  {et~al.}}]{vanDokkum2008}
{van Dokkum} P.~G. {et~al.}, 2008, \apjl, 677, L5

\bibitem[{{van Dokkum} {et~al}\mbox{.}(2010){van Dokkum} P.~G.
  {et~al.}}]{vanDokkum2010profiles}
{van Dokkum} P.~G. {et~al.}, 2010, \apj, 709, 1018

\bibitem[{{Vazdekis} {et~al}\mbox{.}(2012){Vazdekis} A.
  {et~al.}}]{Vazdekis2012}
{Vazdekis} A., {Ricciardelli} E., {Cenarro} A.~J., {Rivero-Gonz{\'a}lez} J.~G.,
  {D{\'{\i}}az-Garc{\'{\i}}a} L.~A., {Falc{\'o}n-Barroso} J., 2012, \mnras,
  424, 157

\bibitem[{{Wake}, {van Dokkum} \& {Franx}(2012){Wake} D.~A., {van Dokkum}
  P.~G., {Franx} M.}]{Wake2012}
{Wake} D.~A., {van Dokkum} P.~G., {Franx} M., 2012, \apjl, 751, L44

\bibitem[{{Wegner} {et~al}\mbox{.}(2012){Wegner} G.~A. {et~al.}}]{Wegner2012}
{Wegner} G.~A., {Corsini} E.~M., {Thomas} J., {Saglia} R.~P., {Bender} R., {Pu}
  S.~B., 2012, \aj, 144, 78

\bibitem[{{Weijmans} {et~al}\mbox{.}(2013){Weijmans} A.-M.
  {et~al.}}]{Weijmans2013}
{Weijmans} A.-M. {et~al.}, 2013, \mnras\ submitted, (Paper~XXIII)

\bibitem[{{Wilkins} {et~al}\mbox{.}(2008){Wilkins} S.~M.
  {et~al.}}]{Wilkins2008}
{Wilkins} S.~M., {Hopkins} A.~M., {Trentham} N., {Tojeiro} R., 2008, \mnras,
  391, 363

\bibitem[{{Williams}, {Bureau} \& {Cappellari}(2009){Williams} M.~J., {Bureau}
  M., {Cappellari} M.}]{Williams2009}
{Williams} M.~J., {Bureau} M., {Cappellari} M., 2009, \mnras, 400, 1665

\bibitem[{{Williams}, {Bureau} \& {Cappellari}(2010){Williams} M.~J., {Bureau}
  M., {Cappellari} M.}]{Williams2010}
{Williams} M.~J., {Bureau} M., {Cappellari} M., 2010, \mnras, 409, 1330

\bibitem[{{Williams} {et~al}\mbox{.}(2010){Williams} R.~J.
  {et~al.}}]{Williams2010z2}
{Williams} R.~J., {Quadri} R.~F., {Franx} M., {van Dokkum} P., {Toft} S.,
  {Kriek} M., {Labb{\'e}} I., 2010, \apj, 713, 738

\bibitem[{{Wolf} {et~al}\mbox{.}(2010){Wolf} J. {et~al.}}]{Wolf2010}
{Wolf} J., {Martinez} G.~D., {Bullock} J.~S., {Kaplinghat} M., {Geha} M.,
  {Mu{\~n}oz} R.~R., {Simon} J.~D., {Avedo} F.~F., 2010, \mnras, 406, 1220

\bibitem[{{Worthey}(1994){Worthey} G.}]{Worthey1994}
{Worthey} G., 1994, \apjs, 95, 107

\bibitem[{{Wuyts} {et~al}\mbox{.}(2011){Wuyts} S. {et~al.}}]{Wuyts2011}
{Wuyts} S. {et~al.}, 2011, \apj, 742, 96

\bibitem[{{Wuyts} {et~al}\mbox{.}(2009){Wuyts} S. {et~al.}}]{Wuyts2009}
{Wuyts} S., {Franx} M., {Cox} T.~J., {Hernquist} L., {Hopkins} P.~F.,
  {Robertson} B.~E., {van Dokkum} P.~G., 2009, \apj, 696, 348

\bibitem[{{York} {et~al}\mbox{.}(2000){York} D.~G. {et~al.}}]{York2000}
{York} D.~G., {Adelman} J., {Anderson}, Jr. J.~E., {et al.}, 2000, \aj, 120,
  1579

\bibitem[{{Young} \& {Currie}(1994){Young} C.~K., {Currie} M.~J.}]{Young1994}
{Young} C.~K., {Currie} M.~J., 1994, \mnras, 268, L11

\bibitem[{{Young} {et~al}\mbox{.}(2011){Young} L.~M. {et~al.}}]{Young2011}
{Young} L.~M. {et~al.}, 2011, \mnras, 414, 940 (Paper~IV)

\bibitem[{{Zaritsky}, {Gonzalez} \& {Zabludoff}(2006){Zaritsky} D., {Gonzalez}
  A.~H., {Zabludoff} A.~I.}]{Zaritsky2006}
{Zaritsky} D., {Gonzalez} A.~H., {Zabludoff} A.~I., 2006, \apj, 638, 725

\bibitem[{{Zepf} \& {Silk}(1996){Zepf} S.~E., {Silk} J.}]{Zepf1996}
{Zepf} S.~E., {Silk} J., 1996, \apj, 466, 114

\bibitem[{{Zhu}, {Blanton} \& {Moustakas}(2010){Zhu} G., {Blanton} M.~R.,
  {Moustakas} J.}]{Zhu2010}
{Zhu} G., {Blanton} M.~R., {Moustakas} J., 2010, \apj, 722, 491

\end{thebibliography}
\end{document}